\documentclass[12pt]{article}
\pdfoutput=1
\usepackage{jheppub}
\usepackage{amsmath}
\usepackage{amssymb}
\usepackage{wrapfig}
\usepackage{bbm}
\usepackage{arydshln}
\usepackage{multirow}
\usepackage{mathtools}
\usepackage{blkarray}
\usepackage{setspace}
\usepackage{dsfont}
\usepackage{float}
\usepackage{subcaption}

\allowdisplaybreaks

\title{Webs of W-algebras}

\abstract{We associate vertex operator algebras to $(p,q)$-webs of interfaces in the topologically twisted $\mathcal{N}=4$ super Yang-Mills theory. Y-algebras associated to trivalent junctions are identified with truncations of $\mathcal{W}_{1+\infty}$ algebra. Starting with Y-algebras as atomic elements, we describe gluing of Y-algebras analogous to that of the topological vertex. At the level of characters, the construction matches the one of counting D0-D2-D4 bound states in toric Calabi-Yau threefolds.  For some configurations of interfaces, we propose a BRST construction of the algebras and check in examples that both constructions agree. We define generalizations of $\mathcal{W}_{1+\infty}$ algebra and identify a large class of glued algebras with their truncations. The gluing construction sheds new light on the structure of vertex operator algebras conventionally constructed by BRST reductions or coset constructions and provides us with a way to construct new algebras.  Many well-known vertex operator algebras, such as $U(N)_k$ affine Lie algebra, $\mathcal{N}=2$ superconformal algebra, $\mathcal{N}=2$ super-$\mathcal{W}_\infty$, Bershadsky-Polyakov $\mathcal{W}_3^{(2)}$, cosets and Drinfeld-Sokolov reductions of unitary groups can be obtained as special cases of this construction.}

\author[a]{Tom\'{a}\v{s} Proch\'{a}zka,}
\author[b]{Miroslav Rap\v{c}\'{a}k}
\affiliation[a]{Arnold Sommerfeld Center for Theoretical Physics,\\ Ludwig Maximilian University of Munich,\\
Theresienstr. 37, D-80333 M\"unchen, Germany}
\affiliation[b]{Perimeter Institute for Theoretical Physics,\\
31 Caroline St N, Waterloo, ON N2L 2Y5, Canada}
\emailAdd{Tomas.Prochazka@lmu.de}
\emailAdd{miroslav.rapcak@gmail.com}

\begin{document}
\maketitle

\section{Introduction}

Recently, a four-parameter class of vertex operator algebras $Y_{K,L,M}[\Psi]$ for integral parameters $K,L,M$ and a continuous parameter $\Psi$ were introduced in \cite{Gaiotto:2017euk}.\footnote{See also \cite{Bershtein} for different construction of the same algebras.} These algebras were associated to the configuration of three stacks of D3-branes attached to the trivalent junction of NS5, D5 and (1,1) branes from the three corners. Y-algebras were then identified as algebras of local operators appearing at a trivalent junction of three interfaces between topologically twisted $U(N)$, $U(M)$, and $U(L)$ gauge theories coming from the low energy limit of the theories on D3-branes. The gauge theory of interest is the geometric Langlands twist of the $\mathcal{N}=4$ super Yang-Mills theory \cite{Kapustin:aa}. 

This paper discusses an identification of Y-algebras with truncations of $\mathcal{W}_{1+\infty}$ and uses them as building blocks to construct more complicated vertex operator algebras. At the level of characters it is analogous to the topological vertex and to the problem of counting of D0-D2-D4 bound states in the toric Calabi-Yau manifolds \cite{Aganagic:2003db,Aganagic:2005wn,Jafferis:2006ny}. This construction leads to new insights on the structure of vertex operator algebras and provides us with a physical realization of the algebras using the brane setups.

\subsection{Y-algebras as truncations of $\mathcal{W}_\infty$}

Starting with the Virasoro algebra and additional primary generators of each integral spin $3,4,\dots$, Gaberdiel and Gopakumar \cite{Gaberdiel:2012aa} have shown that there exists a two-parameter family of algebras $\mathcal{W}_{\infty}[c,\lambda]$ satisfying Jacobi identities.\footnote{See also \cite{Hornfeck:1994is} where the algebra was defined under the name $\mathrm{WA}^n$.} One can identify special curves in the parameter space at which the algebra $\mathcal{W}_\infty$ contains an ideal $\mathcal{I}$. If we quotient out the ideal, we obtain a truncation of the algebra $\mathcal{W}_\infty/\mathcal{I}$. Some of the ideals were identified already in \cite{Gaberdiel:2012aa} with $\mathcal{W}_{N}$ algebras. The structure of ideals has further been analyzed in \cite{Prochazka:2014aa,Prochazka:2015aa} where new truncations were discovered. Based on the structure of vacuum characters, it has already been anticipated in \cite{Gaiotto:2017euk} that there exists a relation between Y-algebras and truncations of $\mathcal{W}_{1+\infty} \equiv \mathcal{W}_{\infty}\times U(1)$.

In chapter \ref{sec:Y-algebras}, we review the story behind both Y-algebras (in particular their defition in terms of a BRST reduction) and truncations of $\mathcal{W}_{1+\infty}$. We establish an identification between these two by matching the central charges and the vacuum characters of $Y_{K,L,M}[\Psi]$ and $\mathcal{W}_{1+\infty}/\mathcal{I}_{K,L,M}$. The vacuum characters can be identified with a generating function of 3d partitions (plane partitions) constrained to fit under the corner shifted by a vector $(K,L,M)$. We show that parameters of $\mathcal{W}_{1+\infty}$ algebra are invariant under the shift of all $L,M,N$ by a constant value $L,M,N\rightarrow L+k,M+k,N+k$. All these shifted algebras correspond to the same truncation curve in the two-parameter space of $\mathcal{W}_{1+\infty}$ algebras. Above each truncation curve, one gets a sequence of truncations corresponding to shifts by a positive integer $k$. 

Apart from the match of the central charges and the vacuum characters there are the following arguments supporting the identification of $Y_{L,M,N}$ and truncations of $\mathcal{W}_{1+\infty}$:
\begin{enumerate}
\item The basic examples of the $\mathcal{W}_N\times U(1)$ truncations can be identified from the BRST definition of $Y_{0,0,N},Y_{0,N,0}$ and $Y_{N,0,0}$ that reduces to the standard Drinfeld-Sokolov and coset constructions of the algebra. Note that the BRST definition of a generic $Y_{L,M,N}$ provides a BRST construction of all the other truncations.
\item The triality transformations from \cite{Gaberdiel:2012aa,Prochazka:2014aa,Prochazka:2015aa} can be identified with the triality transformations of \cite{Gaiotto:2017euk}. 
\item There are three families of modules associated to line operators inserted at the interfaces between the three gauge theories. From the point of view of $\mathcal{W}_{1+\infty}$, they are the modules associated to the three asymptotic directions in the parameter space. They are naturally permuted by the action of the triality. These modules play an important role in the gluing construction discussed below.
\end{enumerate}

\subsection{Gluing Y-algebras}

Instead of the simple trivalent junction, one can consider a more general configuration of a web of $(p,q)$-branes \cite{Aharony:1997bh} and stacks of D3-branes attached to them. From the point of view of the theory on D3-branes, this setup gives rise to the junction of interfaces (descending from D3-branes ending on five-branes) between $U(N_i)$ $\mathcal{N}=4$ super Yang-Mills theories living on stacks of $N_i$ D3-branes. If we look at the same system from the IR, the finite segments of five-branes degenerate and the line operators supported at these interfaces become effectively local operators living at the corner. It is natural to add them into the final vertex operator algebra associated to the configuration of branes. The total vertex operator algebra is thus an extension of a tensor product of algebras associated to each trivalent junction by bimodules of these algebras (and their fusions) associated to the finite five-brane segments. It turns out that the bimodules that need to be added have (half-) integral conformal dimension with respect to the total stress energy tensor of the vertex operator algebra and can indeed be added to the algebra. We give a prescription for such gluing in the case when each of the trivalent junctions inside the $(p,q)$-web can be brought to the elementary trivalent junction corresponding to Y-algebra by an $SL(2,\mathbbm{Z})$ transformation.

In examples, we mostly concentrate on configurations of defects descending from D5-branes ending on $(n,1)$-branes. We expect that the path integral of the $\mathcal{N}=4$ super Yang-Mills theories living at the worldvolume of D3-branes localizes to the path integral of the supergroup Chern-Simons theories supported at the $(n,1)$-interfaces \cite{Gaiotto:2017euk,Mikhaylov:2017ngi,Witten:2010aa,Witten:2011aa,Mikhaylov:2014aa}. These Chern-Simons theories are glued together by boundary conditions following from the boundary conditions descending  from D3-branes ending of D5-branes analyzed in \cite{Gaiotto:2008ac,Gaiotto:2008ab,Gaiotto:2008aa}. For some numbers of D3-branes, the resulting algebra can be given by a BRST construction following \cite{Nekrasov:2010aa,Gaiotto:2011nm,Yagi:2014toa,Gaiotto:2017euk}. This provides us with another insight into the structure of the glued algebras. We show in many examples that the vacuum characters of the BRST construction agrees with those obtained by gluing and that the central charge of the total algebra is simply the sum of the central charges of Y-algebras associated to the trivalent vertices. We check that the full algebras decompose correctly in the case of the  $\mathcal{N}=2$ super Virasoro, the Bershadsky-Polyakov $\mathcal{W}_3^{(2)}$ algebra and the $U(N)_k$ Kac-Moody algebra in the way predicted by gluing.

The algebras associated to the configuration of D5-branes ending on $(n,1)$-branes discussed above can be identified with truncations of infinitely generated $\mathcal{W}$-algebras. Fixing a discrete data which is a $(p,q)$-web and a (half-) integral number $\rho_i$ for each internal edge of the web diagram, one gets a family of algebras parametrized by two continuous parameters as in the case of $\mathcal{W}_{1+\infty}$ algebra. Each parameter $\rho_i$ associated to an internal edge gives one linear constraint on the numbers of D3-branes surrounding the corresponding edge and we are free to choose three integer parameters to fully determine the brane configuration. These parametrize truncations of the corresponding infinite algebra in the same way as $L,M,N$ parametrized truncations of $\mathcal{W}_{1+\infty}$. In the case of $\rho_i=0$, these algebras can be identified with those of \cite{Costello:2016nkh} and contain a $U(M|N)$ matrix of generators of each integral spin $1,2,3,\dots$. Turning on the parameters $\rho_i$ shifts the conformal dimensions of the off-diagonal generators in the $U(M|N)$ matrices. For example, the infinitely generated $\mathcal{W}$-algebra associated to the resolved conifold diagram with $\rho=\frac{1}{2}$ can be identified with $\mathcal{N}=2$ super $\mathcal{W}_{\infty}$ of \cite{Candu:2012tr}. One can understand the truncations of more complicated infinite $\mathcal{W}$-algebras as being glued from truncations of the basic building block $\mathcal{W}_{1+\infty}$.

\subsection{What do we learn about VOAs?}

Let us now list a set of insights about vertex operator algebras that gluing construction provides:
\begin{enumerate}
\item There are various ways of constructing VOAs, such as various BRST reductions, coset constructions, free field realizations, or bootstrap for a given spin content. The gluing construction provides us with a new one. To each web diagram with stacks of D3-branes attached, one associates a tensor product of mutually commuting Y-algebras associated to the vertices of the diagram. Their OPE structure can be identified from \cite{luk1988quantization,Prochazka:2014aa} by specializing the parameters to the corresponding truncation curve. One adds bimodules associated to internal edges. These are the universal building blocks for each diagram. In the last step, one needs to find OPEs of the bimodule fields. To our knowledge the structure of such modules and their OPEs have not been constructed yet but their construction should be possible using bootstrap or the Coulomb gas formalism \cite{Dotsenko:1984nm,Dotsenko:1984ad,Dotsenko:1985hi}.
\item It turns out that many well known algebras can be obtained as special cases of the gluing construction, i.e. they are conformal extensions of a product of Y-algebras by bimodules. Some examples discussed in this paper are
\begin{itemize}
\item $\mathcal{N}=2$ superconformal algebra is a conformal extension
\begin{eqnarray}
\mathcal{N}=2 \enskip \mbox{SCA}\times U(1) \supset Y_{1,0,2} \times Y_{0,1,0}
\end{eqnarray}
\item $U(N)_k$ can be decomposed as a conformal extension of
\begin{eqnarray}
U(N)_k \supset Y_{0,0,1} \times Y_{0,1,2} \times \ldots \times Y_{0,N-1,N}.
\end{eqnarray}
Similar expression exists also for the $U(N|M)_k$ super Kac-Moody algebras.
\item Many (non-principal) DS-reductions can be decomposed in a similar way. An example of such a decomposition is the $\mathcal{W}^{(2)}_3$ algebra
\begin{eqnarray}
\mathcal{W}_3^{(2)}\times U(1)  \supset Y_{0,1,3} \times Y_{0,0,1}.
\end{eqnarray}
\item The super Kac-Moody algebra $D(2,1;\alpha)_1$ can be decomposed in two ways
\begin{eqnarray}\nonumber
D(2,1;\alpha)_1 \times U(1) & \supset & Y_{1,1,2} \times Y_{1,1,0} \times Y_{0,0,1} \times Y_{0,0,1} \\
D(2,1;\alpha)_1 \times U(1) & \supset & Y _{0,2,1} \times Y_{2,0,1} \times Y_{0,0,1} \times Y_{0,0,1}.
\end{eqnarray}
\end{itemize} 
We comment on many more examples in the main text. 
\item The total stress-energy tensor of the glued algebra is a sum of stress-energy tensors coming from the trivalent junctions. From this it follows that the central charge of the final VOA is given by the sum of the central charges of the Y-algebras associated to the vertices. This provides us with a simple way to compute the central charge directly from the web diagram. In the case when the algebra can be given a BRST definition, the equality of the central charge of the resulting algebra with the one comming from the sum of the Y-algebra central charges provides us with a non-trivial check of the equivalence of the two constructions.
\item From a given diagram, dual BRST constructions of the algebras and duality actions on the parameter space of the corresponding infinitely generated algebras can be easily discovered. For example, in the same way as the triality symmetry of Y-algebras was discovered in \cite{Gaiotto:2017euk}, one can identify $\mathbbm{Z}_2\times \mathbbm{Z}_2$ duality action on algebras associated to the resolved conifold diagram. This duality (which we expect to be valid for any value of the parameter $\rho$) generalizes the duality of \cite{Candu:2012tr} which is a special case of $\rho=\frac{1}{2}$.
\item The structure of modules can be understood in terms of the web diagrams. In particular, one can associate a family of degenerate modules to line operators supported at each semi-infinite five-brane and ending at junctions (i.e. associated to each external leg of the diagram). Modules from different families should braid trivially and the corresponding highest weights states should be charged only under the Y-algebra associated to the trivalent junction associated to the corresponding external leg. If the configuration admits a duality action, the families of modules should permute accordingly.
\item Fixing a web configuration, one obtains different VOAs for different choices of numbers of D3-branes attached to the fixed five-branes. We can study various limits where the number of D3-branes becomes infinite. In this way we obtain infinite $\mathcal{W}$ algebras parametrized by a combination of continuous and discrete parameters coming from the number of D3-branes before taking the limit and the relative orientations of the vertices. Analogously to the case of $Y_{L,M,N}$, diagrams with finite numbers of D3-branes should correspond to truncations of these infinitely generated $\mathcal{W}$ algebras.
\item In the context of the topological vertex and BPS counting, the flip transitions play an important role. In our setup, such flip transitions correspond to sliding D5-branes along $(n,1)$ branes. We conjecture that algebras related by a flip transition differ only by decoupled fermions and symplectic bosons. We show it is the case on a simple example of the diagram associated to the $\mathcal{N}=2$ super Virasoro algebra, $\mathcal{W}_3^{(2)}$ algebra and the diagram associated to the flip of $U(1)$ and Virasoro algebras. We also conjecture that in the case of vanishing parameter $\rho_i=0$ of a segment at which we perform the flip transition, both algebras are the same. In the case of algebras with BRST definition is our conjecture supported by calculations of the central charge and the vacuum characters.
\item Since the basic building block $Y_{L,M,N}$ can be thought of as an algebra of Yangian type associated to $\hat{\mathfrak{u}}(1)$ \cite{Schiffmann:2012gf,Maulik:2012rm,tsymbaliuk2017affine,Prochazka:2015aa}, we get for free an interesting integrable structure. In particular, we have an infinite sets of commuting charges coming from the subalgebras associated to the vertices.
\item Physical realization of the algebras suggests applications of the algebras in many places in physics and mathematics such as AGT correspondence, action of VOAs on equivariant cohomologies of moduli spaces of instantons, the geometric Langlands program and many others. These relations still remain to be explored.
\end{enumerate}

\subsection{D0-D2-D4 counting}

There exists a natural duality along the lines of \cite{Leung:1997tw} relating our brane configuration to the one used in the context of the D4-D2-D0 brane counting from \cite{Jafferis:2006ny,Aganagic:2005wn,Aganagic:2012si}. Let us first review the configuration relevant to the counting of D0-D2-D4 bound states in a toric Calabi-Yau three-fold.  Consider the type IIA string theory on a manifold $M^{10}=CY^3\times R^4$ where $CY^3$ is the toric Calabi-Yau manifold that can be viewed as a $T^2\times R$ fibration over $R^3$ with various cycles of $T^2$ shrinking at various codimension one loci of the base $R^3$. Let us introduce D4-branes supported at four-cycles, D2-branes supported at two-cycles and D0-branes supported at points of $CY^3$ that are fixed under the $T^2$ action. All the branes are extended along one of the additional four directions. D4-branes intersect at codimension two defects in their world-volumes. The theory on D4-branes are gauge theories coupled together by extra bi-fundamental matter fields at the loci where the branes intersect as discussed in \cite{Nekrasov:2016qym,Nekrasov:2016gud}. From the point of view of the theory on these intersecting D4-branes, D2- and D0- branes modify the gauge bundle of the gauge theories supported at D4-branes. Fixing the numbers of these branes then corresponds to restricting the path integral of the configuration of D4-branes to a particular instanton sector of gauge field configurations. The characters of \cite{Jafferis:2006ny,Aganagic:2005wn,Aganagic:2012si} are functions of the parameter $q$ corresponding to the fugacity for the D0-charge and $Q_j$ corresponding to the fugacities for the D2 charge.

The configuration of D0-D2-D4 branes discussed above has a natural lift to the M-theory on $M^{11}=CY^3\times R^4\times S^1$. Let us discuss what is the lift of various branes. D4-branes become M5-branes wrapping the same holomorphic four cycle inside CY$^3$ but now also wrapping the M-theory circle $S^1$. On the other hand, D2-branes lift to M2-branes supported at the same holomorphic cycles of CY$^3$ as before. D0-branes are the KK-modes on the M-theory circle.

Following \cite{Leung:1997tw}, we can relate the M-theory configuration to our setup. Loci in the base of the $T^2\times R$ fibration of $CY^3$ where various cycles of $T^2$ shrink give rise to a web of five-branes supported at $R^4\times S^1$ where $S^1$ is the original M-theory circle. M5-branes become D3-branes ending on the five-branes and supported at faces of the $(p,q)$-web. M2-branes reduce to D1 supported at $R^1$ inside $R^4\times S^1$ and one of the directions descending from the base of the $T^2\times R^1$ fibration. This is exactly the configuration used for calculations in this paper. The D1-branes supported at the interfaces and ending at junctions correspond to line operators in gauge theory giving rise to modules for the algebras. Summing over all the possible D1-brane charges then corresponds to summing over bimodules associated to line operators supported at the finite five-brane segments.

In the formulas presented here, we recover expressions from \cite{Jafferis:2006ny,Aganagic:2005wn,Aganagic:2012si} for $Q=1$.  We expect the parameter $Q$ to be related to the $U(1)$ charges of the added bimodules associated to the $U(1)$ currents appearing at each trivalent junction. These indeed measure $U(1)$ charges of corresponding line operators that descent  (in the brane picture) from D1-branes as discussed above.

The characters of Y-algebras as atomic elements of the gluing agree with those of Jafferis in \cite{Jafferis:2006ny} who proposed the same box counting interpretation. The gluing proposal at the level of vacuum characters matches the one proposed in \cite{Aganagic:2005wn,Aganagic:2012si}. On the other hand, gluing at the level of full algebras seems to categorify these BPS counting problems. 

\section{Y-algebras and $\mathcal{W}_{\infty}$}
\label{sec:Y-algebras}

\begin{wrapfigure}{l}{0.325\textwidth}
\vspace{-18pt}
  \begin{center}
      \includegraphics[width=0.32\textwidth]{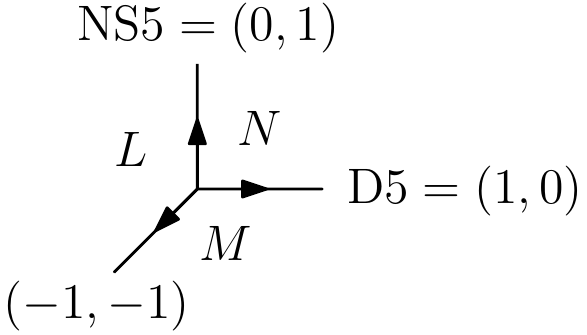}
\end{center}
\vspace{-20pt}
\end{wrapfigure}

In \cite{Gaiotto:2017euk}, $Y_{L,M,N}[\Psi]$ algebras were defined as vertex operator algebras associated to the junction of NS5, D5, and $(1,1)$ branes.\footnote{The configuration might need to be deformed by turning on fluxes.} Parameters $L,M,N$ label numbers of D3-branes attached to the trivalent junction from different sides as shown in the figure. From the point of view of the theory on D3-branes, this leads to $U(L),U(M),U(N)$ gauge theories connected by domain walls descending from five-branes on which D3-branes end. Vertex operator algebras arise as algebras of local operators living at the corner in the Kapustin-Witten twist of the theory. The parameter $\Psi$ is the canonical parameter of \cite{Kapustin:aa} that labels the Kapustin-Witten twisted $\mathcal{N}=4$ SYM theories and plays the role of the level of Kac-Moody algebras used in definition of $Y_{L,M,N}[\Psi]$. From the point of view of the $\mathcal{N}=4$ SYM, $\Psi$ is a combination of the complexified gauge coupling and the twisting parameter $t$. Y-algebras are defined as a BRST reduction of a system of Kac-Moody algebras for super unitary groups and ghost systems. In this section, we review their definition and we identify them with truncations of $\mathcal{W}_{1+\infty}$.

\subsection{Definition of Y-algebras}

Y-algebras were defined as a combination of the Drinfeld-Sokolov reduction and the coset construction of a supergroup Kac-Moody algebra. Schematically, they are defined as\footnote{Throughout the paper, we use the notation $U(N|L;\Psi)=U(1)_{(N-L)\Psi} \times SU(N|L)_{\Psi-N+L}$, where $\Psi-N+L$ is the level of the $SU(N|L)$ Kac-Moody subalgebra, i.e. $\Psi$ is the level relative to the critical level. Although $U(1)$ current algebra does not have any intrinsic level, we use the subscript to indicate the normalization of the $U(1)$ current with respect to which the electric modules have integral dimensions. For more details consult appendix \ref{conventions}.}
\begin{eqnarray}\nonumber
Y_{L,M,N}[\Psi] & = & \frac{\mathcal{DS}_{N-M}[U(N|L;\Psi)]}{U(M|L;\Psi-1)}\qquad \hspace{20pt}\mbox{for}\ N>M\\ \nonumber
Y_{L,N,N}[\Psi] & = & \frac{U(N|L;\Psi)\times \mathcal{S}^{N|L}}{U(N|L;\Psi-1)}\\
Y_{L,M,N}[\Psi] & = & \frac{\mathcal{DS}_{M-N}[U(M|L;-\Psi+1)]}{U(N|L;-\Psi)}\qquad \mbox{for}\ N<M
\end{eqnarray}
where $\mathcal{DS}_{N-M}$ denotes the Drinfeld-Sokolov reduction with respect to $(N-M)\times (N-M)$ diagonal block of $U(N|L)$ and by the division by $U(M|L;\Psi-1)$ we mean the BRST coset to be defined later. $\mathcal{S}^{N|L}$ labels the set of $N$ symplectic bosons and $L$ free fermions that contains a $U(N|L;N-L-1)$ subalgebra formed from the field bilinears (see appendix \ref{conventions}). More concretely, for parameters in the range $N>M$, $Y_{L,M,N}[\Psi]$ is defined as the BRST reduction of the complex
\begin{eqnarray}
U(N|L;\Psi) \times U(M|L; -\Psi+1) \times gh^{(DS)} \times gh^{(coset)}
\end{eqnarray}
by two successive BRST reductions. In the complex above, we have introduced $gh^{(DS)}$ for (super)ghosts needed for the Drinfeld-Sokolov reduction implemented by $Q^{(DS)}_{BRST}$ and $gh^{(coset)}$ for (super)ghosts associated to the BRST coset implemented by $Q^{(coset)}_{BRST}$. 

$Q^{(DS)}_{BRST}$ can be defined in the following three steps (assuming $N>M$):
\begin{enumerate}
\item Pick the principal $SU(2)$ embedding inside the $U(N-M)$ subalgebra associated to the $(N-M)\times (N-M)$ block inside $U(N|L)$. The corresponding Cartan generator of such embedding can be taken to be of the form
\begin{eqnarray}
H=\frac{N-M-1}{2}E_{11} + \frac{N-M-3}{2} E_{22} + \dots + \frac{M-N+1}{2}E_{N-M,N-M}
\end{eqnarray}
where $E_{ij}$ is a generator of the $U(N)$ Lie algebra associated to the matrix with one at the position $i,j$. The generator $H$ provides us with a grading that we use in the next step.
\item Decompose the adjoint representation of $U(N|L)$ into subspaces of $H$-charge greater then, equal to, and smaller than one half: $g_{<\frac{1}{2}} \oplus g_{\frac{1}{2}} \oplus g_{>\frac{1}{2}}$. Introduce fermionic $bc$ ghosts for each bosonic element and bosonic $\beta\gamma$ ghosts for each fermionic element in $g_{>\frac{1}{2}}$ and for half of the elements in $g_{\frac{1}{2}}.$\footnote{This half of the elements needs to to be picked such that they form a Lagrangian subspace inside $g_{\frac{1}{2}}$ with respect to the symplectic pairing given by the standard invariant two-form of $SU(N)$.} This system of (super)ghosts is labeled by $gh^{(DS)}$.\footnote{The conformal dimensions of such ghosts are $h(c^i)=h(\gamma^i)=1-h(b_i)=1-h(\beta_i)=1-h(J_i)=-H(J_i)$ where $H(J_i)$ is the $H$-charge of the element $J_i$. This assignment of conformal dimensions ensures that the BRST charge has degree one with respect to the modified stress-energy tensor of the Drinfeld-Sokolov reduction and it is useful to count the contribution from the ghosts in the total stress-energy tensor.}
\item Define a nilpotent BRST charge $Q_{BRST}^{(DS)}$ constraining $g_{>\frac{1}{2}}$ and half of $g_{\frac{1}{2}}$ generators to a fixed value
\begin{eqnarray}
Q_{BRST}^{(DS)}=\oint dz \left[ (J_i-t^+_i)c^i+\frac{1}{2}f_{ij}^{k}b_k c^i c^j \right]
\end{eqnarray}
where $t^+$ is the raising operator of the $SU(2)$ embedding. In our conventions this generator has zeros and ones above the diagonal and $f_{ij}^{k}$ are the structure constants of the algebra of constraints (restrictions of the structure constants of the $U(N|L)$ Lie algebra).  For an explicit example of such a constraint see appendix \ref{BRSTexamples}.
\end{enumerate}

The coset BRST reduction is then performed by adding (super)ghosts of conformal dimension $h(c^i)=h(\gamma^j)=h(b_i)-1=h(\beta_j)-1=0$, one for each generator of $U(M|L)$. We denote this (super)ghost system by $gh^{(coset)}$ and study the cohomology with respect to
\begin{eqnarray}
Q_{BRST}^{(coset)}=\oint dz \left [(J^{1}_j-J^{2}_j)c^j+\frac{1}{2}f_{jk}^{l} b_l c^j c^k \right].
\end{eqnarray}
Here $J^\alpha_j$ are the currents of the two copies of $U(M|L)$ algebra being sewed and $f_{ij}^{k}$ are the structure constants of $U(M|L)$.\footnote{The upper index $\alpha=1,2$ runs over the two copies of algebra while indices $j,\ldots$ run over the adjoint representation of $U(M|L)$.} For the notational simplicity we wrote the formula as if there were only bosonic generators and fermionic ghosts, but the generalization should be obvious.

In the case when $N-M=1$, the $\mathcal{DS}_1$ is a trivial operation and can be omitted. On the other hand, if $N=M$, one needs to add symplectic bosons $\mathcal{S}^{N|L}$ in the fundamental representation of $U(N|L)$. These are known to contain a conformally embedded $U(N|L;N-M-1)$ Kac-Moody algebra formed by their bilinears. The resulting Y-algebra can be identified with the BRST reduction of the complex
\begin{eqnarray}
U(N|L;\Psi) \times \mathcal{S}^{N|L} \times U(N|L;-\Psi+1) \times gh^{(coset)}
\end{eqnarray}
by the BRST charge
\begin{eqnarray}
Q^{(coset)}_{BRST}=\oint dz\left [c^i(J^1_i-J_i^2-J_i^{\mathcal{S}})+\frac{1}{2}f_{ij}^{k} b_k c^i c^j\right ]
\end{eqnarray}
where $J^{\mathcal{S}}$ are the $U(N|L)$ currents obtained from the bilinears in $\mathcal{S}^{N|L}$ fields. Intuitively, this BRST operator couples the symplectic bosons to the two Chern-Simons theories connected by the interface.

In the following, we will use the unified notation
\begin{eqnarray}
\mathcal{DS}_{N-M}[U(N|L;\Psi)]
\end{eqnarray}
for any non-negative $N-M$ that is defined by the DS-reduction described above for $N-M>1$, that is trivial in the case of $N-M=1$, and that produces
\begin{eqnarray}
\mathcal{DS}_{0}[U(N|L)_\Psi]=U(N|L)_\Psi \times \mathcal{S}^{N|L}
\end{eqnarray}
in the case that $N=M$.

\subsection{$\mathcal{W}_{1+\infty}$ and its truncations}

The vertex operator algebra $\mathcal{W}_{\infty}$ is the algebra obtained by extending the Virasoro algebra by independent primary fields of each integral spin $\geq 3$, so that the generators are
\begin{eqnarray}
T,W_3,W_4,W_5\dots
\end{eqnarray}
Imposing the conditions of associativity, \cite{Hornfeck:1994is,Gaberdiel:2012aa} concluded that there exists a two parameter family of such algebras, one parameter being the central charge $c$ and the other one can be chosen to be
\begin{equation}
x^2=\frac{(C^4_{33})^2 C_{44}^0}{(C_{33}^0)^2}
\label{parameterxsq}
\end{equation}
where $C_{jk}^l$ are the OPE coefficients ($C_{jk}^l$ is the coefficient of primary operator $W_l$ in the OPE of $W_j$ and $W_k$)\footnote{Although starting from spin $6$ the primary operators are not uniquely determined even up to an overall rescaling, there is no such problem with primaries of spin $3$ or $4$.}.

It is convenient to add a decoupled $U(1)$ current into the algebra and define $\mathcal{W}_{1+\infty} \equiv U(1) \times \mathcal{W}_{\infty}$. At special curves in the two-parameter space of such algebras, $\mathcal{W}_{1+\infty}$ develops an ideal $\mathcal{I}$. Quotienting this ideal out, one obtains a truncation of $\mathcal{W}_{1+\infty}$. According to \cite{Gaberdiel:2012aa}, some of such truncations can be identified with $\mathcal{W}_N\times U(1)$ algebras generated by fields up to spin $N$. The structure of truncations of $\mathcal{W}_{1+\infty}$ was further analyzed in \cite{Prochazka:2014aa} where new truncations were discovered. It turns out that Y-algebra can be identified with these more general truncations of $\mathcal{W}_{1+\infty}$.

As pointed out in \cite{Prochazka:2014aa}, there exists an useful parametrization of the structure constants in terms of a triple of parameters $\lambda_i$ satisfying
\begin{eqnarray}
\frac{1}{\lambda_1}+\frac{1}{\lambda_2}+\frac{1}{\lambda_3} = 0
\label{relation1}
\end{eqnarray}
in terms of which the central charge and parameter (\ref{parameterxsq}) are given by
\begin{eqnarray}\nonumber
c_{\infty} & = & (\lambda_1-1)(\lambda_2-1)(\lambda_3-1) \\
x^2 & = & \frac{144(c+2)(\lambda_1-3)(\lambda_2-3)(\lambda_3-3)}{(\lambda_1-2)(\lambda_2-2)(\lambda_3-2)}.
\label{cc2}
\end{eqnarray}
Modifying the stress energy tensor in such a way that the current $J$ has conformal weight one, the central charge get shifted by one $c_{1+\infty}=c_{\infty}+1$. The reason for introducing this parametrization is that for $\lambda_j = N$ where $N$ is any positive integer, the algebra truncates to $\mathcal{W}_N\times U(1)$. Although the structure constants of the algebra in the primary basis are manifestly invariant under $S_3$ transformation permuting the parameters $\lambda_j$, this triality symmetry acts non-trivially on representations. We might as well analytically continue the structure constants of $\mathcal{W}_N\times U(1)$ as a function of the rank parameter $N$ (since with a suitable choice of normalization they are just rational functions of $N$ and $c$) and find following Gaberdiel and Gopakumar \cite{Gaberdiel:2012aa} that for a fixed value of the central charge $c$, there are generically three different values $\lambda_j$ of $N$ for which we get the same structure constants.

\begin{figure}[h]
  \centering
      \includegraphics[width=0.36\textwidth]{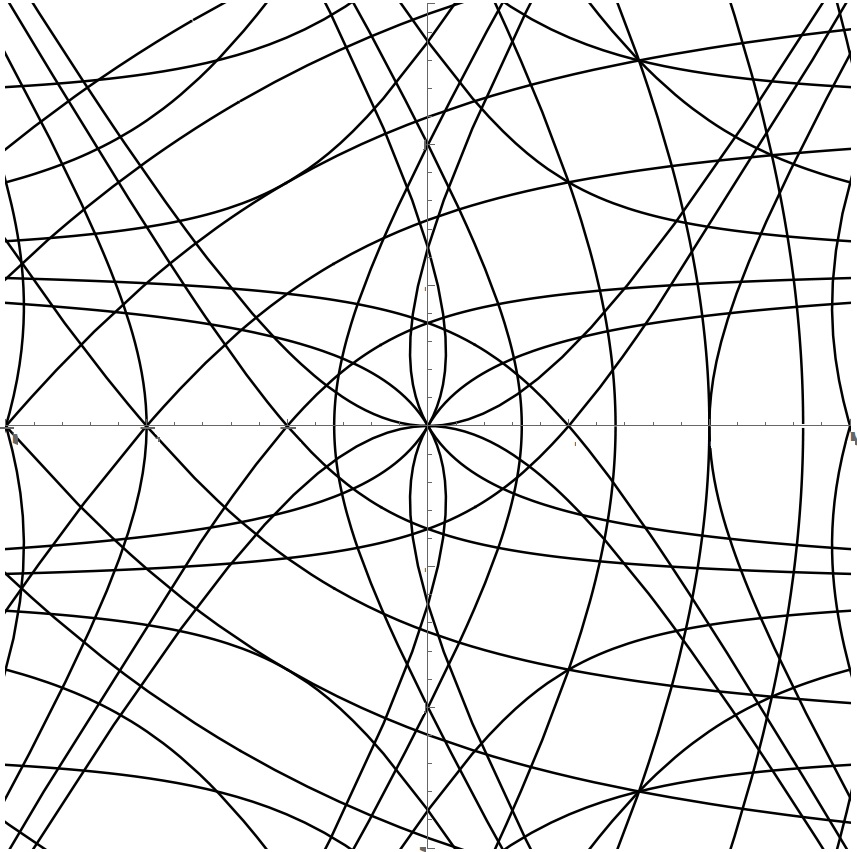}
 \caption{Truncation curves parametrized by $(L,M,N)$ such that $(L+1)(M+1)(N+1) \leq 6$. This restriction means that the first generator that we are quotienting appears at level $\leq 6$ in the vacuum module. We use the parametrization from \cite{Prochazka:2014aa} where the two axes are related to $\lambda_i$ parameters by $x=\frac{1}{3}(2\lambda_1-\lambda_2-\lambda_3)$, $y=\frac{1}{\sqrt{3}}(\lambda_2-\lambda_3)$ which manifestly shows the $S_3$ triality symmetry. At the points where two curves cross, we find the minimal models of $\mathcal{W}_\infty$ algebra if we quotient out by the maximal ideal which in particular contains the two ideals coming from the two curves that meet.}
\label{fig:truncations}
\end{figure}

The local fields of the $\mathcal{W}_{1+\infty}$ algebra can be labeled by 3d partitions where the conformal dimension of the fields is given by the number of boxes of the corresponding partition\footnote{This simple combinatorial interpretation is one of the main reasons for considering the additional $U(1)$ factor instead of restricting purely to $\mathcal{W}_{1+\infty}$.}. At special curves in the two dimensional parameter space of $\mathcal{W}_{1+\infty}$ algebras, the generators associated to 3d partitions having a box at coordinates $(L+1,M+1,N+1)$\footnote{Here we use the convention that the box corresponding to $J_{-1}|0\rangle$ is at position $(1,1,1)$.} form an ideal $\mathcal{I}_{L,M,N}$. In other words, $\mathcal{I}_{L,M,N}$ contains all the configurations, where the boxes do not fit between the corner and its copy shifted by $(L,M,N)$. The curve in the parameter space for which $\mathcal{I}_{L,M,N}$ appears is given by
\begin{eqnarray}
\label{truncations2}
\frac{L}{\lambda_1}+\frac{M}{\lambda_2}+\frac{N}{\lambda_3}=1.
\end{eqnarray}
Note that due to (\ref{relation1}), the ideals $\mathcal{I}_{L,M,N} \supset \mathcal{I}_{L+k,M+k,N+k}$ are associated to the same curve. Derivation of the formula (\ref{truncations2}) along the lines of \cite{Prochazka:2015aa} can be found in the appendix \ref{truncations}. If we quotient by the ideal $\mathcal{I}_{L,M,N}$, we recover an algebra with generators associated to 3d partitions living between the corner at the origin and the corner shifted by $(L,M,N)$. Each truncation curve has a corresponding maximal truncation which we get by quotienting by
\begin{equation}
\mathcal{I}_{(L,M,N)-\max(L,M,N)(1,1,1)},
\end{equation}
or in other words choosing one of $(L,M,N)$ to be zero. These are the truncations discussed in \cite{Prochazka:2014aa} and they correspond to quotients that are irreducible for generic values of the central charge. For illustration, few truncation curves are depicted in figure \ref{fig:truncations}.

\subsection{Identification between $Y_{L,M,N}$ and truncations of $\mathcal{W}_{1+\infty}$}

The same structure of vacuum characters of $Y_{L,M,N}$ was found already in \cite{Gaiotto:2017euk} so we conjecture that Y-algebras are isomorphic to truncations of $\mathcal{W}_{1+\infty}$. In this section we discuss few pieces of evidence supporting this identification.

\paragraph{Vacuum characters}
The vacuum character of $Y_{L,M,N}$ for $N\geq M$ was determined in \cite{Gaiotto:2017euk} to be given by
\begin{eqnarray}
\chi\left [Y_{L,M,N}[\Psi]\right ]=\chi_{\mathcal{W}_{N-M}}(q) \oint dV_{M|L} \chi^{M|L}_{\frac{N-M}{2}}(q,x_i,y_i).
\end{eqnarray}
In this expression, $\chi_{\mathcal{W}_{N}}$ is the character of the $U(1) \times \mathcal{W}_{N}$ algebra, $\oint dV_{L,M}$ is the Vandermonde projector (invariant integration) that projects to $U(L|M)$ invariant combinations of fields, and $\chi^{M|L}_{\frac{N-M}{2}}(q,x_i,y_i)$ is  the character of a system of symplectic bosons in the fundamental representation of $U(M|L)$ and with the level shifted by $\frac{N-M}{2}$ that comes from the DS-reduction of the off-diagonal blocks of $U(N|L;\Psi)$. All of these ingredients are reviewed in appendix \ref{Characters}. This character is expected to have an interesting box counting interpretation: it counts the 3d partitions that fit between the corner at the origin and the corner shifted by $(L,M,N)$. In the limit of large number of D3 branes this simplifies and one finds the famous MacMahon function counting all plane partitions without any additional constraints.

\paragraph{Central charges}

The central charge of $Y_{L,M,N}$ was determined in \cite{Gaiotto:2017euk} to be given by\footnote{Regarding $\Psi$-dependence in the expression, one can identify the pole at 0 with the infinite leg in $(0,1)$ direction, the pole at $\infty$ with the infinite leg in $(1,0)$ direction and the pole at $1$ with the infinite leg in $(1,1)$ direction (these can be thought of as homogeneous and inhomogeneous coordinates on $\mathbbm{CP}^1$). The prefactor in front of each pole is a cubic expression in the difference of numbers of branes attached to the corresponding five-brane from the left and right. In general $(p,q)$-web diagrams, we will recover similar structure of the central charge.}
\begin{align}
&c_{L,M,N}[\Psi]= \frac{1}{\Psi} (L-N)\left((L-N)^2-1\right) + \Psi(M-N)\left((M-N)^2-1\right) + \cr
&+ \frac{1}{\Psi-1}(M-L)((M-L)^2-1)+(2N+M-3L)(N-M)^2+L-N.
\label{cc2}
\end{align}
Note that this is invariant under replacements
\begin{eqnarray}
\nonumber
 &\Psi \leftrightarrow \frac{1}{\Psi} \quad\quad & L \leftrightarrow M \\
 & \Psi \leftrightarrow 1-\Psi \quad\quad & N \leftrightarrow M
\end{eqnarray}
which generate the $S_3$ group of transformations (related to the triality symmetry of $\mathcal{W}_{1+\infty}$). The group acts by permutations on $(L,M,N)$ and on $\Psi$ by fractional linear transformations permuting $(0,1,\infty)$. This motivates us to introduce another parametrization
\begin{eqnarray}
\label{identification1}
\nonumber
\lambda_1 & = & L-(1-\Psi)N-\Psi M \\
\lambda_2 & = & -\frac{L-(1-\Psi)N-\Psi M}{\Psi} \\
\nonumber
\lambda_3 & = & \frac{L-(1-\Psi)N-\Psi M}{\Psi-1}
\end{eqnarray}
satisfying
\begin{eqnarray}
\label{sumlambdainv}
\frac{1}{\lambda_1}+\frac{1}{\lambda_2}+\frac{1}{\lambda_3} = 0
\end{eqnarray}
just like in the case of $\mathcal{W}_{\infty}$. Furthermore, the expression for the central charge (\ref{cc2}) can be rewritten in the form
\begin{eqnarray}
c_{1+\infty}=(\lambda_1-1)(\lambda_2-1)(\lambda_3-1)+1.
\end{eqnarray}
which is equal to the central charge in $\mathcal{W}_{\infty}$ (\ref{cc2}) except for the shift by one due to the $U(1)$ factor. The $S_3$ triality action of \cite{Gaiotto:2017euk} acts simply by permutations of the parameters $\lambda_i$ and the central charge is manifestly triality invariant in this parametrization. It is convenient for what follows to introduce $\epsilon$ parameters $(\epsilon_1,\epsilon_2,\epsilon_3)$ by\footnote{We expect the parameters $\epsilon_i$ to be related to the Nekrasov parameters.}
\begin{eqnarray}
\Psi = -\frac{\epsilon_2}{\epsilon_1}\qquad 0=\epsilon_1+\epsilon_2+\epsilon_3
\end{eqnarray}
(note that they are determined by $\Psi$ only up to an overall scale factor). In terms of these, the parameters $\lambda_i$ can be written in more symmetric form
\begin{eqnarray}
\lambda_i  = \frac{L \epsilon_1 + M \epsilon_2 + N \epsilon_3}{\epsilon_i}
\end{eqnarray}
To summarize, if we identify the parameters $\lambda_i$ from (\ref{cc2}) with those introduced in (\ref{identification1}), we see that the central charges, the vacuum characters and the triality transformations of Y-algebras can be identified with those of the truncations of $\mathcal{W}_{1+\infty}$ algebra. Moreover the parameters (\ref{cc2}) satisfy (\ref{truncations2}). Given the constraining power of the bootstrap analysis this is a strong indication of correctness of our identification of Y-algebras as truncations of $\mathcal{W}_{1+\infty}$.

Note the special points where the two truncation curves intersect. The algebras with such values of parameters contain further null-states that can be factorized. From the point of view of Y-algebras, these points correspond to DS-reduction and coset of Kac-Moody algebras at rational levels. At rational levels the Kac-Moody algebras contain null states and to take them into account one should use Kac-Weyl characters to calculate the characters of the final algebra. At least in the case of $\mathcal{W}_N$ algebras these are known to lead to minimal models \cite{Frenkel:1992ju}. It would be nice to generalize this construction to all Y-algebras. In our considerations we will always consider $\Psi$ to be generic corresponding to a generic points of the truncation curve $(L,M,N)$.

\subsection{Modules of $Y_{L,M,N}$}

\begin{figure}[h]
  \centering
      \includegraphics[width=0.55\textwidth]{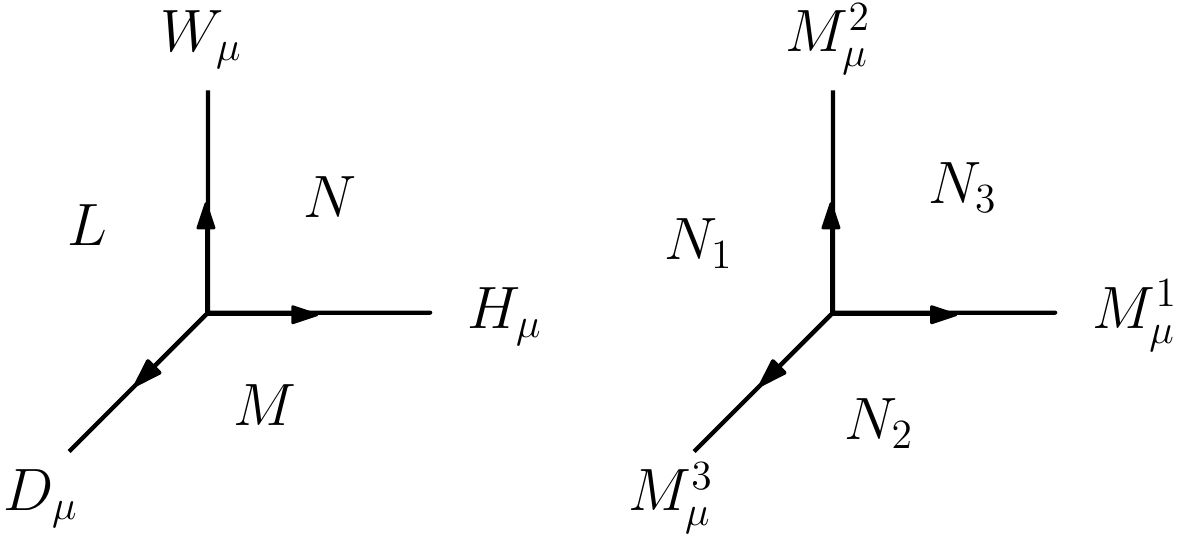}
 \caption{Labeling of modules associated to line operators supported each interface. Modules are labeled by representations of supergroup induced at corresponding interface. For example $D_\mu\equiv M^3_\mu$ modules are labeled by representations of $U(M|L)$.}
\label{fig:modules}
\end{figure}

In order to define the gluing construction, we will need to consider modules for Y-algebras. The most general module needed for the gluing is labeled by three representations of unitary supergroups associated to line operators supported at the three infinite interfaces and ending at the boundary. More concretely, as illustrated in Figure \ref{fig:modules}, the modules for $Y_{L,M,N}$ will be represented by a triple of representations $(\lambda,\mu,\nu)$ of $U(L|M)$, $U(M|N)$ and $U(N|L)$. At present, we have only a partial understanding of the characters and the conformal dimensions of these representations, which we now summarize.

\paragraph{Topological vertex and box counting} In the special case that the three asymptotic representations are contravariant representations (i.e. contained in the tensor power of the fundamental representation), we can use the box counting interpretation of the topological vertex to find the conformal dimensions and the characters \cite{Okounkov:uq,feigin2012quantum,tsymbaliuk2017affine,Prochazka:2015aa}. In this case, the representations $(\lambda,\mu,\nu)$ can be labeled by three Young diagrams. The states in the module of $Y_{L,M,N}$ are then in one-to-one correspondence with the plane partitions which have non-trivial asymptotics given by the Young diagrams $(\lambda,\mu,\nu)$ and further restricted such that the box at coordinates $(N_1+1,N_2+1,N_3+1)$ is not present.

The highest weight state corresponds to the configuration with minimal (but infinite) number of boxes compatible with the asymptotics. The states at level $l$ are in one-to-one correspondence with plane partitions obtained by adding $l$ boxes to this minimal configuration (always in way such that the resulting configuration of boxes is a plane partition). This identification allows us in principle to write down the character purely in terms of a combinatorial counting. The conformal dimension of the module can be similarly computed \cite{Prochazka:2015aa} by first computing the generating function of the conserved charges of $Y_{L,M,N}$ and extracting the eigenvalue of the $L_0$ generator from it (see appendix \cite{appmindim} for details). The result for the representation with an asymptotic Young diagram in the third direction is\footnote{The normalization of $U(1)$ current is discussed later.}
\begin{eqnarray}
j(M^3_{\mu}) & = & \sum_j \mu_j \\
\nonumber
h(M^3_{\mu}) & = & -\frac{\lambda_3}{2\lambda_1} \sum_j \mu_j^2 - \frac{\lambda_3}{2\lambda_2} \sum_j (2j-1) \mu_j + \frac{\lambda_3}{2} \sum_j \mu_j \\
& = & -\frac{\lambda_3}{2\lambda_1} \sum_j \mu_j^2 - \frac{\lambda_3}{2\lambda_2} \sum_j (\mu^T)^2_j + \frac{\lambda_3}{2} \sum_j \mu_j.
\end{eqnarray}
In particular, the conformal dimension of the minimal representation is given by
\begin{equation}
\label{boxdim}
h(\Box_3) = \frac{1+\lambda_3}{2}
\end{equation}
irrespectively of the truncation that we are considering. Its complex conjugate representation has an opposite $U(1)$ charge but the same conformal dimension. By fusing these, we can in principle obtain an arbitrary maximally degenerate representation of the type we need for the gluing procedure.

The main disadvantage of the approach using box counting is that we have only access to representations whose asymptotics are those obtained from the fundamental representation (i.e. contravariant representations) and it is not clear how to generalize these results directly to the case of fusions of both fundamental and anti-fundamental representations. The second disadvantage is the lack of useful closed-form formulas for the characters of the modules, but see \cite{bershtein2015plane} for the case where one of the parameters $(L,M,N)$ vanishes.

\paragraph{Characters from BRST construction}

It was conjectured in \cite{Gaiotto:2017euk} that Y-algebras have three families of degenerate modules associated to line operators supported on the three interfaces and ending at the corner. These three families were labeled as in the figure \ref{fig:modules}. In the following, we will introduce a more uniform notation
\begin{eqnarray}
M^1_{\mu}=H_{\mu},\quad M^2_{\mu}=W_{\mu},\quad M^3_{\mu}=D_{\mu}.
\end{eqnarray}
The parameters $\mu$ in $M^1_{\mu}$ label representations of $U(M|N)$ and analogously for the other two classes  of modules.

In the following we will consider generic representations $M^2_\nu\otimes M^3_\mu$ of the $Y_{0,M,N}[\Psi]$ algebra. These representations are labeled by representations of $U(M)$ and $U(N)$. Characters and conformal dimensions of all the modules we use in this paper can be obtained from the $SL(2,\mathbbm{Z})$ transformations of this setup. Knowledge of these modules will allow us to consider configurations that require gluing along internal lines with one of the stacks on the left or on the right of the leg vanishing. Generic situation requires dealing with technically more complicated representations of $U(N|M)$ Lie superalgebras and is left for future work. 

The representations of $U(M)$ are labeled by a set of integers $(\mu_1,\mu_2,\dots,\mu_M)$, where $\mu_1\geq \mu_2\geq \dots \geq \mu_M$ (note that we do not restrict these to be non-negative and look at all the irreducible representation one can get from the tensor product of the fundamental and the anti-fundamental representation). Choosing a normalization of the $U(1)$ current such that \footnote{There is a freedom of assigning $U(1)$ current algebra factors to vertices and the corresponding $U(1)$ charges to bimodules. Here we fix this freedom in certain way, but in specific examples other choices may be more suitable. See section \ref{appendixun} where this freedom is discussed in detail in the case of $U(N)_\kappa$ affine Lie algebra.}
\begin{eqnarray}
J(z)J(w)\sim \frac{\lambda_1+\lambda_2}{(z-w)^2} = -\frac{\lambda_1\lambda_2}{\lambda_3} \frac{1}{(z-w)^2},
\label{u1norm}
\end{eqnarray}
the charge of $M^3_\mu$ and its conformal dimension are given by \footnote{Checks of these formulas for $Y_{N,0,0}$, $Y_{0,1,0}$, $Y_{0,1,1}$, $Y_{0,2,1}$, $Y_{0,1,2}$ can be found in appendix \ref{ap:dimensions}. This formula was checked for either $M$ or $N$ vanishing.} \footnote{To fully specify the irreducible highest weight representations of $\mathcal{W}_N$, we should specify $N-1$ independent charges. For the maximally degenerate representations these can be determined in terms of the Young diagram labels. For this reason we don't need to know the explicit values of the higher spin charges. An example for the generating function the higher spin charges for the $(\Box,\cdot,\cdot)$ representation see appendix \ref{appmindim}.}
\begin{eqnarray}
\label{intcharges}
j(M^3_\mu) & = & \sum_{j=1}^{M}\mu_j,\\
\label{dimension}
h(M^3_\mu) & = & -\frac{\lambda_3}{2\lambda_1}\sum_{j=1}^{M}\mu_j^2-\frac{\lambda_3}{2\lambda_2}\sum_{j=1}^{M} (2j-M-1)\mu_j + \frac{N}{2}\sum_{j=1}^{M}|\mu_j|.
\end{eqnarray}

The characters of $M^2_{\nu}$ and $M^3_{\mu}$ modules of $Y_{0,M,N}[\Psi]$ can be calculated according to \cite{Gaiotto:2017euk} in a similar way as the vacuum character. The only modification is to insert a corresponding Schur polynomial $s_\mu (x_i)$ and $s_\nu (x_i)$  into the formula, i.e. in the case of $N>M$ and $M_\mu^3$ representation, the character is given by
\begin{eqnarray}
\chi_{0,M,N}(M^3_\mu)=\chi_{\mathcal{W}_{N-M}}(q) \oint dV_{M} \chi^{M|0}_{\frac{N-M}{2}}(q,x_i) s_\mu(x_i).
\end{eqnarray}
In the case of $M_\nu^2$ modules, one needs to first perform the DS reduction by substituting $x_j=q^{\frac{1}{2}(2j-M-1)}$ for $j\leq N-M$ and then insert into the integral
\begin{eqnarray}
\label{charm2}
\chi_{0,M,N}(M^2_\nu)=\chi_{\mathcal{W}_{N-M}}(q) \oint dV_{M}  \chi^{M|0}_{\frac{N-M}{2}}(q,x_i) s_\nu\left (x_j\rightarrow q^{\frac{1}{2}\left (2j-N_2-1\right )},x_i\right).
\end{eqnarray}
One can similarly calculate the characters of modules with two asymptotics $M^2_\mu \otimes M^3_\nu$ by first doing the DS reduction substitution and then inserting into the integral formula both characters. An example of $Y_{0,1,2}[\Psi]$ is given in \ref{Characters}. 

Similarly for the character of the $Y_{L,0,0}[\Psi]$ representation with two asymptotics, no DS-reduction is required and one needs to simply insert both Schur polynomials into the corresponding integral formula. 

For positive values of $\mu_j$, these characters have a nice box-counting interpretation that was discussed above. The conjugate representations have the sign and the order of $\mu$'s reversed and they have the same character and the same conformal dimension. Furthermore, if we split $\mu$ into positive and negative parts, $\mu = \mu_+ + \mu_-$, we see that both the $U(1)$ charge and the dimension are additive under this splitting. As far as the character goes, a generic representation does not seem to have a known simple combinatorial interpretation unless we are dealing with $Y_{0,0,N}=\mathcal{W}_N$ algebra for which the character is invariant (up to overall factor of $q$ to some power) under the shift of all $\mu_i$ by a constant and we can make all of them positive and then deduce corresponding box counting interpretation.

\section{Gluing construction}

\begin{figure}[h]
  \centering
   \includegraphics[width=0.35\textwidth]{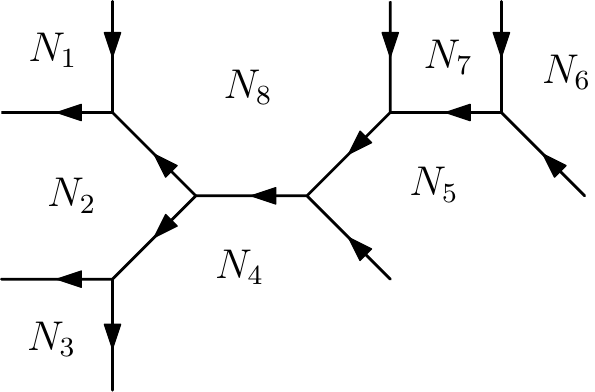}
  \caption{A generic $(p,q)$-web with stacks of $N_i$ D3 branes attached. The gluing construction associates a vertex operator algebra to such a diagram. To each vertex in the diagram, one associates a Y-algebra and to each finite line segment one associates a class of bimodules for the two Y-algebras that are connected by the corresponding line segment. The final vertex operator algebra is a conformal extension of the product of Y-algebras by such bimodules and their fusions.}
	\label{figglue}
\end{figure}

It was already suggested in \cite{Gaiotto:2017euk} that one can use a construction analogous to the topological vertex \cite{Aganagic:2003db,Aganagic:2005wn,Jafferis:2006ny} to produce more complicated vertex operator algebras by gluing $Y$-algebras. Consider a web of $(p,q)$-branes with stacks of D3-branes attached to them from different sides as in Figure \ref{figglue}.\footnote{Throughout the paper, we consider only webs corresponding to toric diagrams of Calabi-Yau three-folds without compact four-cycles, i.e. tree-like diagrams. The construction  should be possible in general but in the presence of the closed faces, generic modules associated the Gukov-Witten defects \cite{Gukov:2006jk} stretched within the internal faces can also be added to the VOA.}

This configuration gives rise to a web of domain walls in the $\mathcal{N}=4$ super Yang-Mills theory. In the topological twist of the theory, local operators inserted at trivalent junctions of the diagram give rise to Y-algebras. Looking at the configuration from the IR, the finite segments of five-branes become infinitely small and the whole configuration can be thought of as a resolution of a single star shaped junction of more complicated domain walls. According to this picture, the line operators supported at finite segments and ending at the two trivalent junctions play the role of local operators of the IR junction and should be added to the final vertex operator algebra. The line operators living at interfaces and ending at their junctions will be associated to modules for Y-algebras. Operators one needs to add to the collection of Y-algebras correspond to bimodules associated to such line operators and their fusions. It turns out that these bimodules have (half-) integral conformal dimension with respect to the total stress-energy tensor (sum of the stress-energy tensors associated to each trivalent junction) and can indeed be added to the vertex operator algebra. In this section, we explain this construction in detail.

\subsection{The vertex}

We start with the description of the basic building blocks of our construction. The algebra of local operators associated to the trivalent junction of D5, NS5 and (1,1) brane\footnote{Note that we identify $(1,0)$ with D5-brane and $(0,1)$ with NS5-brane in the contrast with \cite{Gaiotto:2017euk}.} can be identified with $Y_{L,M,N}[\Psi]$ algebra reviewed above. In order to allow more general gluing, it proves useful to consider a larger family of trivalent junctions that will then serve as building blocks in the gluing construction. Luckily, one can obtain a larger class of such vertices by applying S-duality transformations to the basic D5-NS5-(1,1) junction. In the topological vertex literature, this operation is related to the change of framing.

S-duality acts on an $A^T \equiv (p,q)^T$ five-brane by a left multiplication by an $SL(2,\mathbbm{Z})$ matrix
\begin{eqnarray}
M = \begin{pmatrix} a & b \\ c & d \end{pmatrix}
\qquad \mbox{for} \quad \ ad-bc=1.
\end{eqnarray}
The corresponding transformation of the coupling parameter $\Psi$ is
\begin{eqnarray}
\Psi \rightarrow \frac{a\Psi+b}{c\Psi+d}.
\end{eqnarray}
In terms of $\epsilon$ parameters, the transformation is implemented by the left multiplication of $(\epsilon_1,\epsilon_2)^T$ by matrix
\begin{eqnarray}
\left (M^{-1}\right )^T=\begin{pmatrix}
d&-c\\
-b& a
\end{pmatrix}
\end{eqnarray}
such that the combination
\begin{equation}
\epsilon^T A \equiv \begin{pmatrix} \epsilon_1 & \epsilon_2 \end{pmatrix} \begin{pmatrix} p \\ q \end{pmatrix}
\end{equation}
stays invariant.

\begin{figure}[h]
  \centering
      \includegraphics[width=0.7\textwidth]{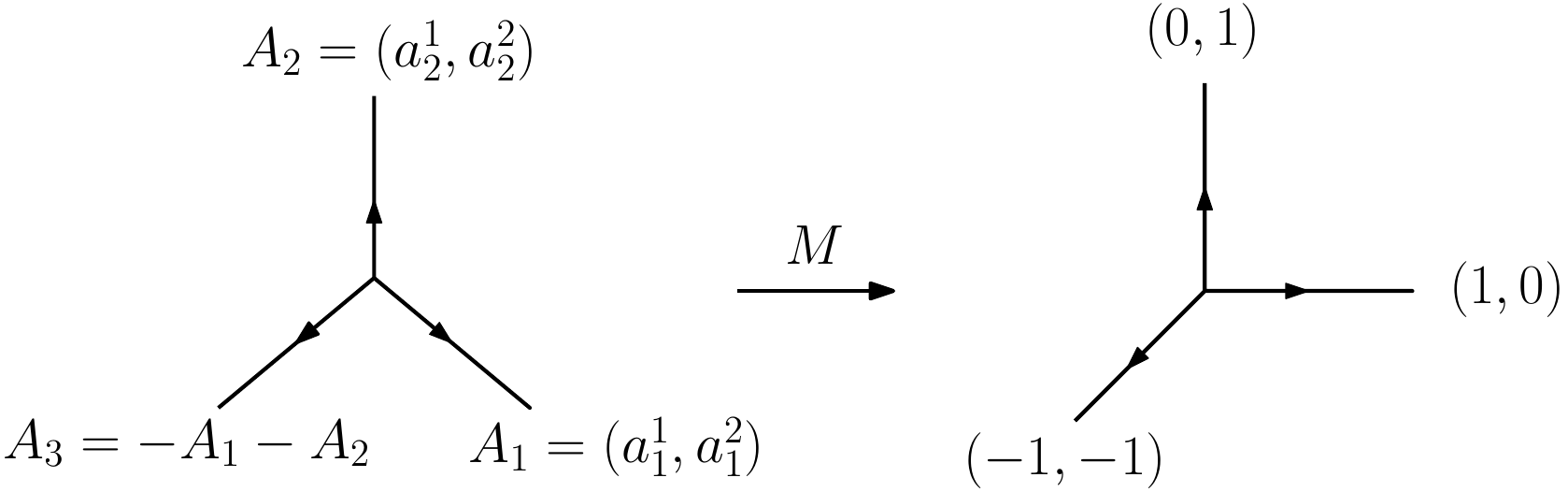}
 \caption{Transformation relating generic vertex of interest with the one used in identification of Y-algebras.}
\label{fig:transformation}
\end{figure}

Using these $SL(2,\mathbbm{Z})$ transformations, one can map a trivalent junction of $A_j=(p_j,q_j), \,j=1,2,3$ defects (satisfying conservation of charges and the condition that ensures existence of such a transformation)
\begin{eqnarray}
0 & = & A_1 + A_2 + A_3 \\
\nonumber
1 & = & A_1 \wedge A_2 \equiv p_1 q_2 - p_2 q_1
\end{eqnarray}
to the configuration used in the definition of Y-algebras by
\begin{eqnarray}
M = \begin{pmatrix} q_2 & -p_2 \\ -q_1 & p_1 \end{pmatrix}.
\end{eqnarray}
In other words, to each such trivalent junction of $A_1$, $A_2$, and $A_3$ defects as shown in the figure \ref{fig:transformation} and the coupling parameter $\Psi$, one associates the algebra
\begin{eqnarray}
Y^{A_1,A_2,A_3}_{L,M,N}\left [\Psi \right ] = Y^{\tiny \begin{pmatrix}1\\0\end{pmatrix},\begin{pmatrix}0\\1\end{pmatrix},\begin{pmatrix}-1\\-1\end{pmatrix}}_{L,M,N}\left[-\frac{q_2\Psi-p_2}{q_1\Psi-p_1} \right] \equiv Y_{L,M,N}\left[-\frac{q_2\Psi-p_2}{q_1\Psi-p_1} \right].
\label{genericY}
\end{eqnarray}
In terms of $\epsilon$ parameters $\epsilon = (\epsilon_1,\epsilon_2)^T$ and the five-brane charges
\begin{eqnarray}
Y^{A_1,A_2,A_3}_{L,M,N}\left[\epsilon_j \right] = Y_{L,M,N}\left[\epsilon^T A_j \right] \equiv Y_{L,M,N}\left[\tilde{\epsilon}_j \right]
\end{eqnarray}
where $\tilde{\epsilon}_j = \epsilon^T A_j = p_j \epsilon_1 + q_j \epsilon_2$. Note that the necessary consistency requirement $\tilde{\epsilon}_1+\tilde{\epsilon}_2+\tilde{\epsilon}_3 = 0$ follows from the charge conservation $A_1+A_2+A_3=0$ at the trivalent junction.\footnote{Note also that identification is possible for any values of $A_1$ and $A_2$ not only those related to the junction of NS5- and D5-branes by S-duality. One is tempted to identify generic vertex with such algebra. This naive guess would not be consistent with gluing proposal since bimodules added in gluing construction would not be (half-) integral.} In terms of the invariant $\lambda$-parameters parametrizing the structure constants of $Y$ (\ref{identification1}) we have
\begin{equation}
\label{lambdafromdiag}
\lambda_j = \frac{L \tilde{\epsilon}_1 + M \tilde{\epsilon}_2 + N \tilde{\epsilon}_3}{\tilde{\epsilon}_j} = \frac{\epsilon^T (L A_1 + M A_2 + N A_3)}{\epsilon^T A_j}.
\end{equation}
This is insensitive to rescalings of $\epsilon$ and $A_j$ parameters and $\lambda_j$ determined in this way satisfy both (\ref{relation1}) and (\ref{truncations2}).

There exists a natural $\mathbbm{Z}_2$ sign of the $SL(2,\mathbbm{Z})$ transformations. By taking a $\mathbbm{Z}_2$ reduction of an $SL(2,\mathbbm{Z})$ transformation matrix, we obtain an element of $SL(2,\mathbbm{Z}_2) \simeq S_3$ and taking the sign of the corresponding permutation gives us a homomorphism $SL(2,\mathbbm{Z}) \to \mathbbm{Z}_2$. Concretely, we can map
\begin{equation}
\begin{pmatrix} a & b \\ c & d \end{pmatrix} \mapsto (-1)^{ac+ad+bd+1}.
\end{equation}
obtaining the required sign. Choosing our canonically oriented vertex to have the $+$ orientation, any other vertex can be assigned an orientation given by the sign of the $SL(2,\mathbbm{Z})$ transformation mapping the canonically oriented vertex to the vertex we are considering. Concretely, the orientation is given by
\begin{equation}
\label{vertexorientation}
\mbox{sgn} \left[ Y^{A_1,A_2,A_3}_{L,M,N} \right] = (-1)^{p_1 p_2 + q_1 q_2 + p_1 q_2 + 1} = (-1)^{p_1 p_2 + q_1 q_2 + p_2 q_1}.
\end{equation}

\subsection{The edge}
\label{sec:gluing}

\begin{figure}[H]
  \centering
      \includegraphics[width=0.52\textwidth]{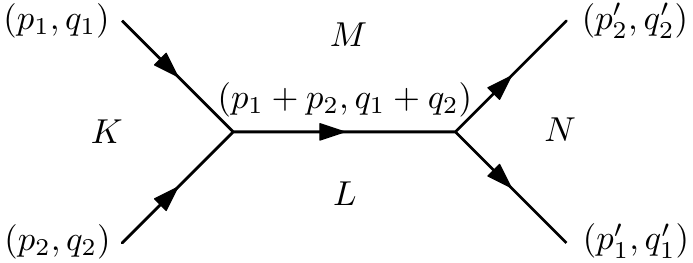}
\caption{Any junction of two Y-diagrams can be put into this form by reversing the orientation of the legs and changing the signs of the corresponding labels. The parameters are subject to the constraints from (\ref{conditions}).}
\label{glue}
\end{figure}

\begin{figure}[H]
  \centering
      \includegraphics[width=0.55\textwidth]{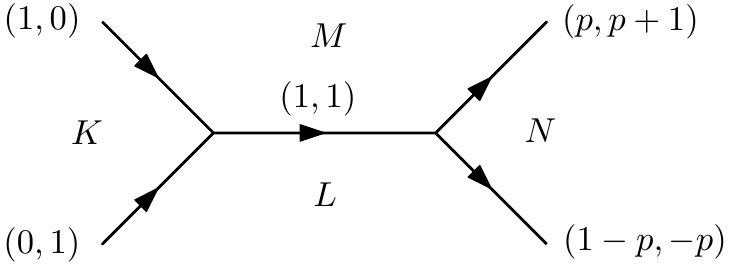}
\caption{By $SL(2,\mathbbm{Z})$ transformation, one can put diagram \ref{glue} to this form where parameter $p$ is given by combination \ref{parameter1}.}
\label{gluing2}
\end{figure}

Let us first consider gluing two vertices as in the figure \ref{glue} where the numbers are subject to constraints
\begin{eqnarray}
\label{conditions}
\nonumber
A_1 + A_2 & = & A_1^\prime + A_2^\prime \\
A_1 \wedge A_2 & = & 1 \\
\nonumber
A_1^\prime \wedge A_2^\prime & = & 1
\end{eqnarray}
The first equation is simply the condition of the conservation of charges and the remaining conditions come from the requirement that both vertices are S-dual to the elementary junction of NS5, D5 and (1,1)-brane. One can always change the orientation of the ingoing and the outgoing legs and change the signs of corresponding $(p,q)$ charges in order to obtain the configuration in \ref{glue}.
Using the S-duality transformation and the fact that all the building blocks are S-dual to the triple junction of D5, NS5, and (1,1)-brane, one can transform our system uniquely to a new configuration depicted in figure \ref{gluing2} by the transformation
\begin{eqnarray}
M=\begin{pmatrix}
q_2 & -p_2\\
-q_1 &p_1
\end{pmatrix}.
\end{eqnarray}
We used the fact that conditions (\ref{conditions}) let us express all pairings in terms of one remaining invariant parameter (measuring the relative framing of the two vertices)
\begin{eqnarray}
p \equiv -A_2 \wedge A_2^\prime = 1 + A_2 \wedge A_1^\prime = -1 + A_1 \wedge A_2^\prime = -A_1 \wedge A_1^\prime.
\label{parameter1}
\end{eqnarray}
The first vertex is by definition positively oriented, while the orientation of the second vertex can be easily read off from (\ref{vertexorientation}) and we find it to be equal to $(-1)^p$.

By looking at the two Y-vertices in diagram \ref{glue} or \ref{gluing2}, one can deduce that the final algebra will be a conformal extension of
\begin{eqnarray}
 Y^{-A_1,-A_2,A_1+A_2}_{L,M,K}[\Psi] \times Y^{A_1^\prime,A_2^\prime,-A_1^\prime-A_2^\prime}_{M,L,N}[\Psi]
\label{Glue_expr1}
\end{eqnarray}
by a collection of $M_{\mu}^3$ bimodules of the two Y-algebras on the right hand side. 

\paragraph{Conformal dimension of gluing fields}
We can now check that dimension of the bimodules are (half-)integral. This can be easily seen from the transformed diagram \ref{gluing2}. Remember that the total stress-energy tensor of the glued algebra is given by a sum os stress-energy tensors associated to the vertices, so in particular the conformal dimension of a bimodule is the sum of the two dimensions coming from the two endpoints,
\begin{equation}
h_\mu = h(M_{\mu}^3)+h^\prime(M_{\mu}^3).
\end{equation}
In the special case that the exchanged representation is the fundamental one, we can use the formula (\ref{boxdim}) and we find
\begin{equation}
h_\Box = 1 + \frac{\lambda_3 + \lambda_3^\prime}{2} = 1 + \frac{K+N-L-M}{2}+\frac{p(M-L)}{2} \equiv 1 + \rho.
\label{dimbibox}
\end{equation}
Note in particular that all the dependence on continuous parameters like $\Psi$ has canceled. The resulting dimension is always (half-)integral. The parameter $\rho$ introduced in this formula will be important in later sections.

Specializing now to the case $L=0$ as in (\ref{dimension}), we can be more general and write the expression for arbitrary line operator in representation $\mu$:
\begin{equation}
h_\mu = \frac{1+p}{2}\sum_{j=1}^{M}\mu_j^2+\frac{1-p}{2}\sum_{j=1}^{M} (2j-M-1)\mu_j+\frac{K+N}{2}\sum_{j=1}^{M}|\mu_j|.
\label{bimodules}
\end{equation}
Analogously, in the case that $N=0$ we have
\begin{equation}
h_\mu = \frac{1-p}{2}\sum_{j=1}^{L}\mu_j^2+\frac{1+p}{2}\sum_{j=1}^{L} (2j-L-1)\mu_j+\frac{K+N}{2}\sum_{j=1}^{L}|\mu_j|
\end{equation}
This is again independent of the continuous parameter $\Psi$ and is (half-)integral.

\paragraph{Gluing in terms of $\lambda$ parameters}
If we fix the discrete parameter $\rho$ which determines the dimension of the gluing matter and the five-brane charges $A_j$ and $A^\prime_j$, we can write explicitly the gluing conditions for Y-algebras directly in terms of $\lambda_j$ and $\lambda^\prime_j$ parameters. Let us first introduce a vector in five-brane charge space characterizing the first vertex
\begin{equation}
\sigma = \frac{A_2}{\lambda_3} - \frac{A_3}{\lambda_2}
\end{equation}
and similarly for the second vertex. Using the five-brane charge conservation
\begin{equation}
A_1 + A_2 + A_3 = 0
\end{equation}
and (\ref{sumlambdainv}) we find that the definition of $\sigma$ is cyclic invariant,
\begin{equation}
\lambda_1 \lambda_2 \lambda_3 \sigma = \lambda_2(\lambda_3 A_3 - \lambda_1 A_1) = \lambda_3(\lambda_1 A_1 - \lambda_2 A_2).
\end{equation}
The usefulness of $\sigma$ lies in the fact that it encodes the $\lambda_j$ parameters of the vertex, i.e.
\begin{equation}
\sigma \wedge A_j = \frac{1}{\lambda_j}.
\end{equation}
Furthermore, eliminating the number of D3-branes and parameter $\Psi$ from the gluing conditions, the gluing condition translates to a simple condition
\begin{equation}
\sigma \wedge \sigma^\prime = 0
\end{equation}
satisfied by $\sigma$ and $\sigma^\prime$ associated to glued vertices. This condition means that the vectors $\sigma$ associated to the neighbouring vertices are parallel. We can use this and the definition of $\rho$ to determine $\lambda^\prime_j$ once we know $\lambda_j$, $\rho, A_j$ and $A_j^\prime$,
\begin{eqnarray}
\label{gluelambda}
\nonumber
2\rho & = & \lambda_3 + \lambda_3^\prime \\
0 & = & (\lambda_1 A_1 - \lambda_2 A_2) \wedge (\lambda_1^\prime A_1^\prime - \lambda_3^\prime A_3^\prime) \\
\nonumber
0 & = & (\lambda_1 A_1 - \lambda_2 A_2) \wedge (\lambda_2^\prime A_2^\prime - \lambda_3^\prime A_3^\prime)
\end{eqnarray}
which is a linear system of equations and can be easily solved for $\lambda^\prime$.

\paragraph{Statistics of the gluing matter}
Gluing fields turn out to have either bosonic or fermionic character depending on the relative $\mathbbm{Z}_2$ sign (\ref{vertexorientation}) of the two vertices that we are gluing (and not whether the dimension of the gluing matter is integral or half-integral). We expect to have fermionic fields if the two vertices have the same sign and bosonic fields if the sign is opposite. In terms of the framing factor $p$ we will have bosons for $p$ odd and fermions for $p$ even. This is indeed consistent examples with values $p=-1,0,1$ that we discuss in greater detail in later chapters.

\paragraph{Summary of gluing}
For a complete gluing prescription for an edge, we need to
\begin{enumerate}
\item Identify Y-algebra bi-modules associated to the line operators supported at the finite five-brane segments. These are labeled by representations of the $U(M|N)$ group where $M$ and $N$ are the numbers of D3-branes attached to the finite edge from the right and left. If one of these vanishes (say $M=0$), they are simply labeled by ordered (both positive and negative) integers $\mu_1\geq \mu_2\geq \dots \geq \mu_N$.
\item To perform the gluing at the level of characters, one needs conformal dimensions and characters of corresponding modules. These can be obtained for example from the BRST reductions of the Kac-Moody algebra modules or box counting for contravariant modules.
\item We expect that the operator product expansions are fixed completely by the discrete data described above. The structure constants of $\mathcal{W}_{1+\infty}$ were found in \cite{luk1988quantization,Prochazka:2014aa} and the structure constants of Y-algebras are simply their restrictions to particular values of the parameters $\lambda$ of the corresponding truncations. The higher spin charges of the modules are determined in terms of the representation data and one can use either the bootstrap approach or the Coulomb gas calculation \cite{Dotsenko:1984nm,Dotsenko:1984ad,Dotsenko:1985hi} to find structure constants associated to both OPEs of the gluing matter with Y-algebra generators and also within gluing matter fields.\footnote{Note that there seems to exist a free field realization of Y-algebras that was constructed in \cite{Bershtein}. It would be nice to explore this connection further.}
\end{enumerate}

\subsection{Gluing in general}

Let us consider an arbitrary of $(p,q)$-webs composed of the trivalent junctions glued by five-brane edges as discussed in the previous sections and let us attach stacks of D3-branes to the faces of the diagram. This configuration gives rise to a web of domain walls in $\mathcal{N}=4$ SYM that we want to associate a vertex operator algebra to. The vertex operator algebra will be a conformal extension of a tensor product of mutually commuting Y-algebras associated to the vertices in the diagram by the bimodules associated to line operators inserted at the finite five-brane segments and their fusions. To each such segment, one can associate a parameter $\rho_i$ as in the case of a single edge.

One can make following conjectures about the resulting algebra:
\begin{enumerate}
\item The total stress-energy tensor of the resulting algebra is the sum of stress-energy tensors of the individual vertices.
\item As consequence of this, the central charge of the resulting algebra is the sum of the central charges associated to all vertices.
\item The characters of modules associated to a collection of edges can be computed as a sum of products of Y-algebra characters, where the sum runs over representations of a tensor product of Lie (super-) algebras associated to the internal edges. For example in the case of two vertices we have
\begin{eqnarray}
\chi= \sum_{\mu} \chi \left [Y^{(1)}\right ](M^3_\mu) \chi \left [Y^{(2)}\right ](M^3_\mu)
\end{eqnarray} 
\item These modules can be obtained by fusion of elementary bimodules associated to the line operators in the fundamental and anti-fundamental representation supported at the internal edge. The dimension of these representations is given by (\ref{dimbibox}).
\item To each external leg, one can associate a family of modules labeled by representations of the supergroup associated to the corresponding leg. Different families associated to non-parallel legs braid trivially, i.e. have conformal dimension that differs by an integer.
\item If the $(p,q)$-web is invariant under a subgroup of $SL(2,\mathbbm{Z})$ transformation, the glued algebra will turn out to have dual BRST realization. If the algebra is realized as a truncation of an infinite algebra, there will be corresponding duality action on the parameter space of the corresponding infinite algebra.
\end{enumerate}
In the following we will illustrate the general discussion of the gluing construction on few concrete examples.

\subsection{Example - $\mathcal{N}=2$ superconformal algebra}
\label{secn2sca}
Let us start with $\mathcal{N}=2$ superconformal algebra. This algebra is obtained by extending the Virasoro algebra by a $U(1)$ current $J$ and two oppositely charged spin $\frac{3}{2}$ fermionic primaries $G^{\pm}$. The $U(1)$ current $J$ generates the $SO(2)$ R-symmetry rotating the supercharges. The operator product expansions are given by
\begin{eqnarray}
\nonumber
T(z) T(w) & \sim & \frac{c/2}{(z-w)^4} + \frac{2T(w)}{(z-w)^2} + \frac{\partial T(w)}{z-w} \\
\nonumber
T(z) G^{\pm}(w) & \sim & \frac{3/2 G^{\pm}(w)}{(z-w)^2} + \frac{\partial G^{\pm}(w)}{z-w} \\
\nonumber
T(z) J(w) & \sim & \frac{J(w)}{(z-w)^2} + \frac{\partial J(w)}{z-w} \\
J(z) J(w) & \sim & \frac{c/3}{(z-w)^2} \\
\nonumber
J(z) G^{\pm}(w) & \sim & \frac{\pm G^{\pm}(w)}{z-w} \\
\nonumber
G^+(z) G^-(w) & \sim & \frac{2c}{3(z-w)^3} + \frac{2J(w)}{(z-w)^2} + \frac{2T(w)}{z-w} + \frac{\partial J(w)}{z-w} \\
\nonumber
G^{\pm}(z) G^{\pm}(w) & \sim & reg.
\end{eqnarray}
The central charge $c$ is the only free continuous parameter.

\begin{wrapfigure}{l}{0.18\textwidth}
\vspace{-15pt}
\begin{center}
\includegraphics[width=0.17\textwidth]{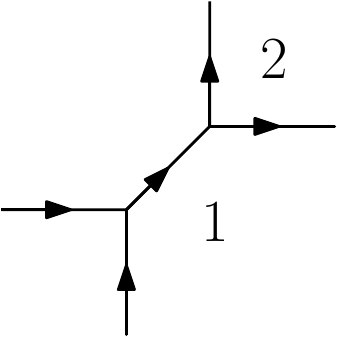}
\end{center}
\vspace{-20pt}
\end{wrapfigure}
From Poincar\'e-Birkhoff-Witt theorem we see that the vacuum character of this algebra at a generic central charge is given by
\begin{equation}
\prod_{n=0}^{\infty} \frac{\left(1+q^{\frac{3}{2}+n}\right)^2}{\left(1-q^{1+n}\right)\left(1-q^{2+n}\right)}
\end{equation}
Up to an $U(1)$ factor, this is exactly what one would obtain from the gluing construction starting from the diagram on the left. We can thus attempt to decompose the $\mathcal{N}=2$ SCA $\times U(1)$ into elementary building blocks that enter the gluing construction. First, we decouple the $U(1)$ currents to isolate the $\mathcal{W}_{\infty}$ stress-energy tensor that lives at the $(2,0,1)$ vertex. The unique combination commuting with $J(z)$ is
\begin{equation}
T_0(z) = T(z) - \frac{3}{2c} (JJ)(z).
\end{equation}
Similarly, we can find spin $3$ and spin $4$ primaries that commute with $J(z)$ (they are determined uniquely up to a rescaling). We can next compute the combination of OPE coefficients
\begin{equation}
\frac{(C_{33}^4)^2 C_{44}^0}{(C_{33}^0)^2} = \frac{12(c+1)(c+9)^2(5c-9)}{(c-1)(c+6)(2c-3)(5c+17)}
\end{equation}
and assuming OPEs to be those of $\mathcal{W}_{\infty}$, this together with $c_{\infty}=c-1$ lets us determine the $\lambda$ parameters associated to $(2,0,1)$ vertex to be
\begin{eqnarray}
\nonumber
\lambda^{(1)}_1 & = & -\frac{2c}{c-3} = \frac{-2\epsilon_1-\epsilon_2}{\epsilon_1} = \Psi-2 \\
\lambda^{(1)}_2 & = & -\frac{c}{3} = \frac{-2\epsilon_1-\epsilon_2}{\epsilon_2} = \frac{2-\Psi}{\Psi} \\
\nonumber
\lambda^{(1)}_3 & = & \frac{2c}{c+3} = \frac{-2\epsilon_1-\epsilon_2}{-\epsilon_1-\epsilon_2} = \frac{2-\Psi}{1-\Psi}
\end{eqnarray}
which is what we could directly read off from the diagram. The identification between parameters is
\begin{equation}
c = 3 - \frac{6}{\Psi}, \qquad \Psi = -\frac{6}{c-3}.
\end{equation}
We can also determine the $\lambda$ parameters of the second vertex
\begin{eqnarray}
\nonumber
\lambda^{(2)}_1 & = & 1 \\
\lambda^{(2)}_2 & = & \frac{c-3}{6} = \frac{\epsilon_1}{\epsilon_2} = -\frac{1}{\Psi} \\
\nonumber
\lambda^{(2)}_3 & = & -\frac{c-3}{c+3} = \frac{\epsilon_1}{-\epsilon_1-\epsilon_2} = \frac{1}{\Psi-1}
\end{eqnarray}
which is consistent with the $U(1)$ degree of freedom coming from the second vertex. 

Finally, let us identify the gluing matter. The fields of the lowest dimension that do not come from the vertices are are the fields $G^{\pm}(z)$. Following the choice of the normalization of $U(1)$ factors (\ref{u1norm}),
\begin{equation}
J^{(1)}(z) J^{(1)}(w) \sim -\frac{c(c+3)}{3(c-3)} \frac{1}{(z-w)^2}, \qquad J^{(2)}(z)J^{(2)}(w) \sim \frac{c+3}{6} \frac{1}{(z-w)^2},
\end{equation}
we know that the basic gluing fields will have charges $\pm 1$ with respect to these. We define a rotated basis of $U(1)$ currents
\begin{equation}
J(z) \equiv \frac{c-3}{3(c-1)} J^{(1)}(z) - \frac{2c}{3(c-1)} J^{(2)}(z), \qquad \tilde{J}(z) \equiv J^{(1)}(z) + J^{(2)}(w)
\end{equation}
such that $J(z)$ is the conventionally normalized R-current in $\mathcal{N}=2$ SCA and that $\tilde{J}(z)$ decouples. The other primary gluing fields are given by the normal ordered products
\begin{equation}
G^{\pm}_{(k)}(z) \equiv (\partial^{k-1} G^{\pm} (\partial^{k-2} G^{\pm} ( \cdots ( \partial G^{\pm} G^{\pm}) \cdots )))(z).
\end{equation}
Their $U(1)$ charges are given by
\begin{equation}
j^{(1)}(G^{\pm}_{(k)}) = \pm k, \quad j^{(2)}(G^{\pm}_{(k)}) = \mp k, \quad j(G^{\pm}_{(k)}) = \pm k, \quad \tilde{j}(G^{\pm}_{(k)}) = 0.
\end{equation}
The conformal dimensions are
\begin{equation}
h_{1+\infty}^{(1)}(G^{\pm}_{(k)}) = \frac{(c-3)k^2+2(c+3)k}{2(c+3)}, \quad h_{1+\infty}^{(2)}(G^{\pm}_{(k)}) = \frac{3k^2}{c+3}, \quad h_{1+\infty}(G^{\pm}_{(k)}) = \frac{k(k+2)}{2}.
\end{equation}
as predicted by (\ref{bimodules})

\paragraph{Truncations}
It is well-known \cite{Zamolodchikov:1986gh,Boucher:1986bh,DiVecchia:1985ief,DiVecchia:1986cdz,DiVecchia:1986fwg} that $\mathcal{N}=2$ SCA has a series of unitary minimal models for
\begin{equation}
c = 3\left(1- \frac{2}{k+2}\right), \quad k=1,2,\ldots
\end{equation}
All of these lie obviously on the truncation curve $(0,1,2)$ of the first vertex as follows from our identification of $\mathcal{N}=2$ SCA. But these minimal models also lie on the truncation curve $(k,0,0)$ which corresponds to the case where the $\mathcal{W}_{1+\infty}$ at the first vertex truncates to $U(1) \times \mathcal{W}_k$. We can thus predict the values of the central charge $c$ for which the $\mathcal{N}=2$ SCA truncates only based on the knowledge of truncation curves of $\mathcal{W}_{1+\infty}$. Analogously, the non-unitary minimal models with
\begin{equation}
c = 3\left(1-\frac{2(p+1)}{p^\prime+2}\right)
\end{equation}
lie on the intersection of the truncation curves $(0,1,2)$ and $(p^\prime,p,0)$ where the irreducible quotient of $Y_{012}$ can also be thought of as a quotient of $Y_{p^\prime p 0}$.

\paragraph{Unifying algebras}
The coset of $\mathcal{N}=2$ SCA with respect to $U(1)$ subalgebra was studied already in \cite{Blumenhagen:1994wg} in the context of unifying algebras. The authors noticed that the resulting algebra, in our notation $Y_{0,1,2}$ or the parafermion algebra, can be coupled in two ways to $U(1)$ current algebra, obtaining as result either $\mathcal{N}=2$ SCA or $U(2)$ affine Lie algebra (depending on the details of the gluing). This is one of the earliest examples of the gluing procedure in the literature, here applied to gluing of $Y_{0,1,2}$ and $Y_{0,0,1}$ vertices.

\subsection{Example - $\mathcal{W}_3^{(2)}$}

As another example, consider the Bershadsky-Polyakov algebra $\mathcal{W}_3^{(2)}$ \cite{Bershadsky:1990bg,Polyakov:1989dm}. It has the same matter content as $\mathcal{N}=2$ SCA except for the fact that the spin $\frac{3}{2}$ fields are bosons instead of being fermions. The operator product expansions are now
\begin{eqnarray}
\nonumber
T(z) T(w) & \sim & -\frac{(2k+3)(3k+1)}{2(k+3)(z-w)^4} + \frac{2T(w)}{(z-w)^2} + \frac{\partial T(w)}{z-w} \\
\nonumber
T(z) J(w) & \sim & \frac{J(w)}{(z-w)^2} + \frac{\partial J(w)}{z-w} \\
\nonumber
J(z) J(w) & \sim & \frac{2k+3}{3(z-w)^2} \\
T(z) G^{\pm}(w) & \sim & \frac{\frac{3}{2} G^{\pm}(w)}{(z-w)^2} + \frac{\partial G^{\pm}(w)}{z-w} \\
\nonumber
J(z) G^{\pm}(w) & \sim & \frac{\pm G^{\pm}(w)}{z-w} \\
\nonumber
G^{\pm}(z) G^{\pm}(w) & \sim & reg. \\
\nonumber
G^{+}(z) G^{-}(w) & \sim & \frac{(k+1)(2k+3)}{(z-w)^3} + \frac{3(k+1) J(w)}{(z-w)^2} - \frac{(k+3)T(w)}{z-w} \\
\nonumber
& & + \frac{3(JJ)(w)}{z-w} + \frac{3(k+1) \partial J(w)}{2(z-w)}.
\end{eqnarray}
\begin{wrapfigure}{l}{0.18\textwidth}
\vspace{-15pt}
\begin{center}
\includegraphics[width=0.17\textwidth]{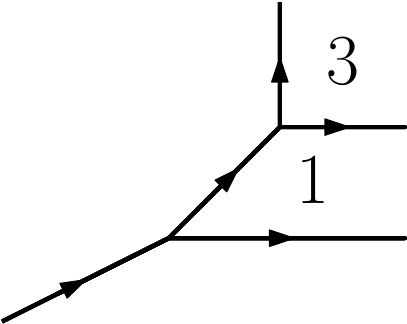}
\end{center}
\vspace{-20pt}
\end{wrapfigure}
From the gluing construct we see that the vacuum character
\begin{equation}
\prod_{n=0}^{\infty} \frac{1}{(1-q^{1+n})^2(1-q^{\frac{3}{2}+n})^2(1-q^{2+n})}
\end{equation}
of $U(1) \times \mathcal{W}_3^{(2)}$ equals that of the diagram on the left. We can try to see if this identification works even at the level of operator product expansions. Similarly to the discussion in the previous section, we can first decouple the $U(1)$ factors and find the stress-energy tensor
\begin{equation}
T^{(1)}_{\infty}(z) = T(z) - \frac{3}{2(2k+3)} (JJ)(z)
\end{equation}
with the central charge
\begin{equation}
c_{\infty}^{(1)} = -\frac{6(k+1)^2}{k+3}.
\end{equation}
Analogously, we can construct primary spin $3$ and spin $4$ currents commuting with $U(1)$ factors,
\begin{eqnarray}
\nonumber
W^{(1)}_3 & = & (G^+ G^-) + \frac{3(k+3)}{2k+3}(TJ)+\frac{k+3}{2} \partial T - \frac{9(k+2)}{(2k+3)^2} (J(JJ)) \\ \nonumber
& & - 3(J\partial J) - \frac{k^2+4k+6}{2k+3} \partial^2 J \\
W^{(1)}_4 & = & (J(G^+ G^-)) + \ldots,
\end{eqnarray}
compute their three-point functions
\begin{eqnarray}\nonumber
C_{33}^0 & = & -\frac{(4k+9)(2k+1)(k+3)(k+1)^2}{2k+3} \\
C_{44}^0 & = & -\frac{(5k+12)(4k+9)(3k+5)(2k+1)(k+1)^2k^2}{3(15k^2+19k-18)} \\ \nonumber
C_{33}^4 & = & \frac{12(k+3)^2}{2k+3}
\end{eqnarray}
and finally find the invariant combination of structure constants
\begin{equation}
\frac{C_{33}^4 C_{44}^0}{(C_{33}^0)^2} = -\frac{48k^2(k+3)^2(3k+5)(5k+12)}{(k+1)^2(2k+1)(4k+9)(15k^2+19k-18)}.
\end{equation}
Equating this to (\ref{cc2}), we can determine the $\lambda^{(1)}_j$ parameters (assuming that the commutant of $U(1)$ currents is $\mathcal{W}_{\infty}$) to be
\begin{eqnarray}\nonumber
\lambda^{(1)}_1 & = & 2k+3 = \frac{\epsilon_2+3\epsilon_3}{\epsilon_1} = 2\Psi-3 \\
\lambda^{(1)}_2 & = & -\frac{2k+3}{k+3} = \frac{\epsilon_2+3\epsilon_3}{\epsilon_2} = \frac{-2\Psi+3}{\Psi} \\ \nonumber
\lambda^{(1)}_3 & = & \frac{2k+3}{k+2} = \frac{\epsilon_2+3\epsilon_3}{\epsilon_3} = \frac{-2\Psi+3}{1-\Psi}.
\end{eqnarray}
We identified
\begin{equation}
\Psi = k+3.
\end{equation}
We can read-off the $\lambda$-parameters of the second vertex from the diagram (again we cannot determine them from the algebra because of no continuous parameters associated to the affine $U(1)$ algebra)
\begin{eqnarray}
\lambda^{(2)}_1 & = & \frac{-2\epsilon_1-\epsilon_2}{-2\epsilon_1-\epsilon_2} = 1 \\
\lambda^{(2)}_2 & = & \frac{-2\epsilon_1-\epsilon_2}{\epsilon_1} = -2+\Psi = k+1 \\
\lambda^{(2)}_3 & = & \frac{-2\epsilon_1-\epsilon_2}{\epsilon_1+\epsilon_2} = \frac{-2+\Psi}{1-\Psi} = -\frac{k+1}{k+2}.
\end{eqnarray}

\paragraph{$U(1)$ currents}
Now we can turn our attention to the identification of the $U(1)$ currents. We take a linear combination of $U(1)$ currents
\begin{eqnarray}
J(z) & = & \frac{k+3}{3(k+2)} J^{(1)}(z) - \frac{2k+3}{3(k+2)} J^{(2)}(z) \\
\tilde{J}(z) & = & J^{(1)}(z) + J^{(2)}(z)
\end{eqnarray}
such that $\tilde{J}(z)$ decouples from $\mathcal{W}_3^{(2)}$.

\paragraph{Gluing matter} The gluing fields in the case of $\mathcal{W}_3^{(2)}$ are given by powers of $G^{\pm}(z)$,
\begin{equation}
G^{\pm}_{(n)}(z) \equiv (G^{\pm} (G^{\pm} ( \cdots (G^{\pm} G^{\pm}) \cdots )))(z).
\end{equation}
Their $U(1)$ charges are
\begin{equation}
j^{(1)}(G^{\pm}_{(n)}) = \pm n, \quad j^{(2)}(G^{\pm}_{(n)}) = \mp n, \quad j(G^{\pm}_{(n)}) = \pm n, \quad \tilde{j}(G^{\pm}_{(n)}) = 0.
\end{equation}
and the conformal dimensions are
\begin{equation}
h_{1+\infty}^{(1)}(G^{\pm}_{(n)}) = \frac{n(3k+6-n)}{2k+4}, \quad h_{1+\infty}^{(2)}(G^{\pm}_{(n)}) = \frac{n^2}{2k+4}, \quad h_{1+\infty}(G^{\pm}_{(n)}) = \frac{3n}{2}.
\end{equation}
which is consistently with (\ref{bimodules}). Note that because of the bosonic nature of the gluing fields, the quadratic terms proportional to $n^2$ in $h_{1+\infty}^{(1)}$ and $h_{1+\infty}^{(2)}$ cancel so that the total dimension of the gluing fields grows linearly with $n$.

\subsection{Example - $D(2,1;-\Psi)_1$}

\begin{figure}[h]
\centering
\includegraphics[width=0.52\textwidth]{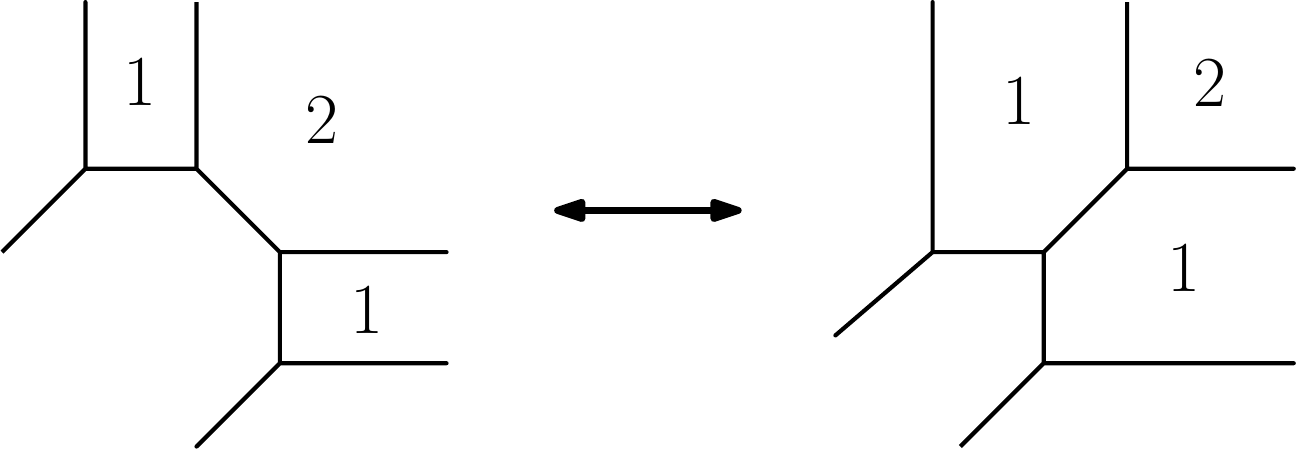}
\caption{Two diagrams associated to $D(2,1;-\Psi)_1\times U(1)_{2\Psi}$ related by a flop.}
\label{GL}
\end{figure}

Let us briefly mention one more exotic example associated to the diagram in the figure \ref{GL}. This configuration playes an important role in \cite{Creutzig:2017uxh} where the corresponding VOA was identified with the exceptional super Kac-Moody algebra $D(2,1;-\Psi)_1\times U(1)_{2\Psi}$ and conjectured to play the role of the quantum geometric Langlands kernel for the $SU(2)$ Group. In \cite{Creutzig:2017uxh}, various decompositions of the algebra as extensions of its subalgebras were discussed. Let us show that the character of $D(2,1;-\Psi)_1\times U(1)_{2\Psi}$  agrees to first few orders in the $q$-expansion with the prediction coming from the conformal extension of the atomic Y-algebras
\begin{equation}
Y_{0,0,1}[\Psi] \times Y_{0,2,1}[\Psi] \times Y_{2,0,1}[\Psi] \times Y_{0,0,1}[\Psi].
\end{equation}
Using formulas, from the appendix \ref{Characters}, one recovers
\begin{equation}
\chi=1+18q+133q^2+730q^3+3284q^4+12868q^5+45441q^6+\dots.
\end{equation} 
Note that we need to sum over modules associated to multiple edges as well as over modules with two asymptotics turned on. One can identify the leading coefficient $18$ with the number of generators of $D(2,1;-\Psi)_1\times U(1)_{2\Psi}$ algebra as expected. At higher levels, null states of the algebra need to be taken into account. Comparing with the formula in the remark 9.9 of \cite{Creutzig:2017uxh}, one recovers the above expansion as expected.

Note also that the diagram on the left of figure \ref{GL} can be transformed to a more symmetric configuration by a flip (to be discussed later) of the middle line segment. The algebra after the flip transition can be identified with a conformal extension of
\begin{equation}
Y_{0,0,1}[\Psi] \times Y_{1,1,2}[\Psi] \times Y_{1,1,0}[\Psi] \times Y_{0,0,1}[\Psi].
\end{equation}
As discussed in later sections, these two diagrams are conjectured to produce the same VOA since the parameter $\rho$ associated to the vertex at which we perform the flip vanishes. Note also that triality properties of $D(2,1;-\Psi)_1\times U(1)_{2\Psi}$ can be understood as a consequence of S-duality properties of the configurations of branes. We expect these configurations of five-branes to produce new families of VOAs with $S_3$ duality action analogous to the triality of Y-algebras.

As a final comment, let us briefly discuss how the presence of $17$ spin $1$ currents of $D(2,1;-\Psi)$ algebra can be seen from the gluing diagram in the right part of the figure \ref{GL}. The diagram contains three resolved conifold subdiagrams and in each case the number of neighbouring $D3$ branes is such that the associated parameter $\rho_j = 0$. This implies that we have $2 \times 3$ fermionic spin $1$ currents associated to three internal edges (each internal edge carries gluing matter field in fundamental and anti-fundamental representation). As usual, we also have the four commuting bosonic spin $1$ currents coming from the vertices. To find the remaining spin $1$ fields, we need to remember that the formula (\ref{boxboxfusion}) tells us that turning on line operators along neighbouring edges produces another spin $1$ field (we are always using the fact that $\rho_j = 0$). In this way we find additional $2 \times 3$ bosonic spin $1$ fields. Turning on the fundamental or anti-fundamental line operators along all three edges at the same time gives us additional two fermionic spin $1$ fields as follows from (\ref{box3fusion}). In total we have $4 + 6 = 10$ bosonic and $6 + 2 = 8$ fermionic spin $1$ fields. One of the bosonic fields is the overall decoupled $U(1)$ current and the remaining $17$ fields are exactly what we need for $D(2,1;-\Psi)$ algebra.

\section{Examples with BRST construction}

In some cases, one can give an alternative definition of the glued algebra in terms of a BRST construction. Gluing construction can then shed new light on the structure of the algebras obtained by such reductions. Configurations we discuss in this section are associated to diagrams with D5-branes ending from both left and right on a linear chain of $(n,1)$ branes. The most general configuration that we will be able to give a BRST definition is such that the diagram can be cut into two halves that satisfy the following condition: The number of D3-branes is non-increasing if we follow the upper part of the diagram from the top to the bottom and the number of D3-branes is  non-increasing if we follow the lower part of the diagram from the bottom up.

Let us motivate the BRST construction. We expect that a proper justification along these lines can be done analogously to \cite{Mikhaylov:2017ngi,Witten:2010aa,Witten:2011aa,Mikhaylov:2014aa}. In the Kapustin-Witten twisted theory, the path integral in this configuration localizes to the path integral of the complexified CS theory supported at $(n,1)$ branes connected by a nontrivial interface descending from D3-branes ending on D5-branes. In the IR, the finite internal five-brane segments shrink and we can view the configuration as a single interface between the upper and the lower CS theory. The half-BPS boundary conditions in the $\mathcal{N}=4$ super Yang-Mills theory descending from D3-branes ending on five-branes were analyzed in \cite{Gaiotto:2008ac}. These boundary conditions can be translated to the boundary conditions of the bosonic blocks of the complexified super Chern-Simons theory. The boundary condition on the off-diagonal blocks (descending from boundary conditions on the 3d bifundamental hypermultiplets supported at the $(n,1)$ interface) requires some guesswork and will be discussed later. We conjecture the corresponding VOA to be a BRST reduction of the super Kac-Moody algebra induced at the interface from the upper and lower CS theories by a BRST charge implementing the boundary conditions.

In the case of a single D5-brane, the BRST reduction that we will consider in the following reduces to the one used in the definition of $Y$-algebras as described in section \ref{sec:Y-algebras}. In this section, we first describe in detail the two reductions associated to two possible configurations with two D5-branes. The first one is associated to D5-branes ending from the opposite sides of a chain of $(n,1)$ branes whereas the second configuration is associated to D5-branes ending from the same side. The gluing construction provides us with various predictions for the structure of the corresponding VOA. We check that the central charge of the algebra obtained by the BRST reduction coincides with the sum of central charges of the two Y-algebras at each corner and in the case of the resolved conifold diagram we check its invariance under the $\mathbbm{Z}_2\times \mathbbm{Z}_2$ transformation which is the symmetry of the diagram. In various examples we find that the vacuum characters indeed match and are invariant under $\mathbbm{Z}_2\times \mathbbm{Z}_2$ transformations. Finally, we give a BRST definition in the presence of more D5-branes and discuss the special examples of the $U(N|M)_\Psi$ Kac-Moody algebras.

\begin{figure}[h]
\centering
\includegraphics[width=1\textwidth]{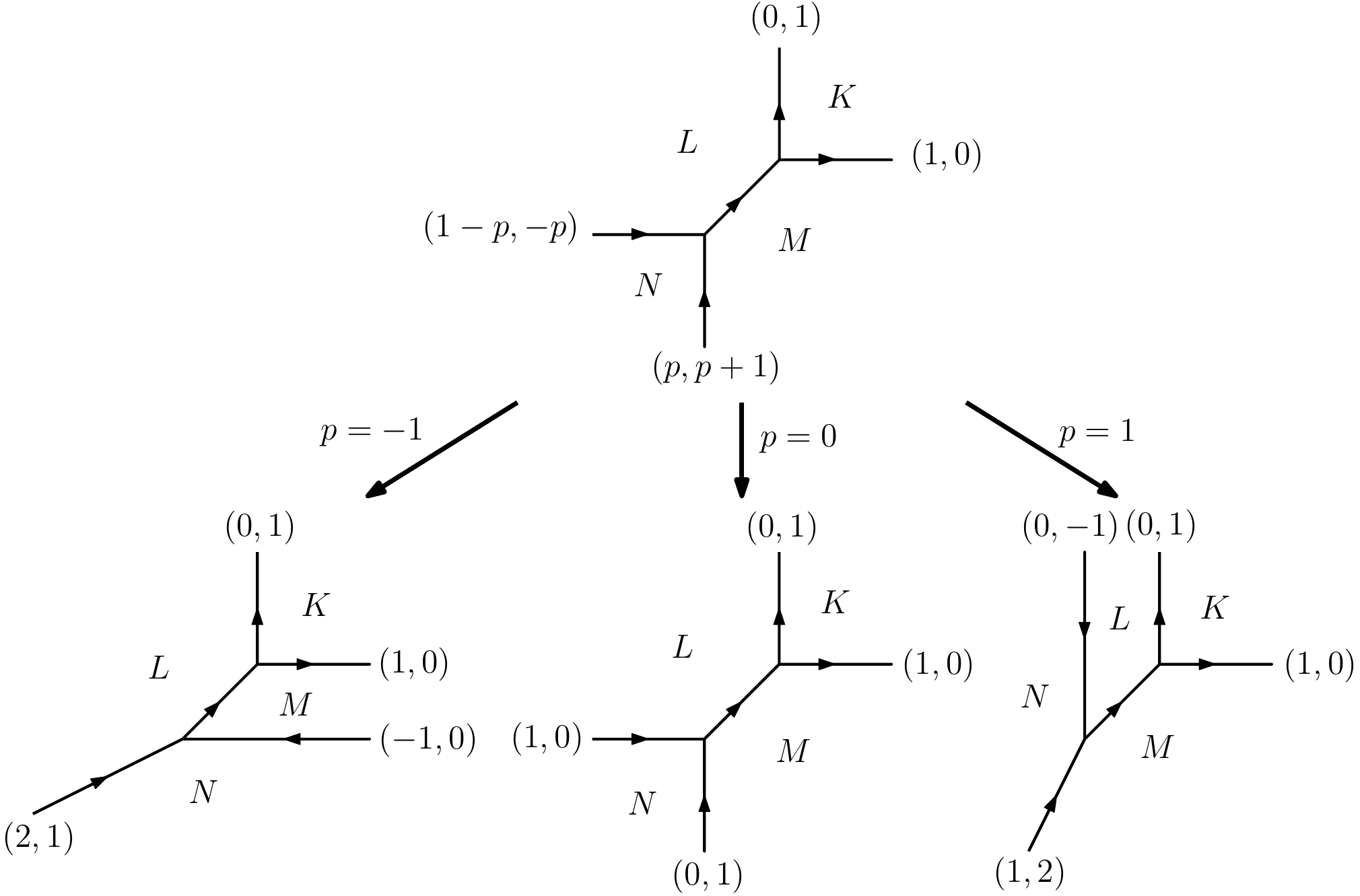}
\caption{Configurations containing a simple finite leg segment. The $p=\pm 1$ cases are related by S-duality and it acts within the family of $p=0$ algebras.}
\label{gluing3}
\end{figure}

One can see that requiring a single internal edge (or equivalently two vertices), only $p=-1,0,1$ are the possible diagrams one might consider. For all the other values of $p$ the branes would intersect. After fixing one of the two vertices to the standard orientation, figure \ref{gluing3} shows all the three possibilities. Note that the S-duality action maps the families $p=-1 \leftrightarrow p=1$. We give a BRST definition for the $p=-1$ case. The $p=1$ algebras can be identified with those by S-duality. On the other hand, $p=0$ example is self-dual under the S-duality action and we expect the corresponding algebras to have dual BRST descriptions in general. This section gives a BRST definition of almost all algebras associated to diagrams with these configurations. The only exceptions are the configurations for $p=0$ with D3-branes satisfying the following four conditions $L>N, M>K,M>N,L>K$.

\subsection{Algebras of type $1|1$ (resolved conifold diagram)}

\begin{wrapfigure}{l}{0.31\textwidth}
\vspace{-15pt}
\begin{center}
\includegraphics[width=0.29\textwidth]{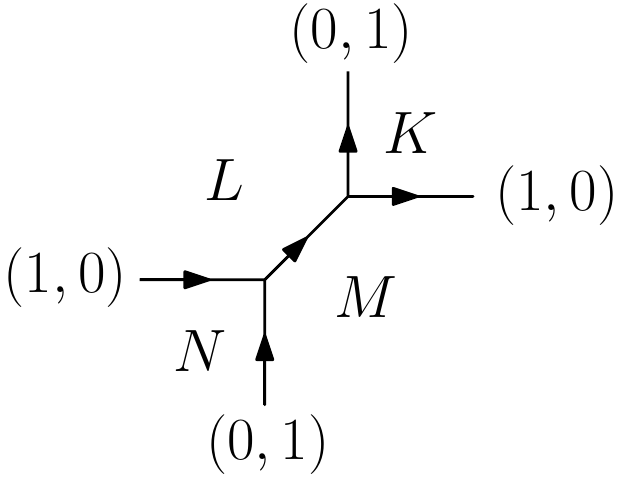}
\end{center}
\vspace{-20pt}
\end{wrapfigure}
In this section we want to discuss the junction of two Y-algebras that corresponds to the resolved conifold diagram as in the figure on the left. We first introduce a convenient notation $\mathcal{L}^{1|1}_{K,\bar{L},M,\bar{N}}[\Psi]$ for these algebras. The label $\mathcal{L}^{1|1}$ refers to algebras associated to a linear chain of $(n,1)$ five-branes to which one D5-brane is attached from the left and the other one is attached from the right. Furthermore, we overline the numbers $\bar{L}$ and $\bar{N}$ of D3-branes ending on the $(n,1)$ branes from the left and we leave $K$ and $M$ for the D3-branes ending from the right. This labeling will be used also for more complicated diagrams with a linear chain.

From the gluing point of view the algebra is a conformal extension of
\begin{eqnarray}
\mathcal{L}^{1|1}_{K,\bar{L},M,\bar{N}}[\Psi] \supset Y_{L,M,K}[\Psi] \times Y_{M,L,N}[\Psi]
\end{eqnarray}
by bimodules labeled by representations of $U(M|L)$. If we specialize the general formulas from the section \ref{sec:gluing}, the scaling dimension of the fundamental gluing field is given by
\begin{eqnarray}
h({\Box})=1+\rho=1+ \frac{K+N-L-M}{2}.
\label{rho_conifold}
\end{eqnarray}
At the level of characters, the BRST construction result must agree with the gluing proposal
\begin{eqnarray}
\chi [\mathcal{L}^{1|1}_{K,\bar{L},M,\bar{N}}]=\sum_\mu \chi [Y_{L,M,K}[\Psi]](M^3_\mu)\ \chi [Y_{M,L,N}[\Psi]](M^3_\mu).
\end{eqnarray}
Another property of $1|1$ algebras is a $\mathbbm{Z}_2\times \mathbbm{Z}_2$ duality action generated by the transformations
\begin{eqnarray}\nonumber
\mathcal{L}^{1|1}_{K,\bar{L},M,\bar{N}}[\Psi] \quad &\leftrightarrow&\quad \mathcal{L}^{1|1}_{K,\bar{M},L,\bar{N}}\left [\frac{1}{\Psi}\right ] \\
\mathcal{L}^{1|1}_{K,\bar{L},M,\bar{N}}[\Psi]\quad &\leftrightarrow&\quad \mathcal{L}^{1|1}_{N,\bar{L},M,\bar{K}}\left [\frac{1}{\Psi}\right ].
\label{Z2Z2}
\end{eqnarray}
These duality transformations can be derived from the S-duality transformation together with rotations and the parity transformation in the same way as the triality action on Y-algebras.

\subsubsection{BRST construction}

In the Kapustin-Witten twisted theory, the path integral of the configuration localizes to the path integral of the complexified $U(K|L)_\Psi$ and $U(M|N)_\Psi$ Chern-Simons theories connected by a nontrivial boundary condition that is a combination of oper boundary condition and continuity condition. The BRST definition of the VOAs is then a reduction that implements the boundary condition on the two $U(K|L)_\Psi$ and $U(M|N)_\Psi$ Kac-Moody factors coming from the restriction of the gauge fields of the upper and lower CS theory to the interface.

Implementing the constraints coming from the boundary conditions for $K>L$ and $N>M$ by a BRST reduction, one expects the final VOA to be a combination of the Drinfeld-Sokolov reduction of $U(K|L;\Psi)$ with respect to the principal $sl_2$ embedding inside the $(K-M)\times (K-M)$ block in the $U(K)$ bosonic part of $U(K|L)$, the Drinfeld-Sokolov reduction with respect to the principal embedding in the $(L-N)\times (L-N)$ block in the other $U(L)$ bosonic part and the coset with respect to the remaining $U(M|N;\Psi)$ Kac-Moody algebra. In analogy with the construction of \cite{Gaiotto:2017euk}, one writes for such a combined BRST reduction
\begin{eqnarray}
\mathcal{L}_{K,\bar{L},M,\bar{N}}^{1|1}[\Psi]=\frac{\overline{\mathcal{DS}}_{L-N}\left [\mathcal{DS}_{K-M}[U(L|K;\Psi)]\right ]}{U(M|N;\Psi)}.
\end{eqnarray}

In expressions of this form, we need to be careful what we mean by a sequence of Drinfeld-Sokolov reductions. There are two natural definitions. The first natural choice would be to pick a grading associated to the sum of the Cartan elements of the two $sl_2$ embeddings and constrain the fields with positive weight with respect to this combined element as in the case of the standard DS-reduction. We can see that this choice would be symmetric with respect to both trivalent junctions of the diagram. This would not match the predictions from the gluing suggesting that this is not the right thing to do (this symmetric variant case will be discussed elsewhere \cite{next2}).

The other possibility is to slightly modify the standard construction by doing the reduction in two steps. Firstly, we need to constrain the components with positive weight with respect to the first embedding associated to the $U(K|M)$ block (DS-reduction associated to the upper vertex as in the case of Y-algberas). Classically (and at the level of characters), this first constraint decomposes $U(K|L;\Psi)$ fields as
\begin{eqnarray}
\mathcal{DS}_{K-M}:U(K|L;\Psi) \ \rightarrow \ \mathcal{W}_{K-M}\times U(M|L;\Psi-1) \times \mathcal{S}^{M|L}_{\frac{K-M}{2}}
\end{eqnarray}
where $\mathcal{W}_{K-M}$ denotes the fields of the $\mathcal{W}_{K-M}$ algebra and $\mathcal{S}^{M|L}_{\frac{K-M}{2}}$ a system of $M$ symplectic bosons and $L$ fermions with the conformal dimension shifted by $\frac{N-M}{2}$. The first reduction produces an algebra containing the $U(M|L;\Psi-1)$ Kac-Moody algebra as a subalgebra coming from the the $U(K|L;\Psi)$ currents modified by off-diagonal ghosts. In the second step, one needs to constrain the fields of the Kac-Moody algebra $U(M|L;\Psi-1)$ with shifted level by setting to zero fields with positive weight\footnote{Remember that only half of the fields with weight $\frac{1}{2}$ need to be constrained as in the case of Y-algebras.} with respect to the Cartan element of the $sl_2$ embedding associated to the second vertex. The algebra decomposes classically (and at the level of characters) as
\begin{eqnarray}
\overline{\mathcal{DS}}_{L-N}: U(M|L;\Psi-1)\ \rightarrow \ \mathcal{W}_{L-N}\times U(M|N;\Psi) \times \overline{\mathcal{S}}^{M|N}_{\frac{L-N}{2}}
\end{eqnarray}
where $\overline{\mathcal{S}}^{M|N}_{\frac{L-N}{2}}$ now contains $M$ fermionic and $L$ bosonic generators that refers to the fact that corresponding D5-brane ends from the opposite direction. The $\mathcal{S}^{M|L}_{\frac{K-M}{2}}$ fields from the first step are left unconstrained but the modification term that needs to be added to the stress-energy tensor in the second step splits them into fields
\begin{eqnarray}
\overline{\mathcal{DS}}_{L-N}:\mathcal{S}^{M|L}_{\frac{K-M}{2}}\ \rightarrow \ \mathcal{S}^{M|N}_{\frac{K-M}{2}} \times \prod^{\rho+L-N-\frac{1}{2}}_{i=\rho+\frac{1}{2}} \mathcal{F}_i
\end{eqnarray}
where $L-N$ components were split into fermionic fields $\mathcal{F}_i$ with dimensions
\begin{eqnarray}
\rho+1,\rho+2,\dots, \rho+L-N.
\end{eqnarray}
An explicit example of the constraints one needs to impose is given in the appendix \ref{BRSTexamples}.

In the case when DS-reduction is with respect to a one dimensional block (i.e. $K-M=1$ or $L-N=1$), no constraints need to be imposed remembering that fields from the off-diagonal block of the first reduction are not constrained in the second step. Similarly if $K-M=0$ or $L-N=0$ vanishes, instead of constraining the fields, one needs to introduce extra $\mathcal{S}^{M|L}$ or $\overline{\mathcal{S}}^{M|N}$ fields into the system, similarly as in the case of the trivalent vertex and use currents modified by bilinears in these extra fields in the BRST reductions of the following steps.

After the two DS-reductions, the algebra still contains a $U(M|N;\Psi)$ factor as a subalgebra and one should take a coset with respect to this factor. By coset, we mean an equivariant BRST reduction with respect to the BRST charge that glues this $U(M|N;\Psi)$ subalgebra with $U(M|N;-\Psi)$ algebra induced from the bottom Chern-Simons theory. Note that the shifted levels of the two factors are opposite which is a necessary condition for the BRST charge to be nilpotent.

An analogous definition of the algebra can be given in the case $M>K$ and $N>L$ with $K\leftrightarrow M$, $N\leftrightarrow L$ and the two DS-reductions interchanged (this correspond to rotation of the diagram by $180^\circ$).

We can also define a similar reduction for the case when the number of D3-branes decreases from the top and from the bottom until the two series of D3-brane numbers meet. In the case of the resolved conifold diagram, this corresponds to $K>M$ and $N>L$. One can then read the boundary conditions from both sides, show that the resulting algebras contain two Kac-Moody algebras of opposite level and then equivariantly glue these factors. The resulting algebra is the BRST reduction of the system 
\begin{eqnarray} 
\mathcal{DS}_{K-M}[U(K|L;\Psi)] \times \overline{\mathcal{DS}}_{N-L}[U(M|N;-\Psi)] \times gh
\label{BRST2}
\end{eqnarray}
that glues the two $U(M|L;\Psi-1) \times U(M|L;-\Psi+1)$ subalgebras. As usual, $gh$ in the expression above denotes the ghosts needed to implement the gluing. Note that combining the fields in the fundamental representation of the remaining $U(M|N)$ factors coming from the two DS-reduction into the $U(M|N)$ invariant combinations  gives rise to fields of dimensions starting with $\rho+1$. This is consistent with the gluing picture and BRST reduction above athough the origin of the fields is slightly different. 

\subsubsection{Central charge and characters}

Having defined the algebras by a BRST reduction, one can calculate the central charge straightforwardly. The result is\footnote{Note that the pole at $\Psi=1$ in the formula disappeared and the poles at $0$ and $\infty$ are multiplied with two factors associated to the two external legs with given asymptotics as expected from the orientation of the infinite five-brane segments.}
\begin{eqnarray}\nonumber
c\left[\mathcal{L}^{1|1}_{K,\bar{L},M,\bar{N}}[\Psi]\right] & = & \Psi\left ( (L-N)((L-N)^2-1)-(K-M)((K-M)^2-1) \right )\\ \nonumber
&&+\frac{1}{\Psi}\left ((L-K)((L-K)^2-1)-(N-M)((N-M)^2-1) \right )\\ \nonumber
&&+(L-N+M-K) (L^2+L N-4 L M+L K-2 N^2\\
&&+N M+2 N K+M^2+M K-2 K^2+1).
\end{eqnarray}
The details of the computation are given in appendix \ref{cc11}.

Having central charge of a general $1|1$ algebra, we can now test the predictions of the gluing construction. We conjectured that the central charge of the glued algebra is simply a sum of the central charges associated to its vertices and indeed
\begin{eqnarray}
c\left [\mathcal{L}^{1|1}_{K,\bar{L},M,\bar{N}}[\Psi]\right ]=c \left [ Y_{L,M,K}[\Psi]\right ]+ c\left [ Y_{M,L,N}[\Psi]\right ]
\end{eqnarray}
so the extension is conformal. Moreover, one can check that the central charge is invariant under the $\mathbbm{Z}_2\times \mathbbm{Z}_2$ duality action (\ref{Z2Z2}).

The vacuum character of the resulting algebra can be also computed straightforwardly following the description outlined above. One finds a general expression
\begin{eqnarray}
\chi \left [\mathcal{L}^{1|1}_{K,\bar{L},M,\bar{N}}\right ]=\chi_{\mathcal{W}_{K-M}}\chi_{\mathcal{W}_{L-N}} \prod_{r=\rho+\frac{1}{2}}^{\rho+L-N-\frac{1}{2}}\chi_{r}^{\mathcal{F}} \oint dV_{M,N}\chi^{N|M}_{\frac{K-M}{2}}(x_j,y_i)\chi^{M|N}_{\frac{L-N}{2}}(y_i,x_j).
\end{eqnarray}
Note that the variables $x_i$ and $y_j$ in the two symplectic boson factors interchange. We can identify the first two factors with the vacuum characters of $\mathcal{W}_{K-M}$ and $\mathcal{W}_{L-N}$ algebras coming from the diagonal blocks of DS-reduction, the factors $\chi_{i}^{\mathcal{F}}$ coming from the $L-N$ fermionic fields with a shifted level and the integral projecting to the $U(M|N)$ invariant combinations of the fundamental fields. Explicit expressions for these building blocks can be found in appendix \ref{Characters}.

To write the characters of more general modules associated to Wilson lines supported at the two NS5-like interfaces one only needs to insert the corresponding Schur polynomials into the formula above in the same way as in the case of Y-algebras. \

\subsubsection{Matching characters}

In this section we discus various examples of $1|1$ algebras at the level of characters to show the match between the results of the gluing and the BRST construction.

\textbf{Example 1: $ \mathcal{L}^{1|1}_{0,\bar{N},0,\bar{0}}$}

\begin{wrapfigure}{l}{0.18\textwidth}
\vspace{-15pt}
\begin{center}
\includegraphics[width=0.17\textwidth]{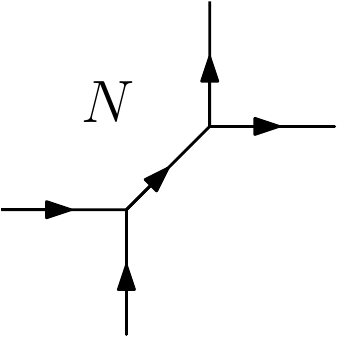}
\end{center}
\vspace{-20pt}
\end{wrapfigure}
The first example is the algebra $\mathcal{L}^{1|1}_{0,\bar{N},0,\bar{0}}$ related to $U(1) \times \mathcal{W}_{N}$ by a flip transition that will be discussed later. The algebra has BRST definition as
\begin{eqnarray}
 \mathcal{L}^{1|1}_{0,\bar{N},0,\bar{0}}[\Psi] = \mathcal{DS}_N\left [U(N;-\Psi) \times \mathcal{F}^{N}\right ].
\end{eqnarray}
The BRST charge implementing the DS-reduction is the one associated to the principal embedding in $U(N)$ and producing $U(1) \times \mathcal{W}_N$ algebra but with the currents $J_{ij}$ of the $U(N;-\Psi)$ Kac-Moody algebra modified by the fermionic bilinears $\hat{J}_{ij}=J_{ij}+\chi_i \psi_j$. These new currents can be shown to produce a new $U(N;-\Psi+1)$ Kac-Moody algebra with shifted level.

Applying the construction described in the previous section, the DS-reduction of the $U(N)$ factor produces the character of $\mathcal{W}_N$ and introduces a shift of the dimensions of the fermionic fields by $-\frac{N-1}{2},\dots, \frac{N-1}{2}$. The character following from the BRST definition is then
\begin{eqnarray}
\chi \left [ \mathcal{L}^{1|1}_{0,\bar{N},0,\bar{0}}\right ]=\prod_{n=0}^{\infty}\prod_{m=1}^N\frac{1}{1-q^{n+m}}\left (1+q^{m+n-\frac{N}{2}}\right )^2.
\label{char1}
\end{eqnarray}
In terms of the gluing, the algebra should be decomposable as a conformal extension of two $U(1) \times \mathcal{W}_N$ algebras
\begin{eqnarray}
\mathcal{L}^{1|1}_{0,\bar{N},0,\bar{0}}[\Psi]=Y_{N,0,0}[\Psi]\boxtimes_N Y_{0,N,0}[\Psi].
\end{eqnarray}
At the level of characters, the above expression for the character (\ref{char1}) should decompose into a sum of the form
\begin{multline}
\chi \left [ \mathcal{L}^{1|1}_{0,\bar{N},0,\bar{0}}\right ] = \\
= \prod_{n=0}^{\infty}\prod_{m=1}^N\frac{1}{(1-q^{n+m})^2}\sum_{\mu}q^{\frac{1}{2}\sum_i \mu^2_i+\frac{1}{2}\sum_i (N-2i+1)\mu_i} s^2_\mu \left (x_i=q^{\frac{1}{2}(N-2i+1)}\right )
\end{multline}
where $s_\mu(x_i)$ is the Schur polynomial. We have checked the equality for $N=1,2,3$.

\textbf{Example 2: $\mathcal{L}^{1|1}_{1,\bar{1},0,0}$ vs. $\mathcal{L}^{1|1}_{1,0,\bar{1},0}$}

\begin{wrapfigure}{l}{0.18\textwidth}
\vspace{-15pt}
\begin{center}
\includegraphics[width=0.17\textwidth]{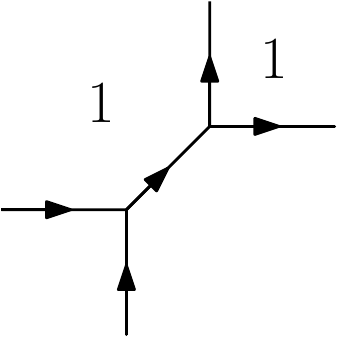}
\end{center}
\vspace{-20pt}
\end{wrapfigure}
The next example is simply the Kac-Moody algebra
\begin{eqnarray}
\mathcal{L}^{1|1}_{1,\bar{1},0,0}[\Psi]=U(1|1;\Psi)
\end{eqnarray}
with the vacuum character
\begin{eqnarray}
\chi\left [\mathcal{L}^{1|1}_{1,\bar{1},0,0}\right ]=\prod_{n=0}^{\infty}\left (\frac{1+q^{1+n}}{1-q^{1+n}}\right )^2.
\label{ref1}
\end{eqnarray}
The S-dual definition of the algebra is
\begin{eqnarray}
\mathcal{L}^{1|1}_{1,0,\bar{1},0}\left [\frac{1}{\Psi}\right ]=\frac{U(1;\frac{1}{\Psi}) \times \mathcal{F}\times \mathcal{B}}{U(1; \frac{1}{\Psi})}.
\end{eqnarray}
By the coset, we really mean the BRST reduction with respect to the BRST charge
\begin{eqnarray}
Q=\oint dz (J^{(1)}-J^{(2)}+\chi \psi+xy)
\end{eqnarray}
where $J^{(1)}$ and $J^{(2)}$ are currents of the two $U(1;\frac{1}{\Psi})$ and $U(1;-\frac{1}{\Psi})$ factors, $\chi, \psi$ are the additional fermions and $x,y$ are the added symplectic bosons. The character of this algebra is given by
\begin{eqnarray}
\chi \left [\mathcal{L}^{1|1}_{1,0,\bar{1},0}\right] = \prod_{n=0}^\infty \frac{1}{1-q^{1+n}}\oint\frac{dx}{x}\prod_{n=0}^\infty \frac{(1+xq^{n+\frac{1}{2}})(1+x^{-1}q^{n+\frac{1}{2}})}{(1-xq^{n+\frac{1}{2}})(1-x^{-1}q^{n+\frac{1}{2}})}
\end{eqnarray}
We can now use the identities from appendix \ref{form} to expand the two factors under the integral sign and perform the integration. One gets
\begin{eqnarray}
\chi \left [\mathcal{L}^{1|1}_{1,0,\bar{1},0}\right] = \prod_{n=0}^{\infty}\frac{1}{(1-q^{n+1})^3}\sum_{n=0}^{\infty}\sum_{m=-n}^{n}(-1)^{n-m}q^{\frac{n(n+1)}{2}}.
\label{ref22}
\end{eqnarray}
This expression can be identified with (\ref{ref1}) using the formulas, in appendix \ref{form} with $z=1$.

We can also construct the same character from the gluing procedure by summing over $M^{3}_\mu$ modules of $Y_{1,0,1}[\Psi]$ and $Y_{0,1,0}[\Psi]$. From the perspective of gluing, it is obvious that the two algebras $X_{1,0,0,1}[\Psi]$ and $X_{0,0,1,1}[-\Psi]$ have the same vacuum character. The expression predicted from the gluing is
\begin{eqnarray}
\chi\left [\mathcal{L}^{1|1}_{1,\bar{1},0,0}\right ] =\prod_{n=0}^{\infty}\frac{1}{(1-q^{n+1})^3}\sum_{n=|m|}^{\infty}(-1)^{n-m}q^{\frac{n(n+1)}{2}}.
\end{eqnarray}
It is simple to show that this expression is equivalent to the one in (\ref{ref22}). We thus conclude that all the three expressions for the character in this simple example agree.

\textbf{Example 3: $\mathcal{L}^{1|1}_{2,\bar{1},0,\bar{0}}$ vs $\mathcal{L}^{1|1}_{2,\bar{0},1,\bar{0}}$}

\begin{wrapfigure}{l}{0.18\textwidth}
\vspace{-15pt}
\begin{center}
\includegraphics[width=0.17\textwidth]{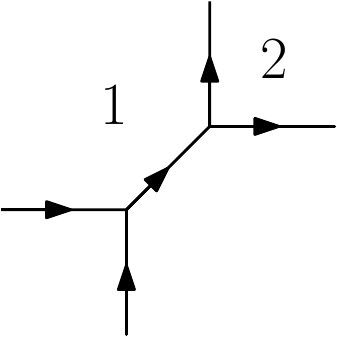}
\end{center}
\vspace{-20pt}
\end{wrapfigure}
The third algebra we consider has the following BRST definition,
\begin{eqnarray}
\mathcal{L}^{1|1}_{2,\bar{1},0,\bar{0}}[\Psi] = \mathcal{DS}_2[U(2|1;\Psi)].
\end{eqnarray}
Explicitly, the $\mathcal{DS}_2$ BRST charge is given by
\begin{eqnarray}
Q=\oint dz \left [ (J_{12}-1)c_{12}+J_{13}\gamma_{13}\right ].
\end{eqnarray}
This BRST reduction was already mentioned in \cite{Gaiotto:2017euk} and is known to be a  realization of $\mathcal{N}=2$ super Virasoro algebra times $U(1)$ current algebra \cite{Kac:qy} with well known vacuum character
\begin{eqnarray}
\chi \left [\mathcal{L}^{1|1}_{2,\bar{1},0,\bar{0}}\right ]=\prod_{n=0}^{\infty}\frac{\left (1+q^{n+\frac{3}{2}}\right )^2}{\left (1-q^{1+n}\right )^2(1-q^{2+n})}.
\label{char4}
\end{eqnarray}

\begin{wrapfigure}{l}{0.18\textwidth}
\vspace{-15pt}
\begin{center}
\includegraphics[width=0.17\textwidth]{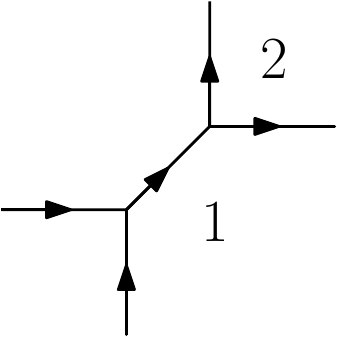}
\end{center}
\vspace{-20pt}
\end{wrapfigure}
The dual BRST construction is given by
\begin{eqnarray}
\mathcal{L}^{1|1}_{2,\bar{0},1,\bar{0}}\left [\frac{1}{\Psi}\right ]=\frac{U(2;\frac{1}{\Psi}) \times \mathcal{F}^{U(1)}}{U(1;\frac{1}{\Psi})}
\end{eqnarray}
by which we mean a BRST reduction with respect to
\begin{eqnarray}
Q=\oint dz (J^{(1)}_{22}+\chi \psi-J^{(2)})
\end{eqnarray}
for $J^{(1)}_{22}$ a component of the $U(2)$ current and $J^{(2)}$ the $U(1)$ current. The character is given by
\begin{eqnarray}
\chi \left [\mathcal{L}^{1|1}_{2,\bar{1},0,\bar{0}}\right ] = \oint \frac{dz}{z}\prod_{n=0}^\infty \frac{(1+zq^{n+\frac{1}{2}})(1+zq^{n+\frac{1}{2}})}{(1-q^{n+1})(1-zq^{n+1})(1-z^{-1}q^{n+1})}.
\label{char2}
\end{eqnarray}
Using the summation and the contour integral formulas from the appendix, one can rewrite the integral as a sum
\begin{eqnarray}
\chi \left [\mathcal{L}^{1|1}_{2,\bar{1},0,\bar{0}}\right ] = \prod_{n=0}^{\infty}\frac{1}{(1-q^{n+1})^4}\sum_{s=-\infty}^\infty \sum_{n=0}^{\infty}(-1)^n(1-q^{n+1})q^{\frac{s^2+n(n+1)}{2}+(n+1)s}.
\label{char3}
\end{eqnarray}
One can analogously expand (\ref{char4}) and check that both expansions agree.

On the other hand, according to the gluing construction, the algebra can be thought of as a conformal extension
\begin{eqnarray}
\mathcal{L}^{1|1}_{2,\bar{1},0,\bar{0}}[\Psi]\supset Y_{1,0,2}[\Psi]\times Y_{0,1,0}[\Psi].
\end{eqnarray}
At the level of characters, one gets
\begin{eqnarray}
\prod_{n=0}^{\infty} \frac{1}{(1-q^{n+1})^4}\sum_{s=-\infty}^\infty q^{\frac{s^2}{2}}\left ( \sum_{n=s}^{\infty}(-1)^{n+s}q^{\frac{n(n+1)-s(s-1)}{2}}+ \sum_{n=s+1}^{\infty}(-1)^{n+s}q^{\frac{n(n+1)-s(s-1)}{2}}\right ).
\end{eqnarray}
This expression can be shown to be equivalent to (\ref{char3}). In the previous section we have already seen that the full $\mathcal{N}=2$ super Virasoro algebra times $U(1)$ decomposes as predicted by the gluing construction.

\textbf{Example 4: $\mathcal{L}^{1|1}_{2,\bar{1},0,\bar{1}}$ vs $\mathcal{L}^{1|1}_{2,\bar{0},1,\bar{1}}$}

\begin{wrapfigure}{l}{0.18\textwidth}
\vspace{-15pt}
\begin{center}
\includegraphics[width=0.17\textwidth]{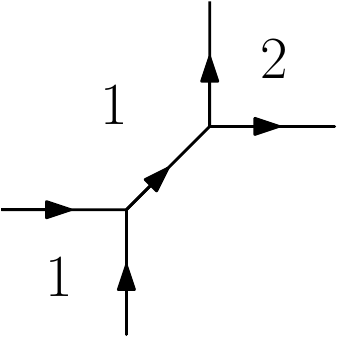}
\end{center}
\vspace{-20pt}
\end{wrapfigure}
The next example tests the BRST construction of type (\ref{BRST2}). Let us start with the conventional BRST reduction associated to $\mathcal{L}^{1|1}_{2,\bar{1},0,\bar{1}}$. The VOA can be identified with
\begin{eqnarray}
\mathcal{L}^{1|1}_{2,\bar{1},0,\bar{1}}[\Psi]=\frac{\mathcal{DS}_2[U(2|1;\Psi)]\times \bar{\mathcal{S}}^{0|1}}{U(0|1;\Psi)}.
\end{eqnarray}
The DS-reduction of the first step is with respect to the standard BRST charge
\begin{eqnarray}
Q = \oint dz \left [ (J_{12}-1)c_{12}+J_{13}\gamma_{13}\right ].
\end{eqnarray}
The cohomology contains the $\hat{J}_{33}$ current containing bilinears in ghosts associated to the off-diagonal components and the coset is identified with the BRST reduction that glues the $\hat{J}_{33}+\chi \psi$ current with the extra $U(1;\Psi)$ current coming from the lower CS theory for $\psi, \chi$ the fermionic fields $ \bar{\mathcal{S}}^{0|1}$.

At the level of characters, the DS-reduction $\mathcal{DS}_2[U(2|1;\Psi)]$ produces Virasoro algebra together with the $U(1)$ current and the fermion with the conformal dimension shifted to $\frac{3}{2}$ and charged under $U(0|1;\Psi)$. The character of the resulting algebra is
\begin{eqnarray}
\chi \left [\mathcal{L}^{1|1}_{2,\bar{1},0,\bar{1}}\right ]=\prod_{n=0}^\infty \frac{1}{(1-q^{n+1})(1-q^{n+2})}\oint \frac{dz}{z}\prod_{n=0}^\infty\frac{(1+zq^{n+\frac{3}{2}})(1+z^{-1}q^{n+\frac{3}{2}})}{(1-zq^{n+\frac{1}{2}})(1-z^{-1}q^{n+\frac{1}{2}})}.
\label{ref4}
\end{eqnarray}

\begin{wrapfigure}{l}{0.18\textwidth}
\vspace{-15pt}
\begin{center}
\includegraphics[width=0.17\textwidth]{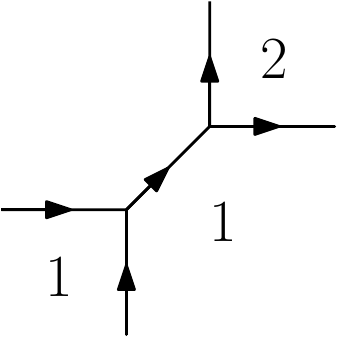}
\end{center}
\vspace{-20pt}
\end{wrapfigure}
The algebra discussed above is related by S-duality to $\mathcal{L}^{1|1}_{2,\bar{0},1,\bar{1}}[\frac{1}{\Psi}]$ which is defined as a BRST complex
\begin{eqnarray}
\left \{U(1|1;-\frac{1}{\Psi}) \times U(2;\Psi) \times \{b,c\},\ Q = \oint \left (J^{(1)}_{22}- J^{(2)}_{22}\right )c \right \}
\end{eqnarray}
where the coset BRST current sews the $J^{(1)}_{22}$ component of the $U(2;\Psi)$ Kac-Moody algebra and the $J^{(2)}_{22}$ component of the $U(1|1;-\Psi)$ Kac-Moody algebra. At the level of the vacuum characters, one gets
\begin{eqnarray}
\chi \left [\mathcal{L}^{1|1}_{2,\bar{0},1,\bar{1}}\right ] = \prod_{n=0}^\infty \frac{1}{(1-q^{n+1})^2}\oint \frac{dz}{z}\prod_{n=0}^\infty\frac{(1+zq^{n+1})(1+z^{-1}q^{n+1})}{(1-zq^{n+1})(1-z^{-1}q^{n+1})}.
\end{eqnarray}
The numerator of the integrand can be expanded and then integrated using the formulas in appendix \ref{form} to get
\begin{eqnarray}
\chi \left [\mathcal{L}^{1|1}_{2,\bar{0},1,\bar{1}}\right ] = \prod_{n=0}^\infty \frac{1}{(1-q^{n+1})^5}\sum_{s=0}^\infty \sum_{s=-m}^m\sum_{n=0}^\infty (-1)^{s+m+n}(1-q^{n+1})q^{\frac{n(n+1)}{2}+\frac{m(m+1)}{2}+(n+1)s}.
\end{eqnarray}
A similar manipulation can be done with the character (\ref{ref4}) and one obtains the same expansion.

Finally, we can write the algebra as a conformal extension of
\begin{eqnarray}
\mathcal{L}^{1|1}_{2,\bar{1},0,\bar{1}}[\Psi]\supset Y_{1,0,2}[\Psi]\times Y_{0,1,1}[\Psi].
\end{eqnarray}
The vacuum character is then given by
\begin{eqnarray}\nonumber
\chi \left [\mathcal{L}^{1|1}_{2,\bar{1},0,\bar{1}}\right ]&=&\prod_{n=0}^\infty \frac{1}{(1-q^{n+1})^5}\sum_{s=-\infty}^{\infty}\left ( \sum_{n=|s|}^\infty (-1)^{n-m}q^{n(n+1)}\right ) \times \\
&& \times \left ( \sum_{n=s}^\infty (-1)^{n+m}q^{\frac{n(n+1)-m(m-1)}{2}} + \sum_{n=s+1}^\infty (-1)^{n+m}q^{\frac{n(n+1)-m(m+1)}{2}}\right )
\end{eqnarray}
that can be shown to be equivalent to the expansion above.

\textbf{Example 5: Self-dual $\mathcal{L}^{1|1}_{2,\bar{0},1,\bar{2}}$}

\begin{wrapfigure}{l}{0.18\textwidth}
\vspace{-15pt}
\begin{center}
\includegraphics[width=0.17\textwidth]{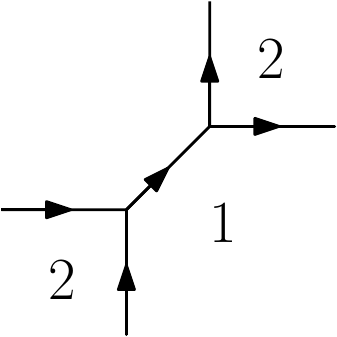}
\end{center}
\vspace{-20pt}
\end{wrapfigure}
In this section, we study an example of decomposition of the algebra $\mathcal{L}^{1|1}_{2,\bar{0},1,\bar{2}}[\Psi]$ into the pair of parafermionic algebras $Y_{0,1,2}[\Psi]$ and $Y_{1,0,2}[\Psi]$. Note that this algebra can be identified by S-duality with $\mathcal{L}^{1|1}_{2,\bar{0},1,\bar{2}}[\frac{1}{\Psi}]$ and the example is actually self-dual. Definition of the algebra using the BRST reduction is in terms of the complex 
\begin{equation}
U(2;\Psi) \times \mathcal{DS}_2[U(1|2;-\Psi)] \times \{ b,c\}
\end{equation}
with BRST charge
\begin{equation}
Q = \oint\left (J^{(1)}_{22}-J^{(1)}_{11}\right )c.
\end{equation}
The BRST charge glues the $J_{11}$ component of $U(1|2;-\Psi)$ with the $J_{22}$ component of $U(2;\Psi)$. The vacuum character of the algebra is
\begin{multline}
\chi \left [\mathcal{L}^{1|1}_{2,\bar{0},1,\bar{2}}\right ] = \\
= \prod_{n=0}^{\infty}\left (\frac{1}{1-q^{1+n}}\right )^2\frac{1}{1-q^{2+n}}\oint \frac{dx}{x}\prod_{n=0}^{\infty}\frac{(1+xq^{\frac{3}{2}+n})(1+x^{-1}q^{\frac{3}{2}+n})}{(1-xq^{1+n})(1-x^{-1}q^{1+n})}.
\label{ref5}
\end{multline}
On the other hand, we expect the algebra to be a conformal extension
\begin{eqnarray}
\mathcal{L}^{1|1}_{2,\bar{0},1,\bar{2}}[\Psi]\supset Y_{0,1,2}[\Psi]\times Y_{1,0,2}[\Psi]
\end{eqnarray}
of the product of two copies of the parafermionic algebra $Y_{0,1,2}$. At the level of characters, one gets
\begin{multline}
\prod_{n=0}^\infty\frac{1}{(1-q^{n+1})^6}\sum_{s=-\infty}^{\infty}q^{\frac{n^2}{2}} \times \\
\times \left (\sum_{n=s}^\infty (-1)^{n+s}q^{\frac{n(n+1)-s(s-1)}{2}}+\sum_{n=s+1}^\infty (-1)^{n+s}q^{\frac{n(n+1)-s(s+1)}{2}}\right )^2
\end{multline}
that can be identified with (\ref{ref5}) after expanding and performing the contour integral.

\textbf{Example 6: $\mathcal{L}^{1|1}_{1,\bar{2},0,\bar{0}}$ vs $\mathcal{L}^{1|1}_{1,\bar{0},2,\bar{0}}$}

\begin{wrapfigure}{l}{0.18\textwidth}
\vspace{-15pt}
\begin{center}
\includegraphics[width=0.17\textwidth]{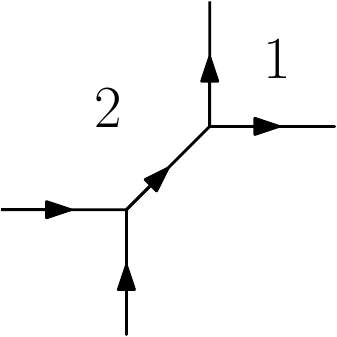}
\end{center}
\vspace{-20pt}
\end{wrapfigure}
The next example is going to probe the gluing proposal in the case when the sum runs over the representations of the non-abelian group $U(2)$. This algebra will later be shown to be related via flip transition to the example 3.

The algebra $\mathcal{L}^{1|1}_{1,\bar{2},0,\bar{0}}$ has the following BRST definition
\begin{eqnarray}
\mathcal{L}^{1|1}_{1,\bar{2},0,\bar{0}}[\Psi] & = & \overline{\mathcal{DS}}_{2}\left[ \mathcal{DS}_1[U(1|2;\Psi)] \right]
\end{eqnarray}
that can be thought of as a BRST reduction of $U(1|2;\Psi) \times \{ b,c\}$ by the BRST charge
\begin{eqnarray}
Q=\oint dz \left (J_{23}-1\right )c_{23}.
\end{eqnarray}
The cohomology will be explicitly constructed later together with its relation to the $\mathcal{N}=2$ super Virasoro algebra. One can immediately write down the expression for the vacuum character of the algebra
\begin{eqnarray}
\chi \left [\mathcal{L}^{1|1}_{1,\bar{2},0,\bar{0}}\right ]=\prod_{n=0}^{\infty}\frac{\left (1+q^{n+\frac{1}{2}}\right )^2\left (1+q^{n+\frac{3}{2}}\right )^2}{\left (1-q^{n+1}\right )^2\left (1-q^{n+2}\right )}.
\end{eqnarray}

\begin{wrapfigure}{l}{0.18\textwidth}
\vspace{-15pt}
\begin{center}
\includegraphics[width=0.17\textwidth]{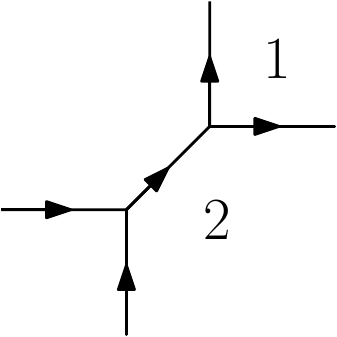}
\end{center}
\vspace{-20pt}
\end{wrapfigure}
There again exists a dual BRST reduction in terms of the coset
\begin{eqnarray}
\mathcal{L}^{1|1}_{1,\bar{0},2,\bar{0}}\left [\frac{1}{\Psi}\right ]=\frac{U(2;-\frac{1}{\Psi}) \times \mathcal{S}^{0|2}}{U(1;-\frac{1}{\Psi})}
\end{eqnarray}
where by this coset, we mean the BRST reduction of the system 
\begin{eqnarray}
U\left (2;-\frac{1}{\Psi}\right )\times U\left (1;\frac{1}{\Psi}\right )\times \{\chi_1, \psi_1\} \times \{\chi_2, \psi_2\} 
\end{eqnarray}
with respect to the BRST charge
\begin{eqnarray}
Q = \oint dz \left (J^{(1)}_{22}+\chi_2 \psi_2-J^{(2)}\right )
\end{eqnarray}
that glues the $J^{(1)}_{22}$ component of the $U(2;-\frac{1}{\Psi})$ algebra with the $U(1;\frac{1}{\Psi})$ current modified by a bilinear in the fermionic fields $\mathcal{S}^{0|2}$. At the level of characters, one gets
\begin{eqnarray}
\chi \left [\mathcal{L}^{1|1}_{1,\bar{0},2,\bar{0}}\right ]=\prod_{n=0}^\infty \frac{\left (1+q^{n+\frac{1}{2}}\right )^2}{(1-q^{n+1})}\oint \frac{dz}{z}\prod_{n=0}^\infty \frac{\left (1+zq^{n+\frac{1}{2}}\right )\left (1+z^{-1}q^{n+\frac{1}{2}}\right )}{\left (1-zq^{n+1}\right )\left (1-z^{-1}q^{n+1}\right )}.
\end{eqnarray} 
The equality of the two expressions follows from the equality of characters already discussed in the example 3. The only difference is the overall contribution from the fermions of conformal weight $\frac{1}{2}$.

It is a non-trivial check to see if the above expressions match the result of gluing. Using our gluing proposal, the algebra is expected to be the conformal extension
\begin{eqnarray}
\mathcal{L}^{1|1}_{1,\bar{2},0,\bar{0}}[\Psi]\supset Y_{2,0,1}[\Psi]\times Y_{0,2,0}[\Psi]
\end{eqnarray}
Using formulas, from appendix \ref{Characters} for modules of $Y_{0,2,1}[\Psi]$ and $Y_{0,0,2}[\Psi]$, we have checked that the character
\begin{eqnarray}
\chi \left [\mathcal{L}^{1|1}_{1,\bar{2},0,\bar{0}}\right ]=\sum_{\mu_1\geq \mu_2}\chi \left [Y_{0,2,1}[\Psi]\right ]\left (D^3_{\mu_1,\mu_2}\right )\chi \left [Y_{2,0,0}[\Psi]\right ]\left (D^3_{\mu_1,\mu_2}\right ).
\end{eqnarray}
agrees with above expressions up to $q^{30}$ term. This provides a non-trivial consistency check for our proposal involving summations over $U(2)$ representations.

\textbf{Example 7: Self-dual $\mathcal{L}^{1|1}_{1,\bar{2},0,\bar{1}}$}

\begin{wrapfigure}{l}{0.18\textwidth}
\vspace{-15pt}
\begin{center}
\includegraphics[width=0.17\textwidth]{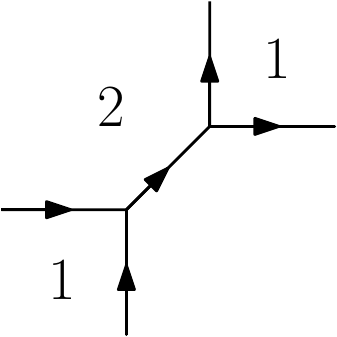}
\end{center}
\vspace{-20pt}
\end{wrapfigure}
Let us now consider the second self-dual example
\begin{eqnarray}
\mathcal{L}_{1,\bar{2},0,\bar{1}}[\Psi]=\mathcal{L}_{1,\bar{2},0,\bar{1}}\left [\frac{1}{\Psi}\right ]=\frac{\overline{\mathcal{DS}}_1\left [\mathcal{DS}_1[U(1|2;\Psi)]\right]}{U(1;-\Psi)}
\end{eqnarray}
that can be thought of as the cohomology of the BRST complex
\begin{eqnarray}
\left \{ U(1|2;\Psi) \times U(1;\Psi) \times \{b,c\},\ Q = \oint dz \left (J^{(1)}_{33}-J^{(2)}\right )c\right \}
\end{eqnarray}
where $J^{(1)}_{33}$ is one of the generators of $U(1|2;\Psi)$ and $J^{(2)}$ is the $U(1;\Psi)$ current. At the level of the characters, we get
\begin{eqnarray}
\chi \left [\mathcal{L}^{1|1}_{1,\bar{2},0,\bar{1}}\right ]=\prod_{n=0}^\infty \frac{\left (1+q^{n+\frac{1}{2}}\right )^2}{(1-q^{n+1})^2}\oint \frac{dz}{z}\frac{\left (1+zq^{z+\frac{1}{2}}\right )\left (1+z^{-1}q^{z+\frac{1}{2}}\right )}{(1-zq^{z+1})(1-z^{-1}q^{z+1})}.
\end{eqnarray}
From the point of view of gluing, this algebra is a conformal extension
\begin{eqnarray}
\mathcal{L}^{1|1}_{1,\bar{2},0,\bar{1}}[\Psi]\supset Y_{2,0,1}[\Psi]\times Y_{0,2,1}[\Psi].
\end{eqnarray}
This example is the second test involving the summation over $U(2)$ representations. Using expressions from the appendix \ref{Characters}, one gets a conjectural equality of the above expression with a double sum
\begin{eqnarray}
\chi \left [\mathcal{L}^{1|1}_{1,\bar{2},0,\bar{1}}\right ]=\sum_{\mu_1\geq \mu_2}\chi[Y_{2,0,1}[\Psi]]\left (D^3_{\mu_1,\mu_2}\right )\chi[Y_{0,2,1}[\Psi]]\left (D^3_{\mu_1,\mu_2}\right )
\end{eqnarray}
that we have checked up to the order $q^{20}$.

\subsection{Algebras of type $0|2$}
\begin{wrapfigure}{l}{0.29\textwidth}
\vspace{-15pt}
\begin{center}
\includegraphics[width=0.27\textwidth]{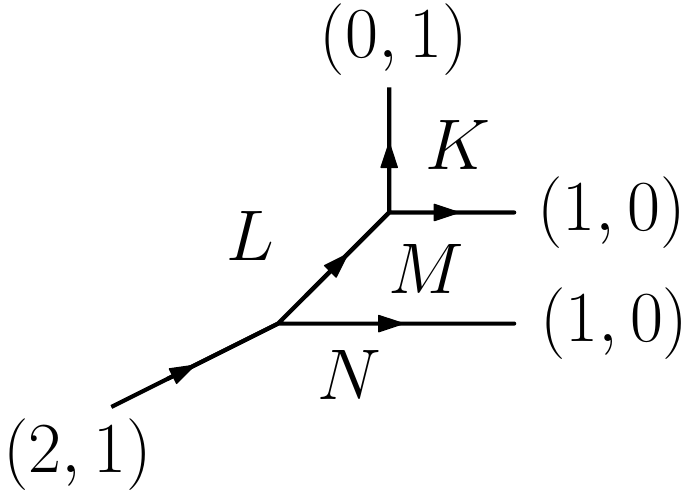}
\end{center}
\vspace{-20pt}
\end{wrapfigure}
In this section, we consider an analogous diagram as the one of the resolved conifold but now with both D5-branes ending from the right as shown in the figure. The discussion will be similar to the one of the previous section but let us highlight few differences.

The glued algebra is a conformal extension of two mutually commuting Y-algebras
\begin{eqnarray}
\mathcal{L}^{0|2}_{K,\bar{L},M,N}[\Psi]\supset Y_{L,M,K}[\Psi]\times Y_{L,N,M}[\Psi-1]
\end{eqnarray}
with gluing matter given by bimodules $M^3_\mu \times M^2_\mu$. Specializing the parameters from the section \ref{sec:gluing} to the case at hand, one finds $p=1$ and
\begin{eqnarray}
h(\Box)=1+\rho =1+ \frac{K+N-2M}{2}
\label{rho2}
\end{eqnarray}
(note that this is independent of $L$). In terms of the characters, we expect the BRST construction to produce
\begin{eqnarray}
\chi\left [\mathcal{L}^{0|2}_{K,\bar{L},M,N}\right ]=\sum_{\mu} \chi[Y_{L,M,K}[\Psi]]\left (M^3_\mu\right )\chi[Y_{L,N,M}[\Psi-1]]\left (M_\mu^2\right )
\end{eqnarray}
and the central charge to be the sum of the central charges of the two Y-algebras.

\subsubsection{BRST construction}

Looking at the system from the IR, the configuration looks like a junction of interfaces between $U(K)$, $U(L)$ and $U(N)$ gauge theories. After a topological twist, the path integral localizes to the path integral of the complexified Chern-Simons theories induced at the NS5 and $(2,1)$ interface glued together by a boundary condition descending from the boundary condition coming from D3-branes ending on five-branes. This boundary condition can be extracted from the boundary conditions discussed in \cite{Gaiotto:2008ac} in the case when $K\geq M \geq N$ or $N \geq M \geq K$ and is a combination of two oper boundary conditions and a continuity condition.

Let us first discuss the $K\geq M \geq N$ case. Imposing the boundary conditions as constraints on the Kac-Moody algebras descending from the upper and the lower CS theories using the BRST procedure leads to the following identification of the VOA
\begin{eqnarray}
\mathcal{L}^{0|2}_{K,\bar{L},M,N}[\Psi] = \frac{\mathcal{DS}_{M-N} \left[\mathcal{DS}_{K-M}[U(K|L;\Psi)]\right]}{U(N|L;\Psi-2)}.
\end{eqnarray}
Note that both DS-reduction are performed in the same block of the bosonic generators of $U(K|L;\Psi)$. Analogously to the resolved conifold algebra, we perform the reduction in three steps. After the first reduction associated to the upper vertex, one obtains an algebra containing the $U(M|L;\Psi-1)$ subalgebra. In the second step one uses the BRST charge implementing the DS reduction associated to the second vertex with the currents of the $U(M|L;\Psi-1)$ algebra with the level shifted by one. Since the second reduction is performed in the same bosonic block of the algebra, the resulting algebra contains subalgebra $U(N|L;\Psi-2)$ with the level shifted by two. In the final step one glues equivariantly the components of the $U(N|L;\Psi-2)$ subalgebra with the extra $U(N|L;-\Psi+2)$ Kac-Moody algebra coming from the lower CS theory.

Under the two DS-reductions, the fields decompose in a similar way as in the case of the $1|1$ algebra. The only difference is that the the $\mathcal{S}_{\frac{K-M}{2}}^{M|L}$ factor from the first reduction decomposes under the second reduction as
\begin{eqnarray}
\mathcal{DS}_{M-N}:\mathcal{S}_{\frac{K-M}{2}}^{M|L}\ \rightarrow \ \mathcal{S}_{\frac{M-N}{2}}^{N|L} \times \prod_{i=\rho+\frac{1}{2}}^{\rho+M-N-\frac{1}{2}}\mathcal{B}_{i}
\end{eqnarray}
producing $M-N$ bosonic fields of the shifted dimension
\begin{eqnarray}
\rho+1,\rho+2,\dots, \rho+M-N.
\end{eqnarray}
Note again the appearance of the parameter $\rho$ from (\ref{rho2}). The fields with shifted dimensions (coming from the off-diagonal bocks containing fields charged under the Cartans of both $sl_2$ embeddings) are now bosonic. The same is true also for the $U(N|L)$ invariant combinations of the symplectic bosons and fermions coming from the two BRST reductions. All the fields of the resulting algebra are bosonic in this case as expected.

An analogous definition can be given in the case of $N\geq M \geq K$ with the factors $K\leftrightarrow N$ and $\Psi \rightarrow -\Psi+2$ exchanged (since this configuration can be obtained from the previous one by an $SL(2,\mathbbm{Z})$ transformation).

One can also define BRST reduction in the case when $K>M$ and $N>M$ by performing the DS-reduction for the upper and to lower vertices independently,
\begin{equation}
\mathcal{DS}_{K-M}[U(K|L;\Psi)] \times \mathcal{DS}_{N-M}U(N|L;-\Psi+2)
\end{equation}
and then gluing the $U(M|L;\Psi-1)$ subalgebra of the first vertex with the $U(M|L;-\Psi+1)$ of the second vertex using BRST (as in the resolved conifold case).

\subsubsection{Central charge and characters}

The central charge is given by (more details of the computation are given in \ref{cc02})\footnote{The structure of poles in $\Psi$ can be again read off from the diagram. Note that the pole at $\Psi=2$ associated to the $(1,2)$ infinite five-brane appeared.}
\begin{eqnarray}
c\left [\mathcal{L}^{0|2}_{K,\bar{L},M,N}[\Psi]\right ]&=&\frac{(N-L)((N-L)^2-1)}{\Psi-2}-\frac{(L-K)((L-K)^2-1)}{\Psi}\\ \nonumber
&&+(((K-M)^2-1)(K-M)+((M-N)^2-1)(M-N))\Psi-M-K\\ \nonumber
&&+(N-M)^2 (-3 L+N+2 M)+(M-K)^2 (-3 L+M+2 K)+2 L\\ \nonumber
&&-\left((N-M)^2-1\right)+(N-M)-M-K(N-M)^2 (-3 L+N+2 M)\\ \nonumber
&&+(M-K)^2 (-3 L+M+2 K)+2 L-\left((N-M)^2-1\right) (N-M).
\end{eqnarray}
It can again be shown to be equal to the sum of the two central charges of the two elementary vertices. The vacuum character is given by an integral formula
\begin{eqnarray}
\chi \left [\mathcal{L}^{0|2}_{K,\bar{L},M,N}\right ]=\chi_{\mathcal{W}_{K-M}}\chi_{\mathcal{W}_{M-N}}\prod_{r=\rho+\frac{1}{2}}^{\rho+M-L-\frac{1}{2}}\chi_r^{\mathcal{B}} \oint dV_{N,M} \chi^{N|M}_{\frac{K-M}{2}}(x_i,y_j)\chi^{N|L}_{\frac{M-N}{2}}(x_i,y_j).
\end{eqnarray}
The characters of the two modules associated to the line defects supported at the NS5 and the $(2,1)$ interface can be computed in a similar way with an extra insertion of the corresponding Schur polynomials.

\subsubsection{Matching characters}

Let us now consider a few familiar examples and match the predictions of the BRST construction with the one coming from the gluing.

\textbf{Example 1: $\mathcal{L}^{0|2}_{0,\bar{L},0,0}$}

\begin{wrapfigure}{l}{0.18\textwidth}
\vspace{-15pt}
\begin{center}
\includegraphics[width=0.17\textwidth]{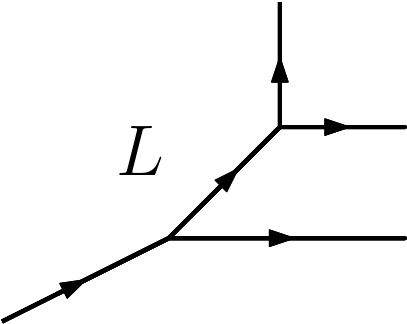}
\end{center}
\vspace{-20pt}
\end{wrapfigure}
The first example is a coset\footnote{This coset has already been discussed in \cite{Creutzig:2017uxh}.}
\begin{eqnarray}
\mathcal{L}^{0|2}_{0,\bar{L},0,0}[\Psi]=\frac{U(L;-\Psi) \times \mathcal{S}^{0|L} \times \mathcal{S}^{0|L}}{U(L;-\Psi+2)}
\end{eqnarray}
again defined as the BRST reduction of the complex
\begin{eqnarray}
U(L)_{-\Psi}\times \mathcal{S}^{0|L}\times  \mathcal{S}^{0|L}\times U(L)_{\Psi-2}\times gh
\end{eqnarray}
with respect to the BRST charge
\begin{eqnarray}
Q_{BRST} = \oint dz (J^i_j-\tilde{J}^i_j+\psi^i\chi_j+\tilde{\psi}^i\tilde{\chi}_j).
\end{eqnarray}
where $(\psi_i,\chi_i)$ and $(\tilde{\psi}_i,\tilde{\chi}_i)$ are the two sets of $L$ fermionic fields. The fields in the cohomology are formed by $U(L)$ invariant combinations of the pair of fermions in the fundamental representation. In particular, the final algebra contains $U(2;1)$ subalgebra generated by derivatives of products of the bosonic bilinears.

One can check that the vacuum character of the BRST construction
\begin{eqnarray}
\chi \left [\mathcal{L}^{0|2}_{0,\bar{L},0,0}\right ]=\frac{1}{K!}\oint \prod_{i=1}^K \frac{dx_i}{x_i} \prod_{i=1}^K \prod_{n=0}^\infty \left (1+x_iq^{n+\frac{1}{2}}\right )^2\left (1+x_i^{-1}q^{n+\frac{1}{2}}\right )^2
\end{eqnarray}
matches the one from the gluing
\begin{eqnarray}
\chi \left [\mathcal{L}^{0|2}_{0,\bar{L},0,0}\right ]=\prod_{n=0}^\infty \prod_{m=1}^L\frac{1}{1-q^{n+m}}\sum_{\mu}q^{\sum_{i=1}^K\mu_i^2+\sum_{i=1}^K(K+1-2i)\mu_i} s^2_{\mu}(x_i=q^{\frac{1}{2}(K-2i+1)}).
\end{eqnarray}
We have checked that they agree for $K=0,1,2$ up to order $q^{30}$, $q^{20}$, and $q^{10}$ respectively.

It is worth mentioning that this diagram has a natural interpretation in the case when $n$ D5-branes end on the $(n,1)$ branes (the diagram we discuss later) and a stack of $L$ D3 branes attached at the face on the left. The corresponding algebra is a cohomology gluing $n$ fermionic fields in the $U(L)$ invariant way. The algebra will include $U(n;n-1)$ subalgebra extended by other fields.\footnote{Note that similar algebras \cite{Nishioka:2011jk,Belavin:2011pp,Belavin:2011tb,Belavin:2011sw,Alfimov:2013cqa} play important role in the discussion of the equivariant cohomology of the moduli space of $U(L)$ instantons on $\mathbb{C}^2/Z_n$. In particular, \cite{Alfimov:2013cqa} discusses construction similar to our gluing.}

\textbf{Example 2: $U(2;\Psi)$ as $\mathcal{L}^{0|2}_{2,\bar{0},1,0}$ }

\begin{wrapfigure}{l}{0.18\textwidth}
\vspace{-15pt}
\begin{center}
\includegraphics[width=0.17\textwidth]{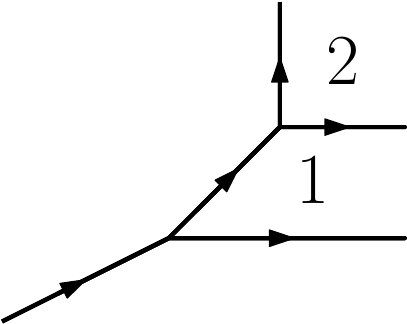}
\end{center}
\vspace{-20pt}
\end{wrapfigure}
Another important example is the Kac-Moody algebra
\begin{eqnarray}
\mathcal{L}^{0|2}_{2,\bar{0},1,0}[\Psi] = U(2;\Psi)
\end{eqnarray}
with the vacuum character
\begin{eqnarray}
\chi \left [\mathcal{L}^{0|2}_{2,\bar{0},1,0}\right ]=\prod_{n=0}^{\infty}\frac{1}{\left (1-q^{n+1}\right )^4}.
\end{eqnarray}

One can see that the above character agrees with the gluing construction and using the formula\ae\, from appendix \ref{Characters}, one gets 
\begin{multline}
\chi \left [\mathcal{L}^{0|2}_{2,\bar{0},1,0} \right ] = \\
= \prod_{n=0}^{\infty}\frac{1}{(1-q^{n+1})^4}\sum_{s=-\infty}^{\infty}\left (\sum_{n=s}(-1)^{n+s}q^{\frac{n(n+1)-s(s-1)}{2}}+\sum_{n=s+1}^{\infty}(-1)^{n+s}q^{\frac{n(n+1)-s(s+1)}{2}}\right ).
\end{multline}
The equality then follows from the identity
\begin{eqnarray}
1=\sum_{s=-\infty}^{\infty}\left (\sum_{n=s}(-1)^{n+s}q^{\frac{n(n+1)-s(s-1)}{2}}+\sum_{n=s+1}^{\infty}(-1)^{n+s}q^{\frac{n(n+1)-s(s+1)}{2}}\right ).
\end{eqnarray}

\textbf{Example 3: $\mathcal{W}_3^{(2)}$ as $\mathcal{L}^{0|2}_{2,\bar{0},1,0}$}

\begin{wrapfigure}{l}{0.18\textwidth}
\vspace{-15pt}
\begin{center}
\includegraphics[width=0.17\textwidth]{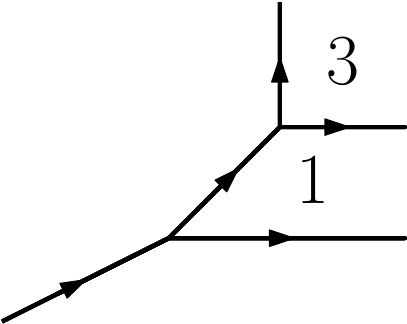}
\end{center}
\vspace{-20pt}
\end{wrapfigure}
The next example can be identified with the $\mathcal{W}_3^{(2)}$ algebra of \cite{Polyakov:1989dm,Bershadsky:1990bg} times the omnipresent $U(1)$ factor. The algebra is given by the BRST reduction\footnote{Note the similarity of the algebra with $\mathcal{N}=2$ super Virasoro coming from the analogous DS-reduction of $U(2|1;\Psi)$.}
\begin{eqnarray}
\mathcal{L}^{0|2}_{2,\bar{0},1,0}[\Psi] = \mathcal{DS}_{2}[U(3;\Psi)].
\end{eqnarray}
The character of the algebra is given by
\begin{eqnarray}
\chi \left[\mathcal{L}^{0|2}_{2,\bar{0},1,0}\right] = \prod_{n=0}^\infty \frac{1}{\left (1-q^{n+1}\right )^{2}\left (1-q^{n+\frac{3}{2}}\right )^2\left (1-q^{n+2}\right )}.
\end{eqnarray}
This expression can be expanded using the formulas in appendix \ref{form} as
\begin{eqnarray}
\chi \left[\mathcal{L}^{0|2}_{2,\bar{0},1,0}\right] =\left (1-q^{\frac{1}{2}}\right ) ^2 \prod_{n=0}^\infty \frac{1}{\left (1-q^{n+1}\right )^{4}\left (1-q^{n+2}\right )}\sum_{n=0}^\infty \sum_{m=-n}^n (-1)^{n+m}q^{\frac{n(n+1)-m^2}{2}}.
\end{eqnarray}
From the gluing perspective, the algebra can be constructed as a conformal extension
\begin{eqnarray}
\mathcal{L}^{0|2}_{2,\bar{0},1,0}\supset Y_{0,1,3}[\Psi]\times Y_{0,0,1}[\Psi-1].
\end{eqnarray}
At the level of characters, one gets
\begin{multline}
\chi \left [\mathcal{L}^{0|2}_{2,\bar{0},1,0}\right ] = \prod_{n=0}^{\infty}\frac{1}{(1-q^{n+1})^4(1-q^{n+2})}\sum_{s=-\infty}^{\infty}\bigg [(1+q)\sum_{n=|s|}^\infty (-1)^{n+s}q^{\frac{n(n+1)-s^2}{2}} \\
+ q^{\frac{1}{2}}\sum_{n=|s+1|}^\infty (-1)^{n+s}q^{\frac{n(n+1)-(s+1)^2}{2}} +q^{\frac{1}{2}}\sum_{n=|s-1|}^\infty (-1)^{n+s}q^{\frac{n(n+1)-(s-1)^2}{2}} \bigg ]
\end{multline}
that is equivalent to the expansion above.

\textbf{Example 4: $\mathcal{L}^{0|2}_{3,\bar{0},2,0}$}

\begin{wrapfigure}{l}{0.18\textwidth}
\vspace{-15pt}
\begin{center}
\includegraphics[width=0.17\textwidth]{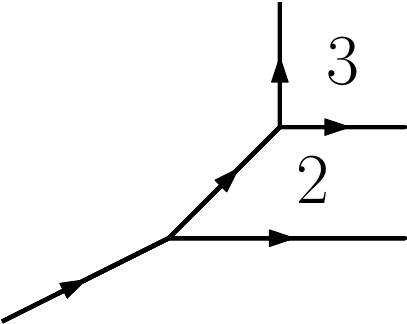}
\end{center}
\vspace{-20pt}
\end{wrapfigure}
The final example deals with a configuration closely related to the previous one. The relation is discussed in detail in the section \ref{flop}. The algebra $\mathcal{L}^{0|2}_{3,\bar{0},2,0}$ is defined as
\begin{eqnarray}
\mathcal{L}^{0|2}_{3,\bar{0},2,0} = \mathcal{DS}_2[\mathcal{DS}_1[U(3;\Psi)]]
\end{eqnarray}
by which we mean the DS-reduction of $U(3;\Psi) \times \{ b_{12},c_{12}\}$ with respect to the BRST charge
\begin{eqnarray}
Q = \oint dz (J_{12}-1)c_{12}.
\end{eqnarray}
The character of the algebra differs from the previous one by the contribution from the free fermion
\begin{eqnarray}
\chi \left [\mathcal{L}^{0|2}_{3,\bar{0},2,0}\right ]=\prod_{n=0}^\infty \frac{1}{(1-q^{n+1})^{2}\left (1-q^{n+\frac{1}{2}}\right )^2\left (1-q^{n+\frac{3}{2}}\right )^2(1-q^{n+2})}.
\label{ref11}
\end{eqnarray}
From the gluing point of view the algebra can be constructed as a conformal extension of the form
\begin{eqnarray}
\mathcal{L}^{0|2}_{3,\bar{0},2,0}[\Psi]\supset Y_{0,2,3}[\Psi]\times Y_{0,0,2}[\Psi-1].
\end{eqnarray}
The vacuum character is
\begin{eqnarray}\nonumber
\chi_{0,0,2,3}&=&\sum_{\mu_1\geq \mu_2}\chi_{Y_{0,2,3}[\Psi]}(W^3_\mu)\chi_{Y_{0,0,2}[\Psi-1]}(W^1_\mu)\\ \nonumber
&=&\frac{1}{2}\prod_{n=0}^{\infty}\frac{1}{(1-q^{n+1})^7}\sum_{\mu_1\geq \mu_2}(q^{-\frac{1}{2}\mu_1+\frac{1}{2}\mu_2}-q^{\frac{1}{2}\mu_1-\frac{1}{2}\mu_2+1}) \times \\ \nonumber
&& \times \sum_{n_1=0}^\infty \sum_{n_1=0}^\infty
(-1)^{n_1+n_2}(1-q^{n_1+1})(1-q^{n_2+1})q^{\frac{n_1(n_1+1)+n_2(n_2+1)}{2}} \times \\ \nonumber
&&\times \big (q^{(n_1+1)\mu_1+(n2+1)\mu_2}+q^{(n_1+1)\mu_2+(n2+1)\mu_1}\\
&&-q^{(n_1+1)(\mu_2-1)+(n2+1)(\mu_1+1)}-q^{(n_1+1)(\mu_1+1)+(n2+1)(\mu_2-1)}\big).
\end{eqnarray}
Multiplying both numerator and denominator of the expression (\ref{ref11}) by the factor of $(1-q)^3$  and using the identities from the appendix to expand the products, one can show that the two proposals for the character match.

\subsection{Algebras of type $M|N$}
\begin{figure}[h]
\centering
\includegraphics[width=0.41\textwidth]{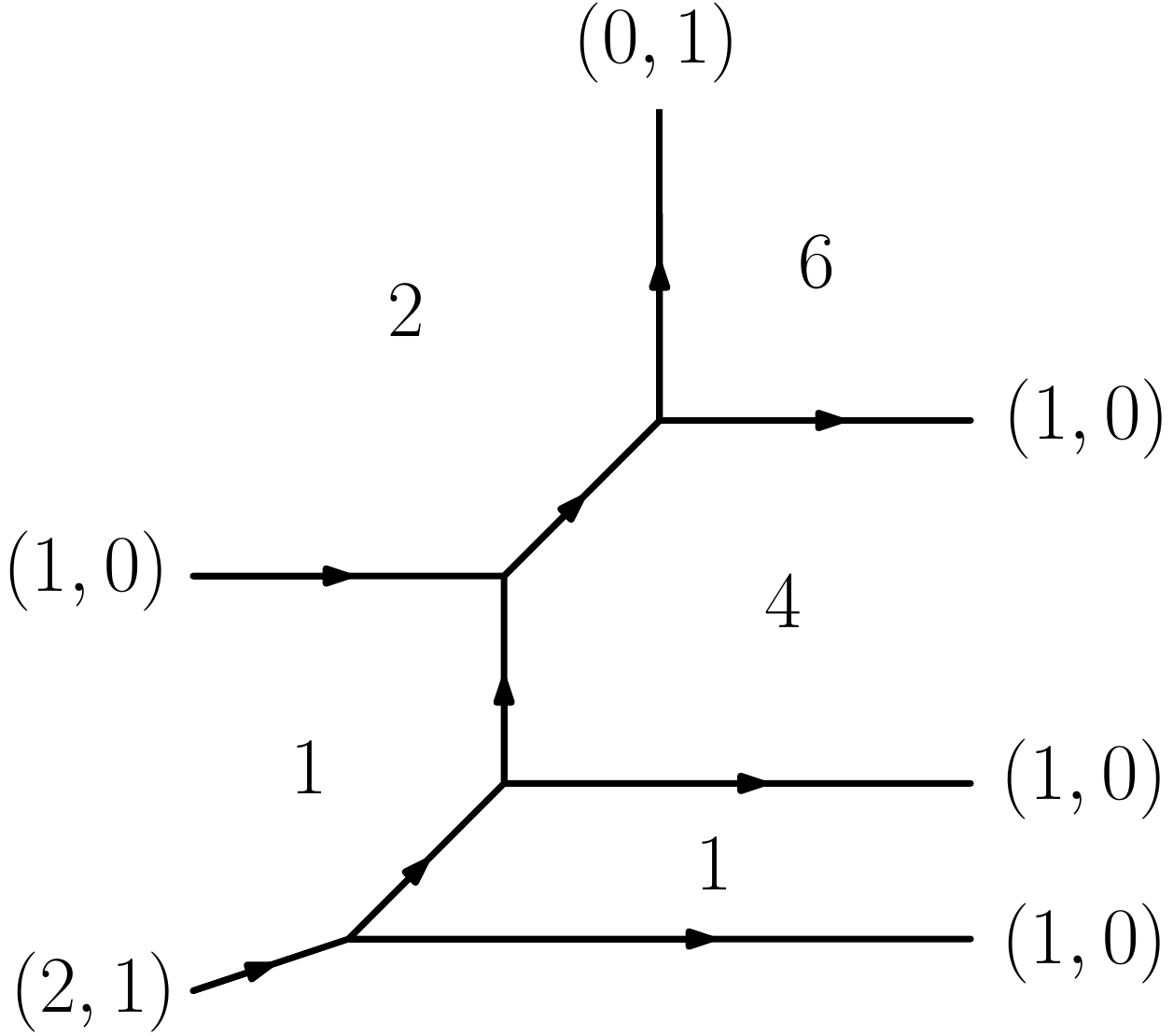}
\caption{Example of a configuration of branes with a BRST definition.}
\label{DSgeneral}
\end{figure}

In this section, we briefly discuss a generalization of the BRST reductions in the case of diagrams with D5-branes ending on $(n,1)$ branes from both left and right. We describe the BRST reduction of a general configuration with monotonic number of D3-branes. Example of such a configuration is given in the figure \ref{DSgeneral}.

\subsubsection{BRST construction}

Let $N_1 \geq N_2 \geq \dots \geq N_n$ be the sequence of D3-branes on the left of the $(n,1)$ branes and $M_1 \geq M_2 \geq \dots \geq M_n$ be the sequence of D3-branes attached from the right. There is a natural generalization of the construction from the previous two sections where $1|1$ and $0|2$ algebras were constructed using a sequence of DS-reductions and one coset construction. To find the expression for the BRST reduction, we follow the diagram from the top to the bottom. We start with the Kac-Moody algebra $U(M_1,N_1;\Psi)$. Each time a D5-brane ends from the right, the Drinfeld-Sokolov reduction $\mathcal{DS}_{M_i-M_{i+1}}$ needs to be performed (where $i$ labels the D5-branes ending from the right). Similarly, each time a D5-brane ends on the chain of $(n,1)$ branes from the left, the Drinfeld-Sokolov reduction $\overline{\mathcal{DS}}_{N_j-N_{j+1}}$ needs to be performed (here $i$ labels the D5-branes ending on the left). Finally, one needs to take a coset with respect to the remaining $U(M_m|N_n)$ super Kac-Moody algebra. For example, the diagram from \ref{DSgeneral} leads to the following algebra
\begin{eqnarray}
\mathcal{L}_{6\bar{2}4\bar{1}10}[\Psi] = \frac{\mathcal{DS}_{1}[\mathcal{DS}_3[\overline{\mathcal{DS}}_1[\mathcal{DS}_2[U(6|2;\Psi)]]]]}{U(1;-\Psi+2)}
\label{complicated}
\end{eqnarray}
where the $\mathcal{DS}_N$ and $\overline{\mathcal{DS}}_N$ are defined as in the case of $1|1$ and $0|2$ diagrams. Note that after each DS-reduction associated to the D5-brane ending from the right, the final algebra contains a Kac-Moody algebra with level shifted by minus one and after each DS-reduction associated to the D5-brane ending from the left, the final algebra contains a Kac-Moody algebra with level shifted by one. The final level one gets after performing all the DS-reductions is opposite to the level of the Kac-Moody algebra induced from the bottom CS theory.

\begin{figure}[h]
\centering
\includegraphics[width=0.85\textwidth]{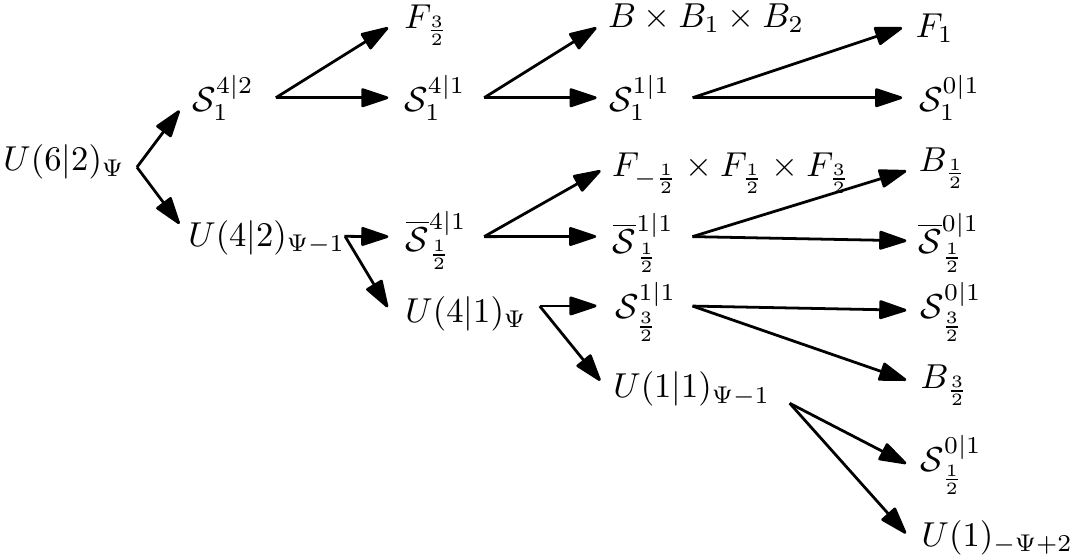}
\caption{The structure of DS-reductions from the example \ref{complicated}.}
\label{branching}
\end{figure}

Let us now summarize how the fields decompose under $\mathcal{DS}_{N-M}$ and $\overline{\mathcal{DS}}_{K-L}$ reductions at each step. The $U(N|K;\Psi)$ Kac-Moody algebra factor decomposes as
\begin{eqnarray}\nonumber
\mathcal{DS}_{N-M}: U(N|K;\Psi) & \rightarrow & \mathcal{W}_{N-M} \times U(M|K;\Psi-1) \times \mathcal{S}^{M|K}_{\frac{N-M}{2}} \\
\overline{\mathcal{DS}}_{K-L}: U(N|K;\Psi) & \rightarrow & \mathcal{W}_{K-L} \times U(N|L;\Psi+1) \times \mathcal{S}^{N|L}_{\frac{K-L}{2}}.
\end{eqnarray}
On the other hand, the fields $\mathcal{S}^{N|K}_{k}$ and $\overline{\mathcal{S}}^{N|K}_{k}$ from the previous steps decompose as
\begin{eqnarray}\nonumber
\mathcal{DS}_{N-M}:\mathcal{S}^{N|K}_{k} & \rightarrow & \mathcal{S}_k^{M|K} \times \mathcal{B}_{k-\frac{N-M-1}{2}} \times \dots \times \mathcal{B}_{k+\frac{N-M-1}{2}}\\ \nonumber
\mathcal{DS}_{N-M}:\overline{\mathcal{S}}^{N|K}_{k} & \rightarrow & \overline{\mathcal{S}}_k^{M|K} \times \mathcal{F}_{k-\frac{N-M-1}{2}}\times \dots \times \mathcal{F}_{k+\frac{N-M-1}{2}}\\ \nonumber
\overline{\mathcal{DS}}_{K-L}:\mathcal{S}^{N|K}_{k} & \rightarrow & \mathcal{S}^{N|L}_{k}\times \mathcal{F}_{k-\frac{K-L-1}{2}} \times \dots \times \mathcal{F}_{k+\frac{K-L-1}{2}}  \\
\overline{\mathcal{DS}}_{K-L}:\overline{\mathcal{S}}^{N|K}_{k} & \rightarrow & \overline{\mathcal{S}}^{N|L}_{k} \times \mathcal{B}_{k-\frac{K-L-1}{2}} \times \dots \times \mathcal{B}_{k+\frac{K-L-1}{2}}.
\end{eqnarray}
The decomposition is shown explicitly for the example above in the figure \ref{branching}.

A similar BRST reduction can be defined in the case when the diagram can be cut into two halves where in the upper half, the number of D3-branes decreases from the top to the bottom and in the lower half it decreases from the bottom to the top. The BRST definition is then given by performing a sequence of DS-reductions on both the upper and the bottom Kac-Moody algebra and then gluing two remaining Kac-Moody subalgebras with opposite level by BRST construction.

\subsection{Kac-Moody algebras}

\begin{figure}[h]
\centering
\includegraphics[width=0.85\textwidth]{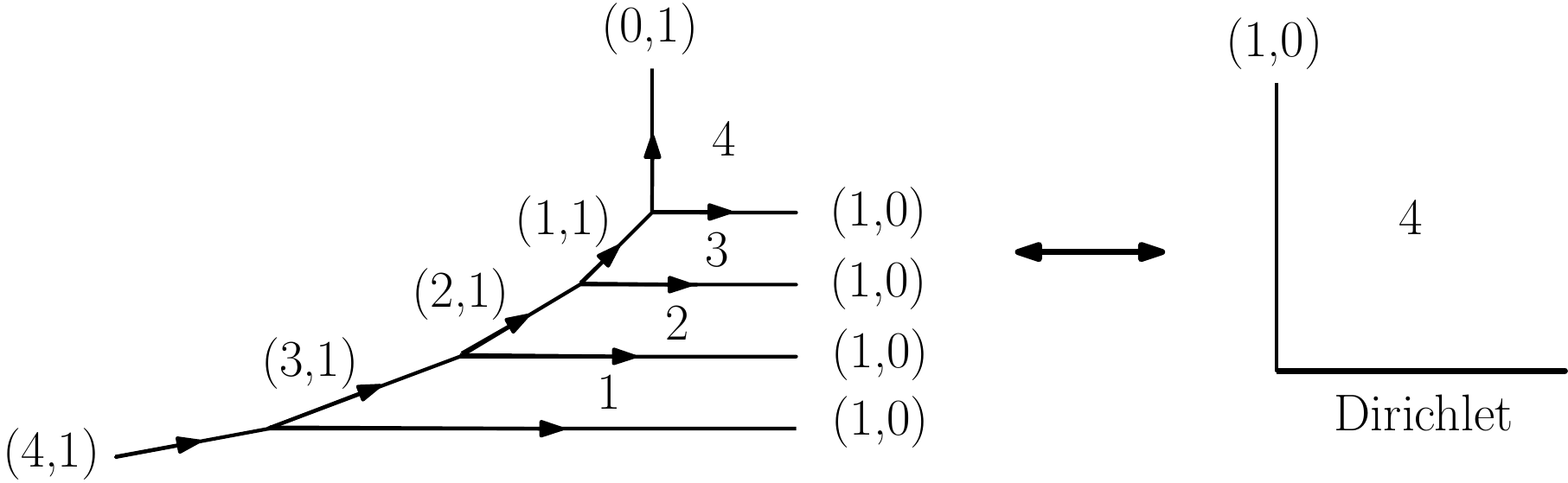}
\caption{Configuration corresponding to $U(4)_{\Psi}$ Kac-Moody algebra.}
\label{Dirichlet}
\end{figure}

It is well known \cite{Witten:1988hf} that imposing $\mathcal{A}_{\bar{z}}=0$ at the boundary of $U(N)$ Chern-Simons theory at level $\Psi$ gives rise to the Kac-Moody algebra $U(N;\Psi)$. A lift of such (rotated Dirichlet) boundary condition can be identified according to \cite{Gaiotto:2008ac} with a configuration of $N$ D3-branes, each of them ending on a single D5-brane with a flux that deforms the standard Dirichlet boundary condition. To obtain the Kac-Moody algebra from the corner configuration, one needs to introduce an extra NS5-like boundary on which the Chern-Simons theory is induced and also to impose a correct boundary condition. Such configuration consists of $N$ D5-branes attached to $(n,1)$ branes with $n$ increasing by one at each junction. The number of D3-branes starts with $N$ in the upper right corner and gradually decreases by one each time we cross a D5-brane. An example of such a configuration for $N=4$ is shown in the figure \ref{Dirichlet}.

Based on this argument about the resolution, one can naturally conjecture that $U(N;\Psi)$ can be constructed as a conformal extension
\begin{eqnarray}
U(N)_\Psi = \times_{n=1}^N Y_{0,n-1,n}[\Psi-N+n].
\end{eqnarray}
This indeed agrees with the BRST definition that contains a sequence of $N$ trivial DS-reductions in one dimensional blocks of $U(N;\Psi)$ in this special case.

The case of $U(1;\Psi)$ is trivially true and the case of $U(2;\Psi)$ was already discussed in the previous section. To support the conjectural relation of the $U(N;\Psi)$ Kac-Moody algebra with gluing of Y-algebras in general case, let us argue that the central charge coming from the gluing construction agrees with the central charge of the Sugawara stress-energy tensor and that the vacuum character for $U(3;\Psi)$ agrees. Indeed, it is easy to see that after summing over the contributions to the central charge from each vertex, one finds the total central charge
\begin{eqnarray}
c\left [U(N;\Psi)\right ]=\sum_{n=0}^{N-1}c\left [Y_{0,N-n-1,N-n}[\Psi-n]\right ]=N^2-\frac{N(N^2-1)}{\Psi}
\end{eqnarray}
that agrees with the central charge of the Sugawara stress-energy tensor for $U(N;\Psi)$.

\begin{wrapfigure}{l}{0.24\textwidth}
\vspace{-15pt}
\begin{center}
\includegraphics[width=0.225\textwidth]{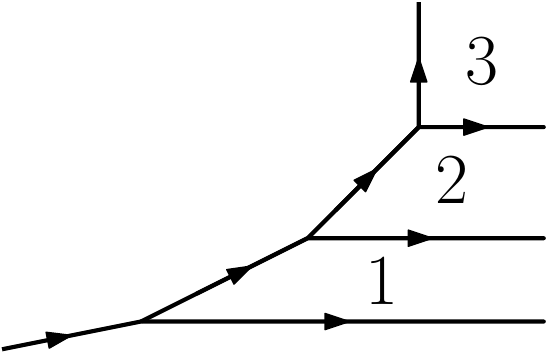}
\end{center}
\vspace{-20pt}
\end{wrapfigure}
Let us check that the gluing construction gives the correct vacuum character also in the case of $U(3;\Psi)$. This algebra is conformal extension of
\begin{eqnarray}
U(3,\Psi)\supset Y_{0,2,3}[\Psi]\times Y_{0,1,2}[\Psi-1]\times Y_{0,0,1}[\Psi-2].
\end{eqnarray}
In this case, one needs to sum over the modules associated to characters with two asymptotic Young diagrams. This calculation is a non-trivial check of our proposals under these conditions. The character can be expressed as
\begin{eqnarray}
\sum_{\nu=-\infty}^{\infty}\sum_{\mu_1\geq \mu_2}\chi[Y_{0,2,3}[\Psi]](M^3_{\mu_1,\mu_2}) \chi[Y_{0,1,2}[\Psi-1]](M^2_{\mu_1,\mu_2}, M^3_{\nu}) \chi[Y_{0,0,1}[\Psi-2]](M^1_\nu).
\end{eqnarray}
Plugging in from the appendix \ref{Characters}, we have checked that the character indeed agrees with the one of $U(3;\Psi)$ up to $q^{10}$.

Similarly, one can conjecture the brane configuration and the corresponding gluing construction giving rise to the $U(N|M;\Psi)$ Kac-Moody algebra. In the case of supergroups, one needs to impose Dirichlet boundary conditions on both sides of the NS5-like boundary. It is unclear at first sight what is the correct ordering of the D5-branes ending from the left and right in the resolution of the configuration in terms of $(p,q)$-webs. But fortunately various configurations are mutually related to each other by a sequence of flip transitions to be discussed later. Parameters $\rho_i$ associated to each finite segment turn out to be zero and as will be discussed in the next section, we expect algebras related by a flip transformation with $\rho=0$ to be equal. Moreover, it is also easy to check that the central charges of algebras related by a flip indeed agree in this case.

\begin{figure}[h]
\centering
\includegraphics[width=0.4\textwidth]{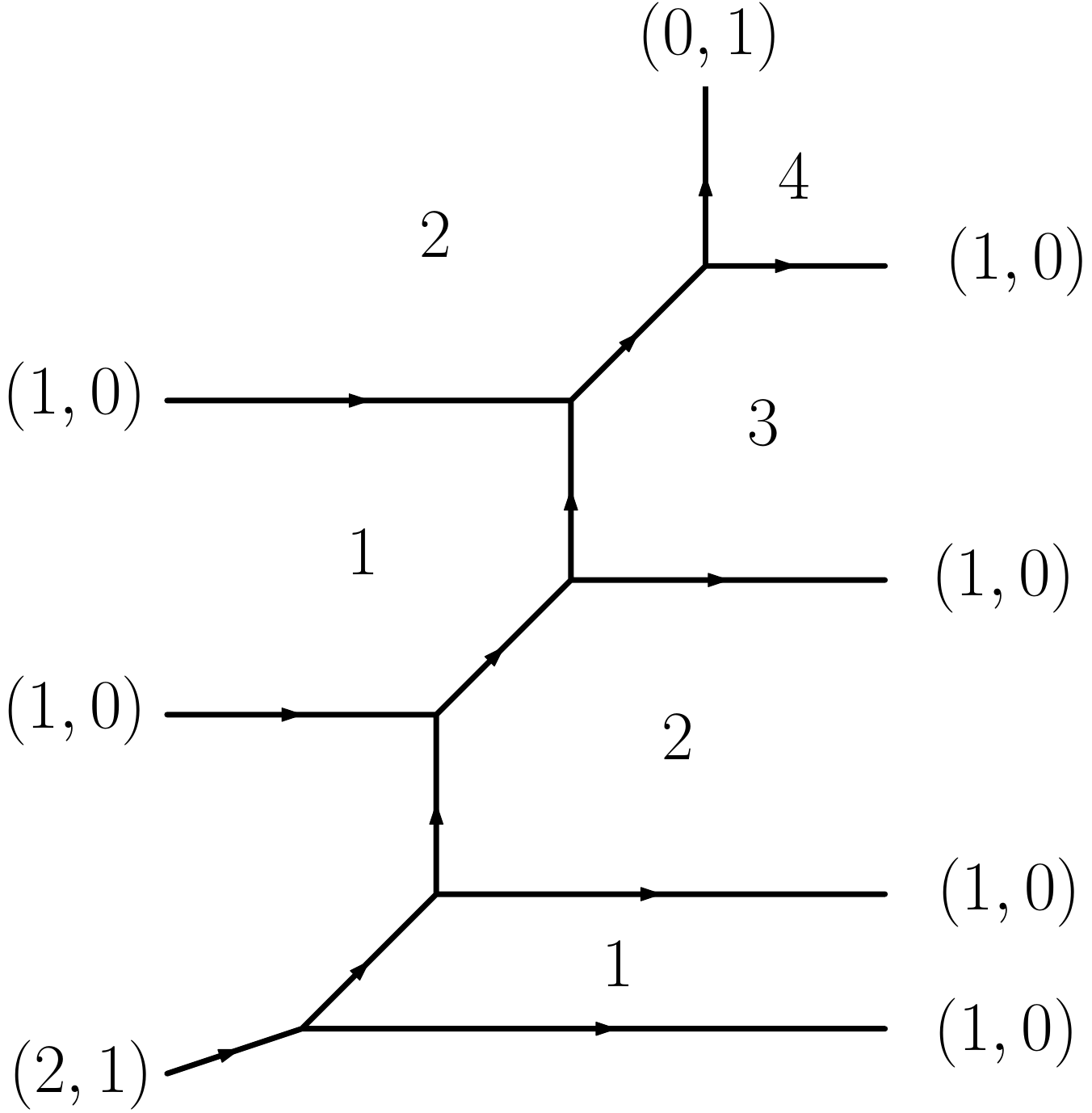}
\caption{Configuration corresponding to $U(4|2)_{\Psi}$ Kac-Moody algebra.}
\label{superpic}
\end{figure}

Let us compute the central charge in the super Kac-Moody algebra case. We start with the ordering of branes such that D5-branes ending on $(n,1)$ branes first alternate and then continue ending from one side after there are no other D5-branes left. Example of such a configuration is shown for $U(4|2;\Psi)$ in figure \ref{superpic}. The contributions to the central charge from all the resolved conifold segments vanish. One is left with the contribution equal to Kac-Moody algebra of difference rank $U(N-M;\Psi)$ for $N>M$ or $U(M-N;-\Psi)$ otherwise. This leads to the correct central charge of the $U(N|L;\Psi)$ Kac-Moody superalgebra.

\subsubsection{Decomposing $U(N)_\kappa$}
\label{appendixun}
Since the affine algebra $U(N)_\kappa$ at level $\kappa$ is an important example of the gluing construction, let us derive the gluing diagram by directly decomposing the algebra itself. We have the conformal decomposition
\begin{equation}
\label{unkdecomp}
U(N)_\kappa \supset \frac{U(N)_\kappa}{U(N-1)_\kappa} \times \frac{U(N-1)_\kappa}{U(N-2)_\kappa} \times \ldots \times \frac{U(2)_\kappa}{U(1)_\kappa} \times U(1)_\kappa.
\end{equation}
From the level-rank duality we know that
\begin{equation}
\frac{U(N)_\kappa}{U(N-1)_\kappa} \simeq U(1) \times \frac{SU(\kappa)_{N-1} \times SU(\kappa)_1}{SU(\kappa)_N} \simeq U(1) \times \mathcal{W}_\kappa,
\end{equation}
i.e. each factor in (\ref{unkdecomp}) can be associated with $Y$ algebra. Let us verify this decomposition by calculating the OPEs. We normalize the generators of $U(N)_k$ such that
\begin{equation}
{J^J}_K(z) {J^L}_M(w) \sim \frac{\kappa \delta^L_K \delta^J_M}{(z-w)^2} + \frac{\delta^L_K {J^J}_M(w) - \delta^J_M {J^L}_K(w)}{z-w}.
\end{equation}
The indices $J,K,\ldots$ run from $1$ to $N$. Now we split the currents according to $N \to (N-1) + 1$ and the lowercase Latin indices $j,k,\ldots$ will run from $1$ to $N-1$. We denote the fields after the splitting
\begin{eqnarray}
\nonumber
X^j(z) & = & {J^j}_N(z) \\
Y_j(z) & = & {J^N}_j(z) \\
\nonumber
Z(z) & = & {J^N}_N(z).
\end{eqnarray}
and ${J^j}_k(z)$. The operator product expansions can now be written as
\begin{eqnarray}
\nonumber
{J^k}_k(z) {J^l}_m(w) & \sim & \frac{\kappa \delta^k_l \delta^j_m}{(z-w)^2} + \frac{\delta^l_k {J^j}_m(w) - \delta^j_m {J^l}_k(w)}{z-w} \\
\nonumber
{J^j}_k(z) X^l(w) & \sim & \frac{\delta^l_k X^j(w)}{z-w} \\
\nonumber
{J^j}_k(z) Y_l(w) & \sim & -\frac{\delta^j_l Y_k(w)}{z-w} \\
\nonumber
{J^j}_k(z) Z(w) & \sim & reg. \\
\nonumber
X^j(z) X^k(w) & \sim & reg. \\
Y_j(z) Y_k(w) & \sim & reg. \\
\nonumber
X^j(z) Y_k(w) & \sim & \frac{\kappa \delta^j_k}{(z-w)^2} + \frac{{J^j}_k(w)}{z-w} - \frac{\delta^j_k Z(w)}{z-w} \\
\nonumber
X^j(z) Z(w) & \sim & \frac{X^j(w)}{z-w} \\
\nonumber
Y_j(z) Z(w) & \sim & -\frac{Y_j(w)}{z-w} \\
\nonumber
Z(z) Z(w) & \sim & \frac{\kappa}{(z-w)^2}.
\end{eqnarray}
We now want to determine the parameters of $\mathcal{W}$-subalgebra commuting with $U(N-1)_{\kappa}$.

\paragraph{$U(1)_k$ current}
We see directly from the operator product expansions that the field $Z(z)$ satisfies the OPE of $U(1)$ current
\begin{equation}
Z(z) Z(w) \sim \frac{\kappa}{(z-w)^2}.
\end{equation}
The corresponding Sugarawa stress-energy tensor with central charge $1$ is
\begin{equation}
\frac{1}{2\kappa} (ZZ)(z).
\end{equation}

\paragraph{Dimension two fields and stress-energy tensor}
We have $7$ dimension $2$ fields $({J^j}_k {J^k}_j)$, $({J^j}_j {J^k}_k)$, ${J^j}_j^\prime$, $(X^j Y_j)$, $(ZZ)$ and $Z^\prime$ and $({J^j}_j Z)$ that commute with the global subalgebra of $U(N-1)_\kappa$ and we want to find a linear combination of these such that it has regular OPE with $Z(z)$ and with ${J^j}_k(z)$. There is a unique field satisfying these conditions,
\begin{eqnarray}
\nonumber
T_{\infty} & = & -\frac{1}{2(N+\kappa)(N+\kappa-1)} ({J^j}_k {J^k}_j) - \frac{1}{2\kappa(N+\kappa)(N+\kappa-1)} ({J^j}_j {J^k}_k) \\
& & + \frac{1}{2(N+\kappa)} ((X^j Y_j) + (Y_j X^j)) - \frac{N-1}{2\kappa(N+\kappa)} (ZZ) + \frac{1}{\kappa(N+\kappa)} (Z {J^j}_j).
\end{eqnarray}
We chose the normalization in such a way that the OPE with $T_{\infty}$ with itself is that of the stress-energy tensor. The central charge is
\begin{equation}
c_{\infty} = \frac{(N-1)(\kappa-1)(N+2\kappa)}{(N+\kappa)(N+\kappa-1)}.
\end{equation}
By construction, the fields $Z$ and ${J^j}_k$ commute with $T_{\infty}$ while the fields $X^j$ and $Y_k$ have OPE
\begin{eqnarray}
\nonumber
T_{\infty}(z) X^j(w) & \sim & \frac{(\kappa-1)(N+2\kappa)}{2\kappa(N-1+\kappa)} \frac{X^j(w)}{(z-w)^2} - \frac{1}{N-1+\kappa} \frac{({J^j}_k X^k)(w)}{z-w} \\
& & - \frac{1}{\kappa(N-1+\kappa)} \frac{({J^k}_k X^j)(w)}{z-w} + \frac{(Z X^j)(w)}{z-w} + \frac{\partial X^j(w)}{z-w}
\end{eqnarray}
This nicely illustrates the general fact that while the gluing fields like $X^j(w)$ are primary with respect to stress-energy tensors at vertices in the sense that the higher than quadratic poles in OPE vanish and that the quadratic pole is proportional to $X^j(w)$, the linear pole is not simply a derivative of $X^j(w)$. In the language of mode operators, the state corresponding to $X^j(w)$ is annihilated by positive modes of the $T_{\infty}$ and is eigenstate of its zero mode, but the action of $L_{-1}$ on fields is not simply the derivative. Only the total stress-energy tensor of the whole algebra generates the spatial translations.

\paragraph{Dimension $3$ and $4$ fields}
In order to find the $\lambda$-parameters of Y algebra, we need to identify the primary spin $3$ and spin $4$ fields. Up to a normalization, there is a unique dimension $3$ primary field
\begin{equation}
W_3(z) = ({J^j}_k({J^k}_l {J^l}_j))(z) + \ldots
\end{equation}
with the OPE structure constant
\begin{equation}
C_{33}^0 = \frac{3(N-1)(\kappa-2)(\kappa-1)(N+\kappa)(N+\kappa-1)(2N+\kappa-2)(N+2\kappa)(2N+3\kappa)}{2\kappa}.
\end{equation}
Similarly, we find a unique spin $4$ primary normalized as
\begin{equation}
W_4(z) = ({J^j}_k ({J^k}_l ({J^l}_m {J^m}_j)))(z) + \ldots
\end{equation}
and with structure constant
\begin{multline}
C_{44}^0 = \frac{4(N-1)N(\kappa-3)(\kappa-2)(\kappa-1)(\kappa+1)(N+\kappa)(N+\kappa-1)}{\kappa^2(5N^2\kappa+17N^2+10N\kappa^2+29N\kappa-17N+12\kappa^2-12\kappa)} \times \\
\times (N+2\kappa)(N+2\kappa-1)(2N+\kappa-2)(2N+3\kappa)(3N+2\kappa-3)(3N+4\kappa).
\end{multline}
The remaining structure constant $C_{33}^4$ needed for identification of $\lambda$ is
\begin{equation}
C_{33}^4 = -9(N+\kappa)(N+\kappa-1).
\end{equation}
\begin{figure}
\centering
\includegraphics[width=0.36\textwidth]{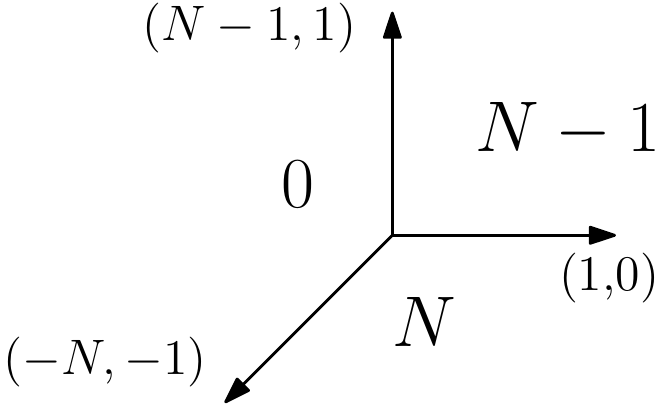}
\caption{A configuration of branes corresponding to the coset $U(N)_\kappa / U(N-1)_\kappa$. The level is encoded in $\Psi = \kappa$.}
\label{figglvert}
\end{figure}
Comparing this with (\ref{cc2}), we can identify
\begin{equation}
\lambda = \left( \kappa, \frac{\kappa}{N+\kappa-1}, -\frac{\kappa}{N+\kappa} \right).
\end{equation}
These parameters are exactly what one would read off from the diagram shown in figure \ref{figglvert} with the value of level $\kappa$ determined as
\begin{equation}
\Psi = -\kappa.
\end{equation}

\paragraph{Charges of gluing fields}
We would like to identify the fields $X^j$ and $Y_k$ as the gluing fields. Unfortunately, their charges with respect to $Z$ don't match with the gluing prescription (\ref{u1norm}). There is however a freedom of redefinition of $U(1)$ currents associated to vertices. Let us first normalize the $U(1)$ currents $J_\alpha(z)$ such that
\begin{equation}
J_\alpha(z) J_\beta(w) \sim \frac{\delta_{\alpha\beta}}{(z-w)^2}.
\end{equation}
The index $\alpha$ labels the individual vertices so in the case of $U(N)_\kappa$ it runs from $1$ to $N$. Having fixed the normalization of the fields in this way, we are still free to make global $SO(N)$ rotations in the space of $U(1)$ currents. The gluing matter fields have charges which follow from (\ref{u1norm}) and (\ref{intcharges}). For example in the case of $U(N)_\kappa$ diagram, the basic bimodule between $\alpha$-th and $(\alpha+1)$-st node has charges described by the charge vector
\begin{equation}
v_{\alpha,\alpha+1} = \sqrt{\frac{1}{\kappa + \frac{\kappa}{\alpha-1+\kappa}}} \, e_\alpha - \sqrt{\frac{1}{\kappa-\frac{\kappa}{\alpha+1+\kappa}}} \, e_{\alpha+1}
\end{equation}
where $e_\alpha$ is the standard Euclidean basis of the charge space. The invariant quantities under orthogonal rotations of $U(1)$ fields are the inner products between the charge vectors of neighbouring bimodules. In our case, we find the non-zero inner products
\begin{equation}
v_{\alpha,\alpha+1} \cdot v_{\alpha+1,\alpha+2} = -\sqrt{\frac{1}{\kappa-\frac{\kappa}{\alpha+1+\kappa}}} \sqrt{\frac{1}{\kappa+\frac{\kappa}{\alpha+\kappa}}} = -\frac{1}{\kappa}.
\end{equation}
This should be compared to diagonal $U(1)$ currents of $U(N)_\kappa$ affine Lie algebra: the normalized currents are in our case the Cartan currents
\begin{equation}
\left( \frac{1}{\sqrt{\kappa}} {J^1}_1, \frac{1}{\sqrt{\kappa}} {J^2}_2, \ldots, \frac{1}{\sqrt{\kappa}} {J^N}_N \right).
\end{equation}
The gluing matter is given by the fields associated to simple roots
\begin{equation}
\left({J^1}_2, {J^2}_3, \ldots, {J^{N-1}}_N \right)
\end{equation}
as well as their conjugates. Their corresponding charge vectors can be read off from the OPE
\begin{equation}
\frac{1}{\sqrt{\kappa}} {J^j}_j(z) {J^k}_{k+1}(w) \sim \sqrt{\frac{1}{\kappa}} (\delta_{j,k} - \delta_{j,k+1}) \frac{{J^k}_{k+1}(w)}{(z-w)}
\end{equation}
and we find the vectors
\begin{equation}
w_{\alpha,\alpha+1} = \sqrt{\frac{1}{\kappa}} e_\alpha - \sqrt{\frac{1}{\kappa}} e_{\alpha+1}.
\end{equation}
The inner products between these are
\begin{equation}
w_{\alpha,\alpha+1} \cdot w_{\alpha+1,\alpha+2} = -\frac{1}{\kappa}.
\end{equation}
as before. We conclude that the basis of $U(1)$ charges described in (\ref{u1norm}) and (\ref{intcharges}) and the natural Cartan basis in $U(N)_\kappa$ are related by an $O(N)$ rotation in a way that is compatible with charge assignments of the gluing matter fields.

\paragraph{Positive roots from composite line operators}
We can now understand the origin of all spin $1$ $U(N)_\kappa$ fields directly from the gluing diagram. The Cartan $U(1)$ fields come from the vertices of the diagram. The simple roots and their negatives come from the spin $1$ gluing matter fields in the fundamental representation. Every time we have $\rho=0$ associated to an edge, it gives rise to such a spin $1$ field. If we turn on the line operators in the fundamental representation along neighbouring edges, the formula (\ref{boxboxfusion}) tells us that we get additional spin $1$ fields which correspond to positive roots. We thus have a one-to-one correspondence between connected chains of edges in the diagram and positive (or negative) roots, simple edges corresponding to simple roots.

\subsubsection{Dual diagram}
\begin{figure}[h]
\centering
\includegraphics[width=0.4\textwidth]{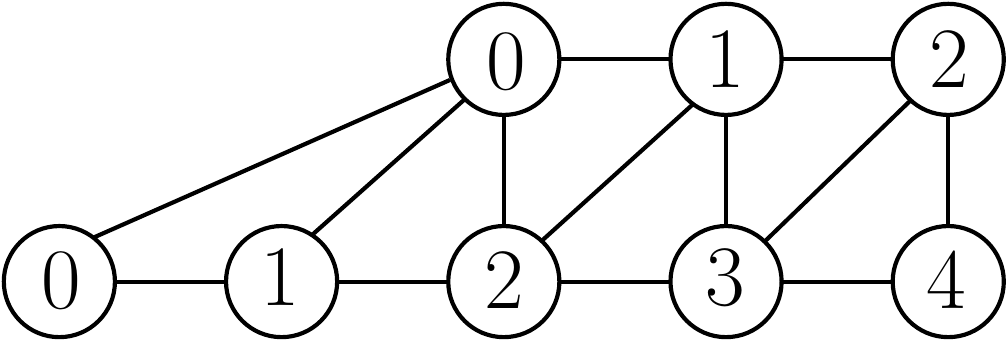}
\caption{Dual diagram representing the $U(4|2;\Psi)$ Kac-Moody algebra. In this picture, the different orderings of five-branes coming from the left and right correspond to different triangulations of the polygon.}
\label{dualu42pic}
\end{figure}

The diagrams drawn so far represent the configuration of D3 and five-branes as they are in type IIB string theory. For some purposes it is useful to draw the dual diagrams, where the nodes correspond to faces and are labeled by the number of D3 branes of the corresponding stack of branes. The slopes of lines correspond to five-brane charges as before, but the five-brane charge conservation condition now translates to the fact that the faces of the diagram are closed triangles. An example of such diagram is shown in figure \ref{dualu42pic}. Very similar diagrams appear in the related context in \cite{Sulkowski:2009rw}.

\section{Large number of D3-branes}

In this section we study what happens as the number of attached D3-branes grows to infinity. We already discussed the simplest case of the algebra $Y_{L,M,N}$. As $L$, $M$ or $N$ goes to infinity, the vacuum character approaches the MacMahon function which counts the plane partitions without any further constraints. Combinatorially, the requirement of the absence of a box at position $(L+1,M+1,N+1)$ which leads to truncation of the algebra disappears. The operator product expansions for the spin content given by MacMahon functions were studied in \cite{Gaberdiel:2012aa} and the result is a two-parametric family of algebras $\mathcal{W}_{1+\infty}$ parametrized by $\lambda_j$ with constraint (\ref{relation1}). The central charge and the OPE structure constants in the primary basis are determined in terms of $\lambda_j$ as in (\ref{cc2}). $Y_{L,M,N}$ algebras can be recovered by recalling that if (\ref{truncations2}) is satisfied the algebra $\mathcal{W}_{1+\infty}$ develops an ideal such that $Y_{LMN}$ is the quotient of $\mathcal{W}_{1+\infty}$ by this ideal. In this section, we generalize this point of view to other algebras that we constructed by the gluing procedure.

\subsection{Resolved conifold - $\mathcal{W}_{1|1\times \infty}^{\rho}$ algebras}
As a first example, let us see what are the possible limits of the conifold $\mathcal{L}^{1|1}_{K,\bar{L},M,\bar{N}}$ algebras as the number of D3 branes approaches infinity. Compared to the $Y_{L,M,N}$ junction, the conifold configuration has another stack of D3 branes so one might naively expect a three-parametric family of algebras. We will see that in the infinite numbers of branes limit, one recovers different characters for each choice of the discrete parameter
\begin{eqnarray}
\rho=\frac{N+L-K-M}{2}
\end{eqnarray}
that we keep fixed as we take the limit. For each choice of $\rho$, there are two continuous independent $\lambda$-parameters from one of the vertices as in the case of $Y_{L,M,N}$. The $\lambda$-parameters of the second vertex can be then determined in terms of the discrete parameter $\rho$ and the gluing conditions (\ref{gluelambda}). We thus obtain a family of algebras $\mathcal{W}_{1|1\times \infty}^{\rho}$ associated to the conifold diagram, labeled by one discrete parameter $\rho$ (associated to the edge) and two continuous parameters parametrizing the structure constants of the algebra. We expect to be able to recover $\mathcal{L}^{1|1}_{K,\bar{L},M,\bar{N}}$ algebras as truncations of the $\mathcal{W}_{1|1\times \infty}^{\rho}$ family.

\subsubsection{Vacuum character from BRST computation}
Let us now explicitly verify the claims of the previous section by computing the vacuum character of $\mathcal{W}_{1|1\times \infty}^{\rho}$,
\begin{eqnarray}
\label{charX}
\chi \left [\mathcal{W}_{1|1\times \infty}^{\rho}\right ]=\prod_{n=1}^\infty \frac{\left (1+q^{n+\rho}\right )^{2n}}{(1+q^{n})^{2n}}.
\end{eqnarray}
Let us first see how the character (\ref{charX}) appears from the BRST definition of the algebra for $K\geq M$ and $L\geq N$. When computing the vacuum character, there are various contributions coming from the different blocks of $U(K|L)$. Firstly, there are characters of $\mathcal{W}_{K-M}$ and $\mathcal{W}_{L-N}$ coming from the two diagonal blocks. Secondly there is a sequence of pairs of $L-N$ fermionic fields with conformal weights
\begin{eqnarray}
\rho+1,\rho+2,\dots \rho+L-N
\end{eqnarray}
coming from the fermionic off-diagonal blocks that are influenced by both DS-reductions. Apart from these, there are $U(M|N)$ invariant combinations of $\mathcal{S}_{\frac{K-M}{2}}^{M|N}$ and $\overline{\mathcal{S}}^{M|N}_{\frac{L-N}{2}}$. These can be identified with products of bilinears of their generators. If we forget about the relations satisfied by the products of bilinears (which is a condition that disappears in the infinite number of branes limit), $\mathcal{S}^{M|N}_{\frac{K-M}{2}}$ fields form an infinite tower of generators of each integral spin starting with $K-M+1$. This sequence continues the one of $\mathcal{W}_{K-M}$ and together they form one factor of $\mathcal{W}_{1+\infty}$. Similarly, $U(M|N)$ invariant combinations of $\overline{\mathcal{S}}^{M|N}_{\frac{L-N}{2}}$ continues the sequence of fields of $\mathcal{W}_{L-N}$ to produce the second factor of the $\mathcal{W}_{1+\infty}$ vacuum character. Finally, the bilinears mixing $\mathcal{S}^{M|N}_{\frac{K-M}{2}}$ and $\overline{\mathcal{S}}^{M|N}_{\frac{L-N}{2}}$ form an infinite tower starting at conformal dimension $\rho+L-N+1$. Note that these fields are fermionic since a bosonic field gets combined with a fermionic field and these combinations continue the $L-N$ fermionic fields discussed above. One can see that total character is indeed given by (\ref{charX}).

The BRST proposal for $K\geq M$ and $N\geq L$ produces the same character by a slightly different argument. The two $\mathcal{W}_{K-M}$ and $\mathcal{W}_{N-L}$ blocks get extended by bilinears of $\mathcal{S}^{M|L}_{\frac{K-M}{2}}$ and $\overline{\mathcal{S}}^{M|L}_{\frac{K-M}{2}}$ respectively. In this case there are no off-diagonal blocks that would be influenced by both DS-reductions but $U(M|K)$ invariant combinations combining the fields coming from both $\mathcal{S}^{M|L}_{\frac{K-M}{2}}$ and $\overline{\mathcal{S}}^{M|L}_{\frac{K-M}{2}}$ give rise to fermions with each integral spin starting at spin $1+\rho$ and we can draw the same conclusion as in the previous case.

\subsubsection{Vacuum character from gluing}

From the point of view of gluing, the character formula (\ref{charX}) can be obtained by a small modification of the standard sums used in topological vertex calculations. In the limit of infinite numbers of branes $K,L,M,N\rightarrow \infty$, the relevant tensor representations of $U(\infty)$ decouple into contravariant representations (contained in tensor powers of the fundamental representation) times covariant representations (contained in tensor powers of the anti-fundamental representation). Moreover, the pit conditions truncating the two trivalent vertex algebras disappear to infinity and the characters involved considerably simplify.

We can use this example to illustrate the gluing at the level of $\mathcal{W}_{1+\infty}$ algebras. First of all, the $\lambda$ parameters associated to two vertices are connected via
\begin{equation}
\lambda_j^\prime = \lambda_j \frac{\lambda_3}{2\rho-\lambda_3}
\end{equation}
as follows from (\ref{gluelambda}). We want to sum over all characters of $\mathcal{W}_{1+\infty}$ labeled by the representations of the line operators stretched along the edge. In the limit of large number of D3 branes, these are parametrized by a pair of Young diagram labels $(\mu,\nu)$, the first labeling the contravariant part and the second labeling the covariant part of the $U(\infty)$ representation. The corresponding $\mathcal{W}_{1+\infty}$ character factorizes and is equal to
\begin{equation}
\chi_{(\mu,\nu)} = q^{h_\mu+h_\nu} \prod_{n=1}^{\infty} \frac{1}{(1-q^n)^n} P_\mu(q) P_\lambda(q)
\end{equation}
where the power of $q$ in the prefactor is the conformal dimension of the representation and where $P_\mu(q)$ is the quantum dimension of the representation (normalized to be a polynomial in $q$ starting with $1 + \ldots$),
\begin{equation}
P_\lambda(q) = \prod_{\Box \in \lambda} \frac{1}{1-q^{\mathrm{hook}(\Box)}}
\end{equation}
(see \cite{Aganagic:2003db,Okounkov:uq,Prochazka:2015aa}). The full vacuum character for the conifold algebra is now obtained by summing over all the representations of the line operators and taking the product of characters of algebras associated to both vertices
\begin{eqnarray}
\chi_{\mathcal{W}_{1|1\times \infty}^{\rho}} & = & \sum_{\mu,\nu \geq 0} z^{|\mu|-|\nu|} \chi_{(\mu,\nu)} \chi^\prime_{(\mu,\nu)} \\
& = & \prod_{n=1}^{\infty} \frac{1}{(1-q^n)^{2n}} \times \sum_{\mu,\nu \geq 0} \left( q^{h_\mu+h^\prime_\mu} z^{|\mu|} P^2_\mu(q) \right) \times \left( q^{h_\nu+h^\prime_\nu} z^{-|\nu|} P^2_\nu(q) \right).
\end{eqnarray}
We turned on the fugacity parameter $z$ for the $U(1)$ current associated to one of the two Y-algebra vertices which refines the character. Now we need to evaluate
\begin{equation}
\sum_{\mu \geq 0} \left( q^{h_\mu+h^\prime_\mu} z^{|\mu|} P^2_\mu(q) \right) = \sum_{\mu \geq 0} \left( q^{\frac{1}{2} \sum_j \mu_j^2 + \frac{1}{2} \sum_j (2j-1) \mu_j + \rho \sum_j \mu_j} z^{\sum_j \mu_j} P^2_\mu(q) \right).
\end{equation}
This sum is a typical example of sums studied in the topological vertex computations and we find
\begin{equation}
\sum_{\mu \geq 0} \left( q^{h_\mu+h^\prime_\mu} z^{|\mu|} P^2_\mu(q) \right) = \prod_{n=1}^{\infty} (1+z q^{\rho+n})^n.
\end{equation}
This again reproduces the formula (\ref{charX}), this time with the additional fugacity parameter $z$.


Let us now consider two special values of the parameter $\rho$. In the case when $\rho=0$, one gets in the large $N$ limit the character of $\mathcal{W}_{1+\infty}^{1|1}$ (an algebra generated by $2 \times 2$ matrix of generators for each integral spin). This algebra appeared in \cite{Costello:2016nkh} as an example in the context of categorified Donaldson-Thomas invariants and corresponding counting of D0-D2-D6 bound states. We devote the next section to the example of $\rho=\frac{1}{2}$ that can be identified with $\mathcal{N}=2$ super-$\mathcal{W}_{\infty}$.

\begin{figure}
\centering
\begin{subfigure}[b]{0.3\textwidth}
\includegraphics[width=\textwidth]{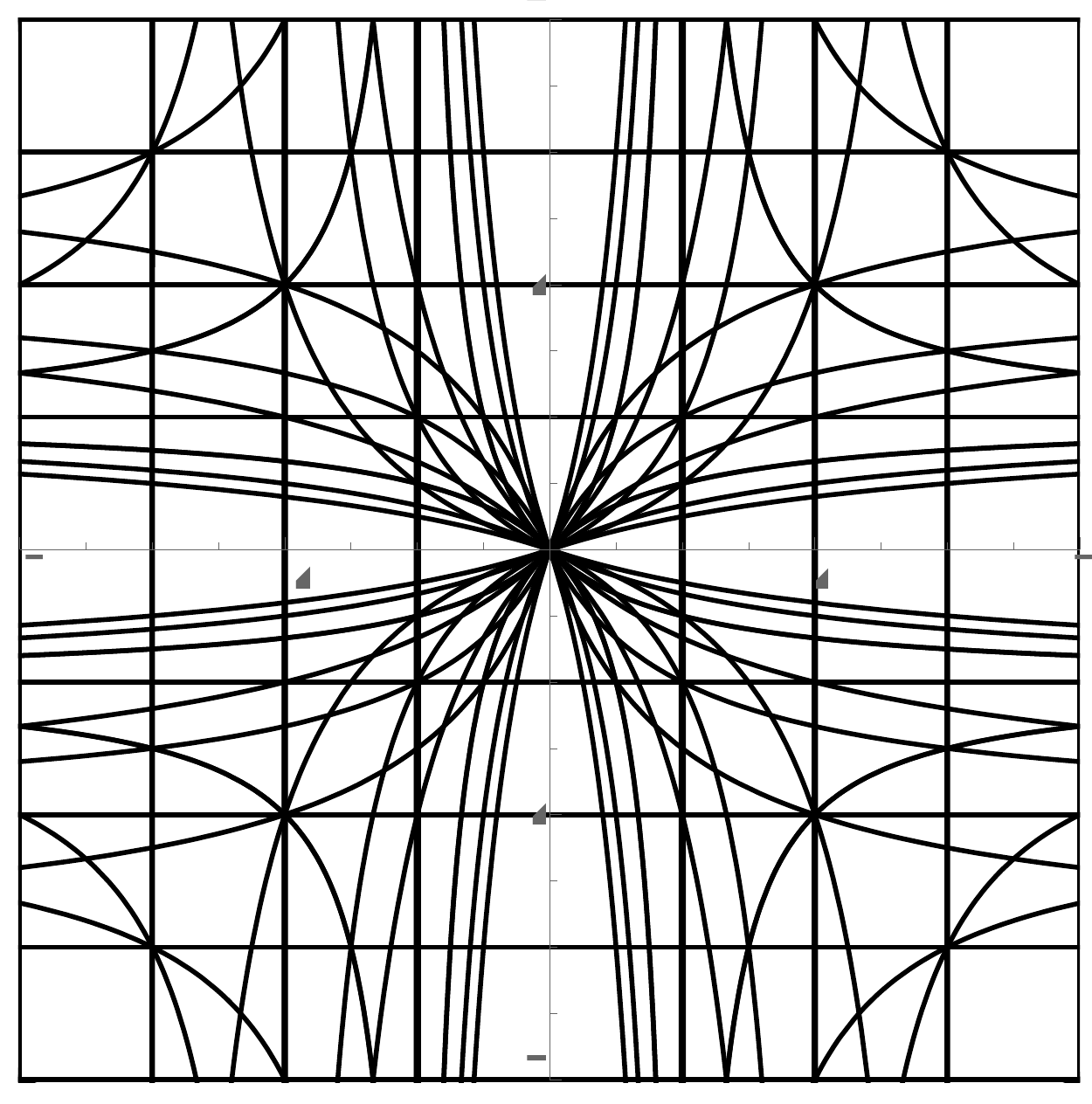}
\caption{$\rho=0$}
\end{subfigure}
\begin{subfigure}[b]{0.3\textwidth}
\includegraphics[width=\textwidth]{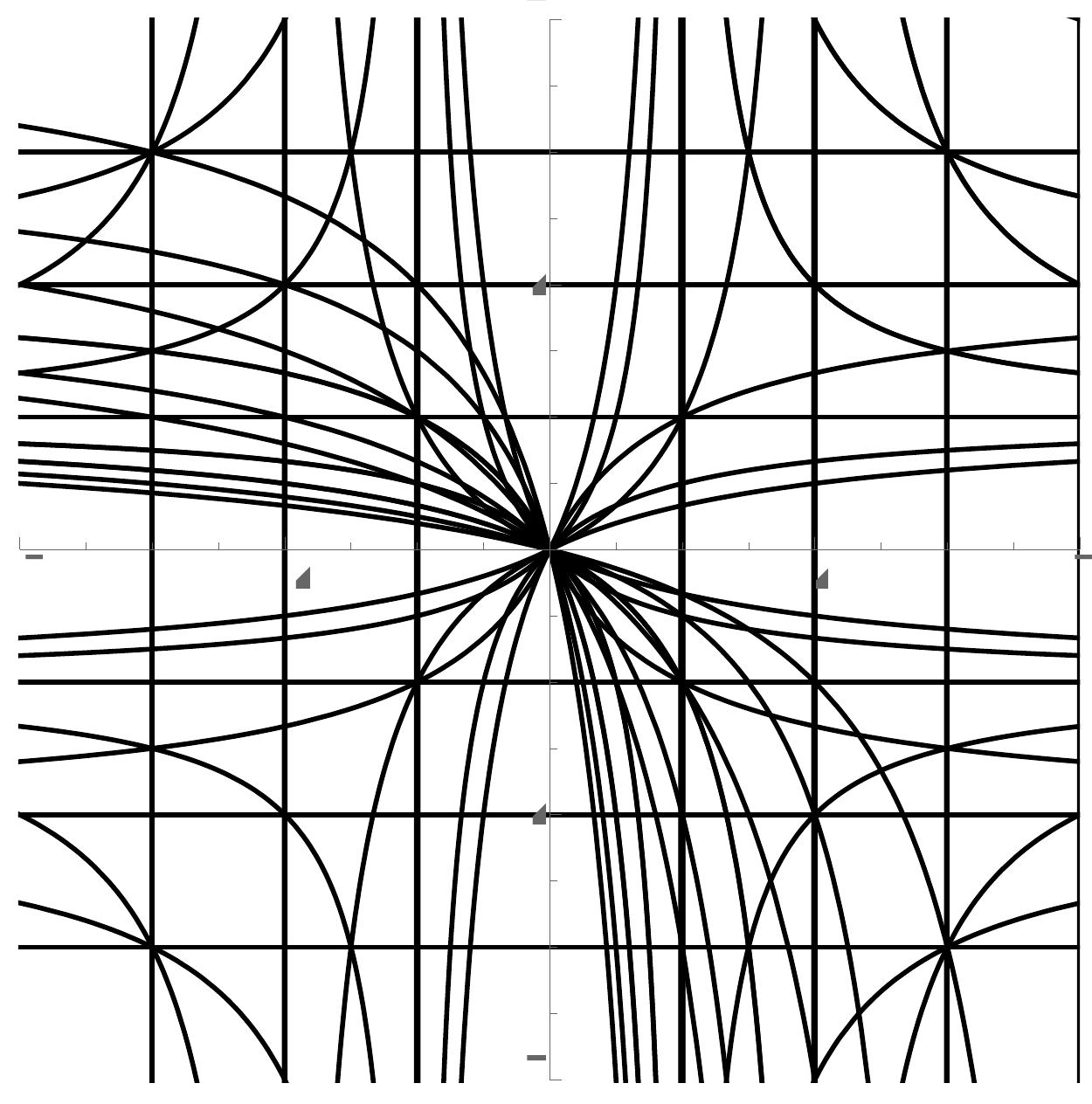}
\caption{$\rho=\frac{1}{2}$}
\end{subfigure}
\begin{subfigure}[b]{0.3\textwidth}
\includegraphics[width=\textwidth]{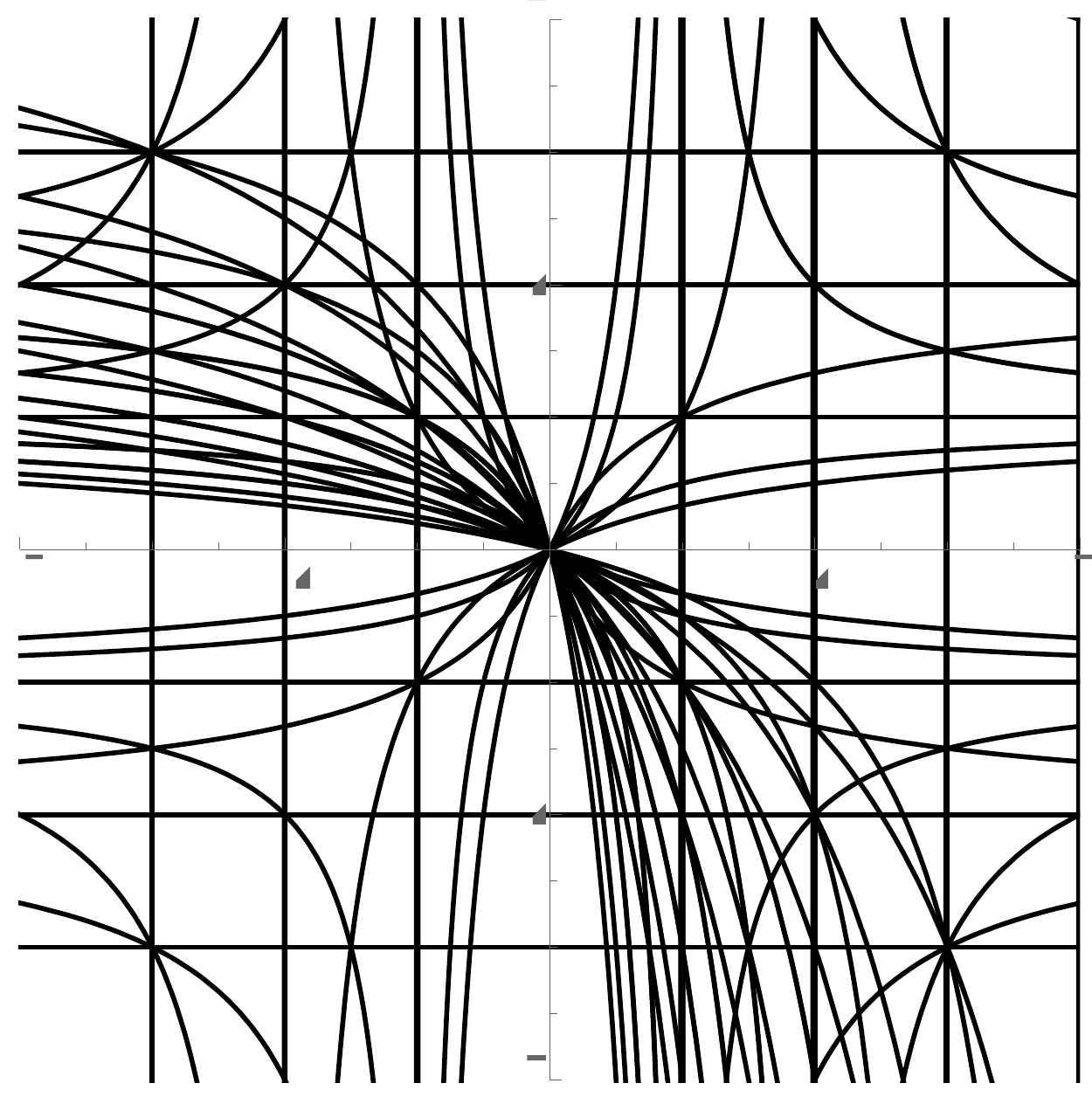}
\caption{$\rho=1$}
\end{subfigure}\\
\begin{subfigure}[b]{0.3\textwidth}
\includegraphics[width=\textwidth]{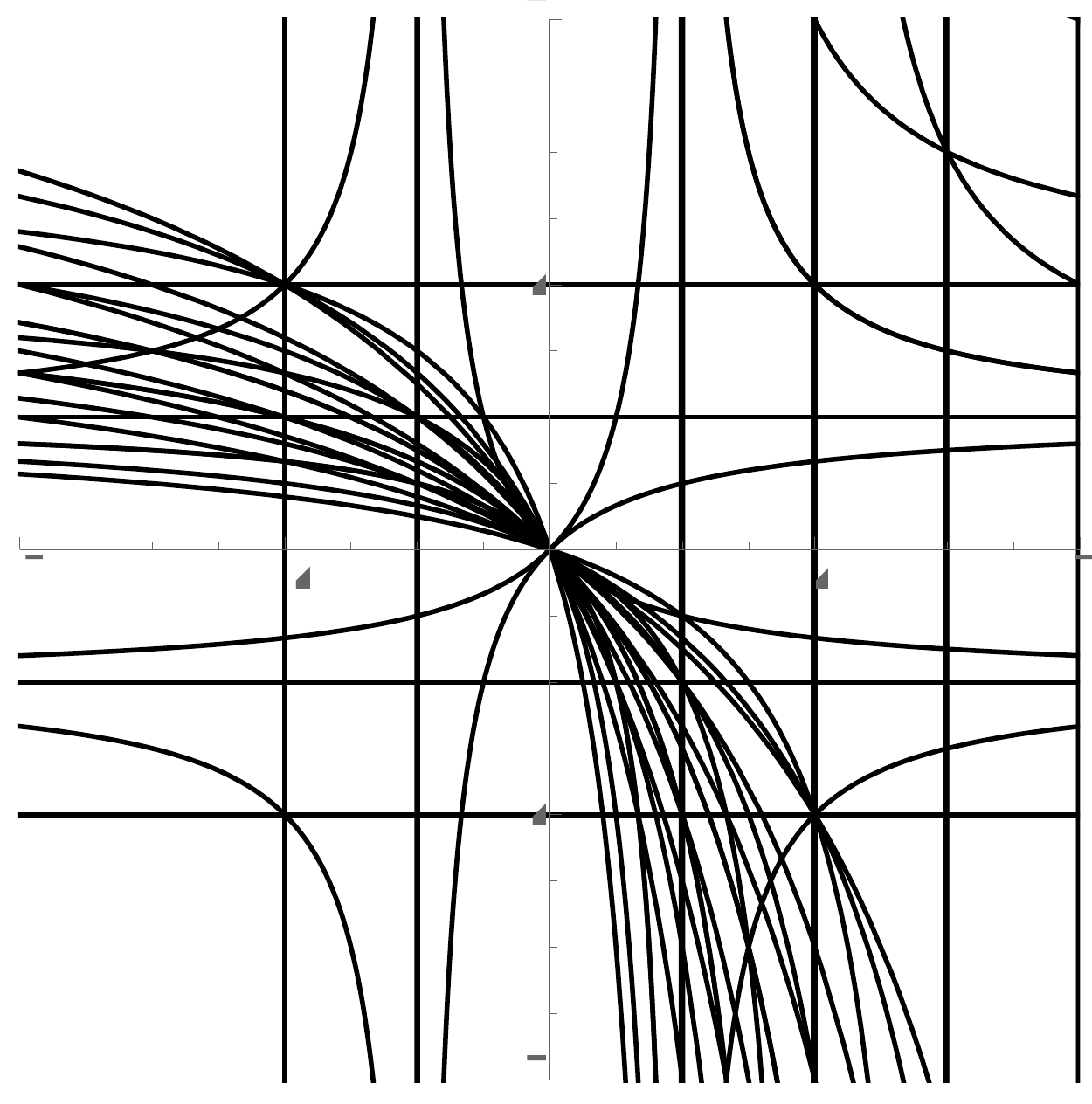}
\caption{$\rho=\frac{3}{2}$}
\end{subfigure}
\begin{subfigure}[b]{0.3\textwidth}
\includegraphics[width=\textwidth]{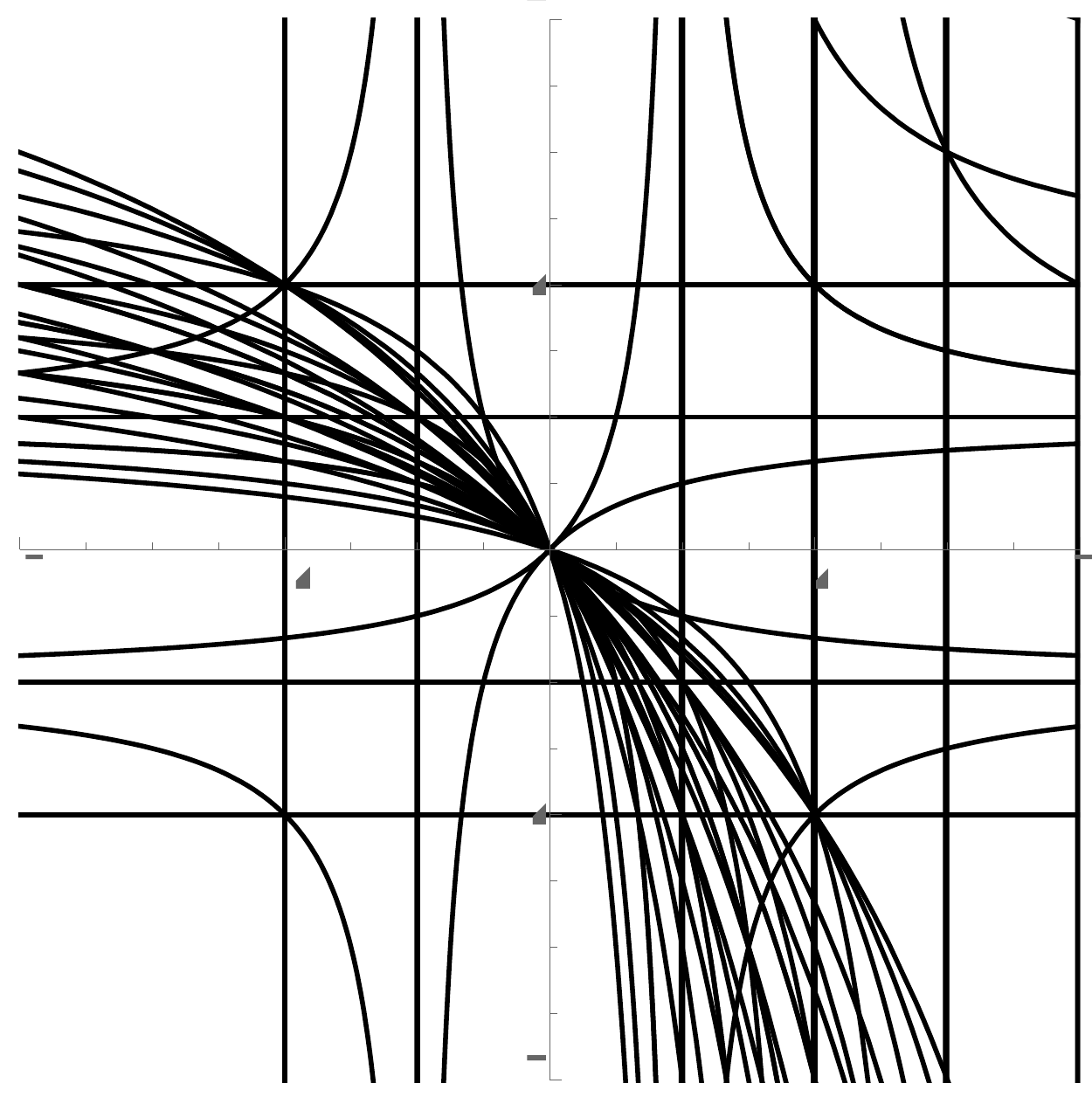}
\caption{$\rho=2$}
\end{subfigure}
\begin{subfigure}[b]{0.3\textwidth}
\includegraphics[width=\textwidth]{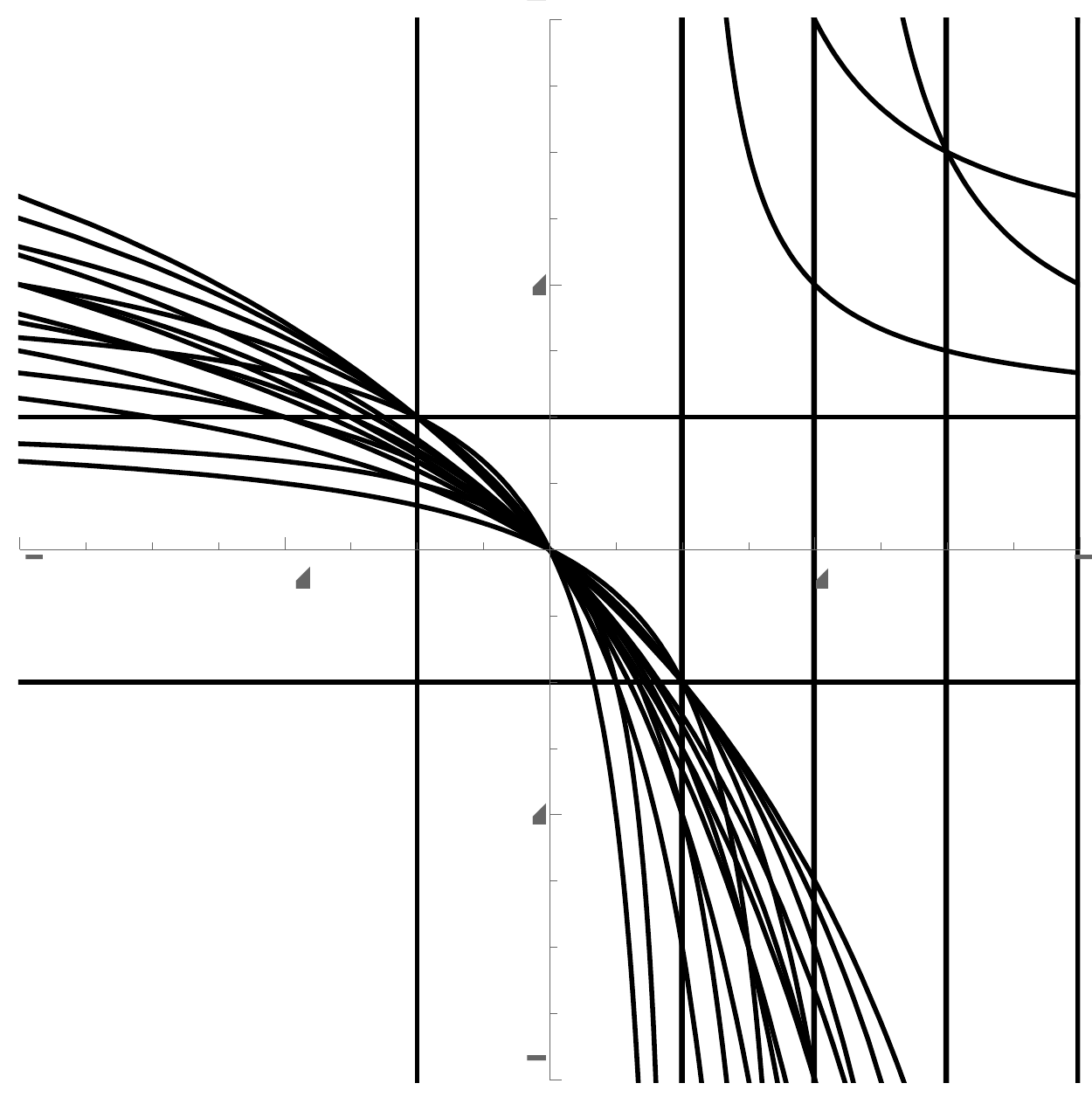}
\caption{$\rho=\frac{5}{2}$}
\end{subfigure}
\caption{First few truncation curves in the $(\mu_1,\mu_3)$ parametrization for $0\leq\rho\leq \frac{5}{2}$ and the number of branes $K+L+M+N\leq 8$. Note that the figures are invariant under the reflection $\mu_1 \leftrightarrow \mu_2$ corresponding to the S-duality action.}
\label{Xtrunc}
\end{figure}

\subsubsection{$\mathcal{N}=2$ super-$\mathcal{W}_\infty$}

 In the case of $\rho=\frac{1}{2}$, the character can be identified with the vacuum character of $\mathcal{N}=2$ super-$\mathcal{W_{\infty}}\times U(1)$ of \cite{Candu:2012tr} (times our standard $U(1)$ factor).\footnote{See also \cite{Bergshoeff:1990yd} for the special case of parameters where the algebra becomes linear.} The authors extended the $\mathcal{N}=2$ superconformal algebra by a simple tower of higher spin $\mathcal{N}=2$ supermultiplets with spins of lowest components being $2,3, \ldots$. Imposing the Jacobi identities, a two-parameter family of such algebras was found. For special values of parameters, a truncation of this algebra admits a coset construction using the Kazama-Suzuki coset
\begin{eqnarray}
\frac{U(N+1;\Psi^{-1}) \times \mathcal{S}^{N|0}}{U(N;\Psi^{-1})}
\end{eqnarray}
and a construction using the Drinfeld-Sokolov reduction of $U(N+1|N)$. Both of these realizations can be identified with the BRST constructions that we propose for a special choice of discrete parameters $K,L,M,N$ and turn out to be related by S-duality of our diagram 
\begin{eqnarray}
\mathcal{L}^{1|1}_{N+1,\bar{N},0,\bar{0}}[\Psi]\quad \leftrightarrow \quad X_{N+1,\bar{0},N,\bar{0}}\left[\frac{1}{\Psi}\right].
\end{eqnarray}

In particular, for $N=1$, we get the $\mathcal{N}=2$ superconformal algebra that is an extension of the Virasoro algebra by a spin $1$ current and two oppositely charged spin $\frac{3}{2}$ fermions. Together with the stress-energy tensor, these four fields form a $\mathcal{N}=2$ supermultiplet with lowest component having spin $1$. In \cite{Romans:1991wi}, the $\mathcal{N}=2$ SCA was extended by adding a $\mathcal{N}=2$ supermultiplet with lowest spin $2$. In appendix \ref{secn2w3} this algebra is discussed from the gluing point of view. The algebra is again associated to the conifold diagram $N=2$. It is natural to conjecture that in general the $\mathcal{N}=2$ extension of $\mathcal{W}_N$ is given by the $\mathcal{L}^{1|1}_{N+1,\bar{N},0,\bar{0}}$ algebra and that all the other configurations with $\rho=\frac{1}{2}$ are other truncations of $\mathcal{N}=2$ $\mathcal{W}_{\infty}$.

Candu and Gaberdiel introduce a parameter $\mu$ with the property that setting $\mu=-N$, we recover the truncations discussed above and parametrize the full algebra in terms of $\mu$ and the central charge $c$. Analogously to the triality symmetry of $\mathcal{W}_{1+\infty}$, at each generic fixed value of the central charge $c$ there are four different values of $\mu$ which give identical OPEs in the primary basis. These values of $\mu$ are \cite{Candu:2012tr}
\begin{eqnarray}\nonumber
\mu_1 = \mu, \qquad \mu_2 = \frac{(c-1)\mu}{c+3\mu}, \qquad \mu_3 = \frac{c+3\mu}{3(\mu-1)}, \qquad \mu_4 = -\frac{c}{3\mu}.
\end{eqnarray}
Since $\mathcal{N}=2$ $\mathcal{W}_{\infty}$ has apart from the stress-energy tensor an extra spin $2$ primary field commuting with the $U(1)$ factor, we can find a linear combination of the spin $2$ fields which give us two independent commuting Virasoro subalgebras. Their central charges are
\begin{eqnarray}
\nonumber
c_1 & = & \frac{c(\mu+1)(c+6\mu-3)}{3(c+3\mu)^2} \\
c_2 & = & -\frac{(c-3\mu)(c(\mu-2)-3\mu)}{3(c+3\mu^2)} \\
\nonumber
c & = & c_1 + c_2 +1.
\end{eqnarray}
Note that we have
\begin{eqnarray}\nonumber
c_1 & = & \frac{(1+\mu_1)(1+\mu_3)(\mu_1\mu_3-\mu_1-\mu_3)}{\mu_1+\mu_3}, \\
c_2 & = & \frac{(1+\mu_2)(1+\mu_4)(\mu_2\mu_4-\mu_2-\mu_4)}{\mu_2+\mu_4}
\end{eqnarray}
so defining
\begin{eqnarray}
\mu_1=-\lambda_1,\quad \mu_2=-\lambda^\prime_1,\quad \mu_3=-\lambda_2,\quad \mu_4=-\lambda^\prime_2
\end{eqnarray}
we can rewrite the partial central charges $c_1$ and $c_2$ in the standard form
\begin{eqnarray}\nonumber
c_1 & = & (\lambda_1-1)(\lambda_2-1)(\lambda_3-1)\\
c_2 & = & (\lambda^\prime_1-1)(\lambda^\prime_2-1)(\lambda^\prime_3-1).
\end{eqnarray}
We can thus identify the parameters $\lambda_j$ and $\lambda^\prime_j$ with the $\lambda$-parameters associated to the two vertices. This hints that $\mathcal{N}=2$ super $\mathcal{W}_\infty$ algebra indeed contains two mutually commuting $\mathcal{W}_\infty$ algebras as subalgebras and gives a picture consistent with the gluing\footnote{A similar observation was made by \cite{gaberdieln2yang} where the authors study the gluing of $\mathcal{N}=2$ $\mathcal{W}_{\infty}$ from the Yangian point of view.}. The duality transformations of the algebra that can be identified with the $\mathbbm{Z}_2\times \mathbbm{Z}_2$ duality action given by transformations
\begin{equation}
\Psi \leftrightarrow \Psi, \qquad K \leftrightarrow M, \qquad L \leftrightarrow N
\end{equation}
and
\begin{equation}
\Psi \leftrightarrow \frac{1}{\Psi},\qquad K\leftrightarrow M
\end{equation}
that can be identified with permutation of parameters
\begin{eqnarray}\nonumber
&\mu_1 \leftrightarrow \mu_2,\qquad \mu_3 \leftrightarrow \mu_4&\\
&\mu_1 \leftrightarrow \mu_3, \qquad \mu_2 \leftrightarrow \mu_4.&
\end{eqnarray}
Note that the parametrization and the whole construction works for arbitrary value of $\rho$ and we expect an existence of a two continuous parameter families of $\mathbbm{Z}_2\times \mathbbm{Z}_2$ algebras for each choice of $\rho$ such that $\mathcal{L^{1|1}_{K,\bar{L},M,\bar{N}}}$ are their truncations. The structure of these truncations in the $(\mu_1,\mu_2)$ parameter space is shown in the figure \ref{Xtrunc} for various values of $\rho$. You can see that figures are indeed invariant under $\mu_1\leftrightarrow \mu_2$. The points where two truncation curves intersect correspond to the BRST reductions at rational levels and we expect them to correspond to minimal models of $\mathcal{W}_{1|1\times \infty}^{\rho}$ algebras.

\subsection{Truncations of $\mathcal{W}_{2\times \infty}^{\rho}$}

Let us now consider the case  of  algebras of the type $0|2$ with the corresponding parameter $\rho$ fixed. Sending parameters $K,L,M,N \rightarrow \infty$ to infinity, relations satisfied by product of bilinears  in the BRST calculation of the character disappear and one gets a character analogous to $\mathcal{W}_{1|1\times \infty}^\rho$. The only difference is that all the invariant combinations are bosonic and the final character is given by
\begin{eqnarray}
\chi \left [\mathcal{W}_{0|2\times \infty}^{\rho}\right ]=\prod_{n=1}^{\infty}\frac{1}{(1-q^{n})^{2n}(1-z q^{n+\rho})^n(1-z^{-1}q^{n+\rho})^n}.
\end{eqnarray}
The same formula can be obtained from the gluing construction in the same way as in the resolved conifold case. We just need to use
\begin{equation}
\sum_{\mu \geq 0} \left( q^{h_\mu+h^\prime_\mu} z^{|\mu|} P^2_\mu(q) \right) = \sum_{\mu \geq 0} \left( q^{\sum_j (2j-1) \mu_j + \rho \sum_j \mu_j} z^{\sum_j \mu_j} P^2_\mu(q) \right) = \prod_{n=1}^{\infty} \frac{1}{(1-z q^{\rho+n})^n}.
\end{equation}
Algebras discussed in this section can be identified with truncations of $\mathcal{W}_{0|2\times \infty}^{\rho}$. Note that $\rho=0$ case again coincides with algebras studied in \cite{Costello:2016nkh} in the context of counting D6-D2-D0 bound states on the resolution of $\mathds{C}^2/\mathbbm{Z}_2 \times \mathds{C}$. 

\subsection{Truncations of $\mathcal{W}_{M|N\times \infty}^{\rho_i}$}

All examples discussed so far in this section can be identified with truncations of some infinite algebra. In the BRST reduction described in above, one generates $\mathcal{W}_N$ algebra and symplectic bosons in fundamental representation of the reduced group associated to each vertex. Moreover, at each vertex, symplectic bosons generated in the previous step decomposes into the fields of shifted dimensions and symplectic bosons in fundamental representation of the reduced group. Example of such process for first three reductions from the example (\ref{complicated}) is diagrammatically captured in \ref{branching}. After projecting to the coset invariant combinations, one can argue that in the infinite number of branes limit, one obtains the character of the form
\begin{eqnarray}
\chi [\mathcal{W}_{M|N\times \infty}^{\rho_i}]=\left(\prod_{n=1}^{\infty}\frac{1}{(1-q^n)^n}\right)^{N+M}\prod_{i=1}(1\pm q^{n+\rho_i})^{\pm 2n}\prod_{i> j}(1\pm q^{n+\rho_i+\rho_j})^{\pm 2n}
\end{eqnarray}
where the products run over all internal edges and one chooses the $+$ sign if both branes of corresponding finite segment ends from the same side and the $-$ sign otherwise. The same character follows from the gluing construction: for example, in the $U(3)$ case we use the fact that
\begin{multline}
\prod_{n=1}^\infty \frac{1}{(1-\frac{z_1}{z_2}q^n)^n (1-\frac{z_1}{z_3}q^n)^n (1-\frac{z_2}{z_3}q^n)^n} = \\
= \sum_{\mu,\nu} q^{||\mu||^2 + ||\nu||^2 - (\mu,\nu)} \left(\frac{z_1}{z_2}\right)^{|\mu|} \left(\frac{z_2}{z_3}\right)^{|\nu|} P(\cdot,\mu) P(\mu,\nu) P(\nu,\cdot)
\end{multline}
where $P(\mu,\nu)$ are the the box-counting functions \cite{Prochazka:2015aa} related up to an overall factor to the topological vertex $C(\mu,\nu,\cdot)$. For the total character we thus find
\begin{equation}
\chi^{\rho=0}_{3 \times \infty} = \prod_{n=1}^{\infty} \prod_{j,k=1}^{3} \frac{1}{(1-z_j z_k^{-1} q^n)^n}
\end{equation}
as expected for $\mathcal{W}_{3 \times \infty}$.

Algebras coming from gluing or BRST construction for finite number of branes can be identified with truncations of $\chi_{N|M\times \infty}^{\rho_i}$. For fixed values $\rho_i$, there are three integral parameters left unfixed. These parameters parametrize truncation lines of $\chi_{N|M\times \infty}^{\rho_i}$ inside the conjecturally two parameter family of algebras. Shifting all the numbers of branes by a constant value again corresponds to a different truncation above the same truncation curve.

For $\rho_i = 0$ we already know a large three-parameter family of truncations coming from the coset algebras
\begin{equation}
\frac{U(K+M|L+N)_k}{U(K|L)_k}.
\end{equation}
By construction these cosets contain $U(M|N)_k$ as subalgebra but for $K \neq 0 \neq L$ also other fields extending it.

\section{Flip transition}
\label{flop}

In this section, we comment on the flip transition that plays an important role in the literature related to the BPS counting. At the level of diagrams the flip transition corresponds to the sliding of five-branes. We conjecture that the algebras associated to diagrams related by a flip transformation differ by a trivial algebra of decoupled fermions or symplectic bosons. 

As a test of this conjecture we argue that central charges of the algebras related by a flip differ only by a contribution of free fermions and symplectic bosons. In the case of the linear diagrams with a BRST definition, we show that vacuum characters differ by a contribution of such free fields. We explicitly construct algebras $\mathcal{L}^{0|2}_{3,\bar{0},2,0}$, $\mathcal{L}^{1|1}_{0,\bar{1},0,\bar{0}}$ and $\mathcal{L}^{1|1}_{0,\bar{2},0,\bar{0}}$ and show that they differ from the flipped algebra by a free field contribution. 

\subsection{Flip of algebras of type $1|1$}

\begin{figure}[h]
\centering
\includegraphics[width=0.62\textwidth]{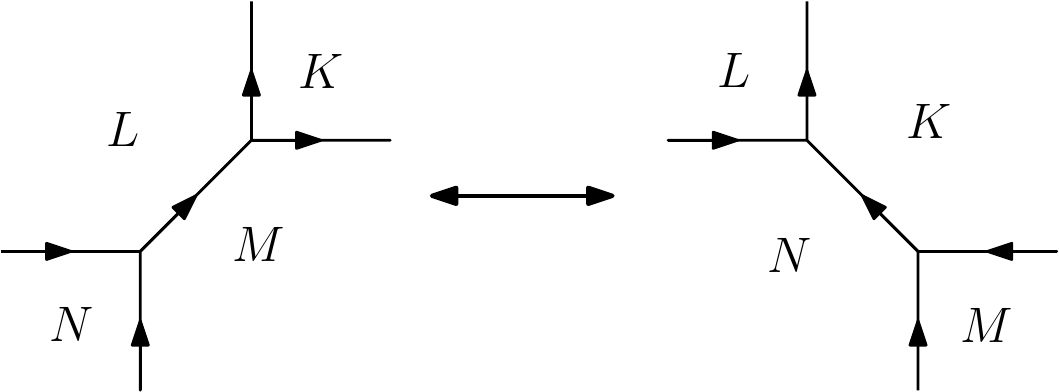}
\caption{Flip transition for type $1|1$ (resolved conifold) algebras.}
\label{con_flop}
\end{figure}

Let us start with a flip in the resolved conifold diagram. This transition exchanges the order of the two D5-branes as shown in the figure \ref{con_flop}. The related algebras are
\begin{eqnarray}
\mathcal{L}^{1|1}_{K\bar{L},M,\bar{N}}[\Psi]\quad \leftrightarrow \quad \mathcal{L}^{1|1}_{L,\bar{K},N,\bar{M}}[-\Psi]
\end{eqnarray}
where the minus sign is a consequence of the parity transformation relating the right hand side of the figure \ref{con_flop} to the standard resolved configuration. 

The central charges of the flipped algebras differ by a $\Psi$-independent factor
\begin{eqnarray}
c\left [\mathcal{L}^{1|1}_{K\bar{L},M,\bar{N}}[\Psi]\right ]-c\left [\mathcal{L}^{1|1}_{L,\bar{K},N,\bar{M}}[-\Psi]\right ]=4\rho(2\rho^2-1).
\end{eqnarray} 
This is exactly the contribution coming from free $(b,c)$ systems with conformal dimensions
\begin{eqnarray}
(\rho+1,-\rho), (\rho+2,-\rho),\dots, (-\rho,\rho+1)
\end{eqnarray}
since (for $\rho<0$)
\begin{eqnarray}
\sum_{h=\rho+1}^{-\rho}c[h]=4\rho(2\rho^2-1)
\end{eqnarray}
where $c[h]$ is the central charge of the stress-energy tensor of the $b,c$ ghost system with respect to which $h_c=h$ and $h_b=1-h$ are the conformal weights of the $c,b$ fields.

At the level of characters, the difference is by a factor of (for $\rho$ negative)
\begin{eqnarray}
\prod_{m=\rho+\frac{1}{2}}^{-\rho+\frac{1}{2}}\chi_m^{\mathcal{F}}=\prod_{n=0}^{\infty}\prod_{m=\rho+1}^{-\rho+1}\left (1+q^{n+m}\right )
\end{eqnarray}
as can be most easily seen from the BRST construction. The only difference at the level of the BRST reduction is in the two off-diagonal blocks whose elements are charged under the Cartan elements of both $sl_2$ embeddings. The contributions from the $\mathcal{W}_{K-M}$ and $\mathcal{W}_{L-N}$ factors are present in the characters of both algebras. The integral projecting on the $U(N|M)$ invariant combinations is the same as well since all the fields under the integral originate from the off-diagonal blocks charged with respect to only one of the two $sl_2$ embeddings. The only difference is thus in the product of $\chi^\mathcal{F}_i$ factors given above.

\subsubsection{Example - Flip of $U(1;\Psi)$}
The flip of the $U(1;\Psi)$ algebra is the simplest but also trivial example of a flip transition since the algebra is simply
\begin{eqnarray}
\mathcal{DS}_{0}[U(1;\Psi)]=U(1;\Psi)\times \mathcal{F} 
\end{eqnarray}
from the definition. It automatically contains a decoupled fermion $\mathcal{F}$.

\subsubsection{Example - Flip of Virasoro $\times$ $U(1;\Psi)$}

A non-trivial example is the flip of the $\mathcal{L}^{1|1}_{2,\bar{0},0,\bar{0}}[\Psi]$ algebra, i.e.  $\mathcal{L}^{1|1}_{0,\bar{2},0,\bar{0}}[-\Psi]$. The BRST definition of the algebra is in terms of a reduction of
\begin{eqnarray}
U(2;\Psi)\times \{ \chi_1,\psi_1\}\times \{ \chi_2,\psi_2\}\times \{ b,c\}
\end{eqnarray}
implemented by the BRST charge
\begin{eqnarray}
Q=\oint dz (J_{12}+\chi_1\psi_2-1)c
\end{eqnarray}
that can be identified with the BRST reduction associated to the principal $sl_2$ embedding inside $U(2)$ but with the current modified by a fermionic bilinear.

The cohomology is generated by fields
\begin{eqnarray}
\nonumber
J & = & J_{11}+J_{22}\\ \nonumber
\tilde{\psi}_1 & = & \psi_1+cb\psi_2+J_{11}\psi_2\\ \nonumber
\tilde{\chi}_1 & = & \chi_1\\ \nonumber
\tilde{\psi}_2 & = & \psi_2\\
\tilde{\chi}_2 & = & \chi_2-cb\chi_1+J_{11}\chi_1\\ \nonumber
T & = & \frac{1}{2\Psi}\left ( J_{11}J_{11}+2J_{12}J_{21}+J_{22}J_{22}\right ) \\
\nonumber
& & -c\partial b-\psi_1\partial \chi_1+\partial \psi_2 \chi_2+\frac{\Psi-1}{2\Psi}(J_{11}'-J_{22}')
\end{eqnarray}
where $T$ is the stress-energy tensor with the central charge
\begin{eqnarray}
c = 10-6\Psi-\frac{6}{\Psi}
\end{eqnarray} 
as expected. The fields $(\tilde{\psi}_1,\tilde{\chi}_1)$ and $(\tilde{\psi}_2,\tilde{\chi}_2)$ have OPEs of a pair of free fermions. Their OPEs with $J$ are 
\begin{eqnarray}
J(z)\tilde{\psi}_1\sim -\frac{\Psi \tilde{\psi}_2}{(z-w)^2},\quad J(z)\tilde{\chi}_1\sim \frac{\Psi \tilde{\chi}_2}{(z-w)^2}.
\end{eqnarray}
$\tilde{\chi}_1$ and $\tilde{\psi}_2$ are primaries of conformal dimension $0$ while $\tilde{\chi}_2$ and $\tilde{\psi}_1$ having OPEs with the stress-energy tensor of the form
\begin{eqnarray}\nonumber
T(z)\tilde{\chi}_2(w)&\sim& \frac{(1-\Psi)\tilde{\chi}_1}{(z-w)^3}+\frac{\tilde{\chi}_2(w)}{(z-w)^2}+\frac{\partial \tilde{\chi}_2(w)}{z-w}\\
T(z)\tilde{\psi}_1(w)&\sim& -\frac{(1-\Psi)\tilde{\psi}_2}{(z-w)^3}+\frac{\tilde{\psi}_1(w)}{(z-w)^2}+\frac{\partial \tilde{\psi}_1(w)}{z-w}
\end{eqnarray}
Note that one can modify both the $J$ current and the stress-energy tensor $T$ as
\begin{eqnarray}\nonumber
J \ &\rightarrow& \ J+\Psi \partial \tilde{\chi}_1\tilde{\psi}_2\\
T \ &\rightarrow& \ T-\partial \tilde{\chi}_1\tilde{\psi}_1-\partial \tilde{\psi}_2\tilde{\chi}_2-\Psi \partial \tilde{\chi}_1\tilde{\chi}_1\tilde{\psi}_2\partial \tilde{\psi}_2+\frac{\Psi-1}{2}\partial^2 (\tilde{\chi}_1\tilde{\psi}_2).
\end{eqnarray} 
After such a modification, the free fields decouple and one is left with the Virasoro algebra times a $U(1)$ algebra. The central charge of the modified stress-energy tensor 
\begin{eqnarray}
14-6\Psi-\frac{6}{\Psi}
\end{eqnarray}
is the same as the central charge of the algebra before the flip.

\subsection{Flip of algebras of type $0|2$}

\begin{figure}[h]
\centering
\includegraphics[width=0.7\textwidth]{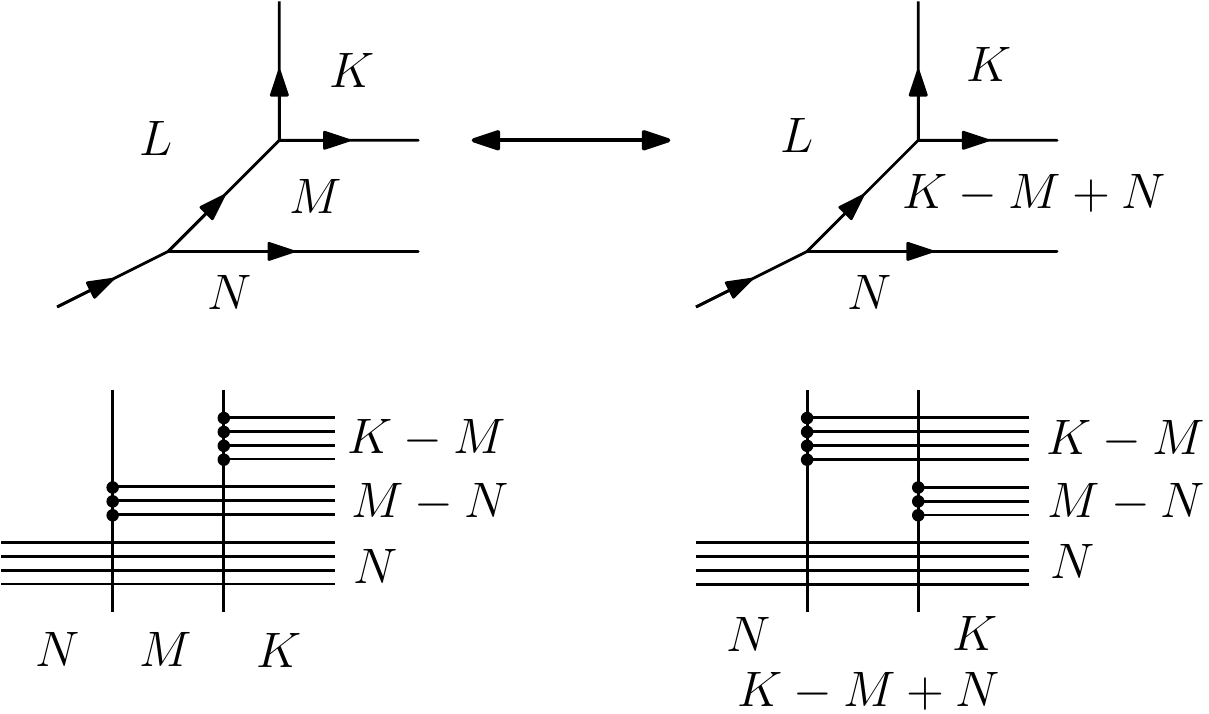}
\caption{Flip transition for type $0|2$ algebras in the case of $K\geq M\geq N$. A fixed number of D3-branes is attached to each D5-brane. The crossing of D5-branes acts non-trivially on the number of D3-branes at the internal face. }
\label{F_flop}
\end{figure}
Flipping D5-branes in the $0|2$ diagram results in the change of numbers of D3-branes between the two five-branes as shown in the figure \ref{F_flop}. The algebras related by such a flip are
\begin{eqnarray}
\mathcal{L}^{0|2}_{K,\bar{L},M,N}[\Psi]\ \leftrightarrow \ \mathcal{L}^{0|2}_{K,\bar{L},K-M+N,N}[\Psi].
\end{eqnarray}
One can show analogously to the resolved conifold diagram that the central charges and characters again differ by a contribution of $2|\rho|$ copies of the $(\beta, \gamma)$ systems with correct conformal dimensions. The only difference in this case is the different expression for the parameter $\rho$ and the bosonic nature of the decoupled fields.

\subsubsection{Example $\mathcal{L}^{0|2}_{0,\bar{0},2,3}$}

Let us show that the algebra $\mathcal{L}^{0|2}_{0,\bar{0},2,3}$ associated to the flip of the $\mathcal{W}_3^{(2)} \times U(1)$ algebra contains $\mathcal{W}_3^{(2)} \times U(1)$ as a subalgebra together with a decoupled free fermion.

The algebra $\mathcal{L}^{0|2}_{0,\bar{0},2,3}$ is defined as a BRST reduction of $U(3;\Psi) \times \{ b_{12},c_{12}\}$ by the BRST charge
\begin{eqnarray}
Q_2=\oint dz (J_{12}-1)c_{12}.
\end{eqnarray} 
The cohomology contains the currents
\begin{eqnarray}\nonumber
J_1&=&\frac{J_{11}+J_{22}}{\Psi}-\frac{1-\Psi}{\Psi}J_{33},\\
J_2&=&J_{11}+J_{22}
\end{eqnarray} 
that are mutually local and they are normalized  according to (\ref{dimension}) as
\begin{eqnarray}
J_1(z)J_1(w)\sim \frac{\Psi+\frac{3}{\Psi}-4}{(z-w)^2},\qquad J_1(z)J_1(w)\sim \frac{2(\Psi-1)}{(z-w)^2}.
\end{eqnarray}
Apart from these currents, the reduced algebra contains generators of dimension $\frac{1}{2}$ given by
\begin{eqnarray}\nonumber
G_1^+&=&J_{13},\\
G_1^-&=&J_{32}
\end{eqnarray}
and of dimension $\frac{3}{2}$ of the form
\begin{eqnarray}\nonumber
G_2^+&=&J_{23}+(\Psi-2)J_{11}J_{13}+(\Psi-3)J_{13}J_{22}+J_{13}bc,\\
G_2^-&=&J_{31}+(\Psi-3)J_{11}J_{32}+(\Psi-2)J_{22}J_{32}-J_{32}bc
\end{eqnarray}
together with the stress-energy tensor
\begin{eqnarray}
T=\frac{1}{2\Psi}\sum_{i,j=1,2,3}J_{ij}J_{ji}+\frac{1}{2}\partial J_{11}-\frac{1}{2}\partial J_{22}+\partial b_{12}c_{12}
\end{eqnarray}
with the central charge
\begin{eqnarray}
c=25-\frac{24}{\Psi}-6\Psi.
\end{eqnarray}
The superscripts $\pm$ in the expressions above denote the charge of the gluing fields with respect to $J_1$ and $J_2$ currents. 

Let us discuss OPEs of the algebra. $G_1^\pm$ form subalgebra of symplectic bosons with OPE
\begin{eqnarray}
G_1^+ (z)G^-_1(w)\sim \frac{1}{z-w}.
\end{eqnarray}
The operator product expansions between $J_i$ currents and $G_2^\pm$ fields are
\begin{eqnarray}
J_1(z)G_2^\pm(w)\sim \pm \frac{G_2^\pm}{z-w},\qquad J_2(z)G_2^\pm(w)\sim \frac{(\Psi-1)(2\Psi-5)G^\pm_1}{(z-w)^2}\pm \frac{G_2^\pm}{z-w},
\end{eqnarray}
OPEs between $G_1^\pm$ and $G_2^\pm$ fields are
\begin{eqnarray}\nonumber
G_1^{\pm}(z)G_2^{\pm}(w)&\sim& \pm \frac{(2-\Psi)G_1^\pm G_1^\pm }{z-w}\\
G_1^{\pm}(z)G_2^{\mp}(w)&\sim& \pm \frac{1}{z-w}\left ((3-\Psi)G^-_1G^+_1+\frac{\Psi}{\Psi-1}J_1+\frac{1-3\Psi+\Psi^2}{\Psi-1}J_2\right ).
\end{eqnarray}
Finally for OPEs between $G_2^\pm$ fields we find
\begin{eqnarray}\nonumber
G_2^\pm(z)G_2^\pm(w)&\sim&\pm\left (2\Psi^3-13\Psi^2+27\Psi-17\right )\left (\frac{G_1^\pm G_1^\pm }{(z-w)^2}+\frac{G_1^\pm \partial G_1^\pm}{z-w}\right ) \\ \nonumber 
G_2^+(z)G_2^-(w)&\sim&\frac{(\Psi-1)(\Psi-3)(2\Psi-5)}{(z-w)^3}\\ \nonumber 
&+&\frac{1}{(z-w)^2}\left ( \frac{2(2\Psi-\Psi^2)}{\Psi-1}J_1+\frac{(\Psi-3)(\Psi-2)}{\Psi-1}J_2+(2\Psi^3-11\Psi^2+16\Psi-3)G^+_1G^-_1\right )\\ \nonumber
&+&\frac{1}{z-w}\bigg ( \Psi T+2(2\Psi-5)J_2G_1^+G_1^- -\frac{\Psi^2}{2(\Psi-1)^2}J_1J_1+\frac{\Psi (6-7\Psi+2\Psi^2)}{(\Psi-1)^2 }J_1J_2\\ \nonumber
&+&\frac{(2\Psi^4-12\Psi^3+21\Psi^2-8\Psi-4)}{2(\Psi-1)^2}J_2J_2+(1-\Psi)G_1^+G_2^- \\ \nonumber
&+&2(\Psi^3-6\Psi^2+10\Psi-3)\partial G_1^+G_1^- +(\Psi^2-4\Psi+3)G_1^+\partial G_1^- \\
&+&\frac{2\Psi-\Psi^2}{\Psi-1}\partial J_1-\frac{\Psi^2-5\Psi+6}{2(\Psi-1)}\partial J_2\bigg ).
\end{eqnarray}
If we redefine the generators of the algebra as
\begin{eqnarray}\nonumber
J_1 &\rightarrow&  J_1+G_1^+G_1^- \\ \nonumber
J_2 &\rightarrow& J_2+G_1^+G_1^- \\  \nonumber
G_2^+ &\rightarrow&G_1^+ -(G_1^+)^2G_1^-+\frac{\Psi J_1G_1^+}{1-\Psi} +\frac{(1-3\Psi+\Psi^2)J_2G_1^+}{1-\Psi}+(\Psi-1)\partial G_1^+ \\ \nonumber
G_2^-&\rightarrow& G_1^- -G_1^+(G_1^-)^2+\frac{\Psi J_1G_1^-}{1-\Psi} +\frac{(1-3\Psi+\Psi^2)J_2G_1^-}{1-\Psi}-(\Psi-1)\partial G_1^- \\
T &\rightarrow& T-\frac{1}{2}\partial (G_1^+G_1^-)
\end{eqnarray}
we discover that the currents $G_1^\pm$ form a symplectic boson pair and they decouple. The remaining algebra can be identified with the algebra $\mathcal{W}_{3}^{(2)} \times U(1)$ with the stress-energy tensor of the correct central charge
\begin{eqnarray}
c=26-\frac{24}{\Psi}-6\Psi.
\end{eqnarray}

\subsection{Flip in a general diagram}

In a general tree diagram, the flip is a local transition that influences only the vertices associated to the flipped leg and the edge along which the leg flipped. In particular, this means that both the vacuum character and the central charge of the full algebra differ by a contribution of free fields that can presumably be decoupled and the change in these quantities can be read-off locally. 

Note also that both the central charge and the vacuum character remain the same in the case of vanishing $\rho=0$. At the level of BRST reduction, one can indeed see that the two reductions can be related by a unitary transformation of the current algebra generators. It is natural to expect that the two algebras related by a flip are equal in arbitrary diagram as soon as $\rho=0$ as we mentioned already in the discussion of the $D(2,1;-\Psi)_1\times U(1;2\Psi)$ algebra. This situation also applies for the $U(M|N;\Psi)$ Kac-Moody algebras where for symmetry reasons we expect the algebra to be independent of the order of reductions in the first and second diagonal block (see figure \ref{superpic}).

\section{Outlook}

\paragraph{Summation over supergroup representations}
From the technical point of view, we lack full understanding of summation over $Y_{L,M,N}$ representations with all three asymptotics non-trivial and the most general pit condition. At the level of characters, it would be very useful to know a formula for the character of representations of $Y_{L,M,N}$ with three non-trivial asymptotics with both covariant and contravariant labels (boxes and antiboxes) and with the general truncation condition.  From the BRST point of view, the main obstacle at this level seems to be related to understanding the relevant representations of supergroups and their characters. From the point of view of $\mathcal{W}_{1+\infty}$, one needs to understand the fusion ring of the box and anti-box representations, in particular the mixed covariant and contravariant representations.

\paragraph{Double truncations}
In most of this article, we considered one-parameter truncations of the algebras, i.e. keeping $\Psi$ generic. An interesting situation appears at the double truncation points, i.e. rational points in the space of parameters. These double truncations are associated to the points where two (an thus infinitely many) truncation curves in the $\lambda_i$ parameter space of $\mathcal{W}_{1+\infty}$ intersect. From the point of Y-algebras, these correspond to BRST reductions of Kac-Moody algebras at rational levels that contain null-vectors.  Such BRST reductions are known to lead (at least in the case of $\mathcal{W}_N$ algebras) to minimal models \cite{Frenkel:1992ju}. Note that at rational values of $\Psi$, the Kapustin-Witten thery contains bulk line operators and it might be possible to give a gauge theory interpretation of the identification of various modules by moving the line operators to the bulk.  Double truncations include in particular all the (unitary or not) $\mathcal{W}_N$ minimal models and lead to nice combinatorial problems counting the states in the highest weight representations. The doubly-truncated algebras are also the ones considered by \cite{Beem:2013sza,Beem:2014rza} which is another motivation to study them better.

\paragraph{Web diagrams with loops}
So far, all examples that we considered were associated to tree-level diagrams. The gluing procedure does not seem to lead to any inconsistency when applied also to diagrams with closed loops. However, it is not a priori clear if application of the gluing procedure straightforwardly would describe degrees of freedom associated to the corresponding brane configuration or if there are other degrees of freedom coresponding to the Gukov-Witten defects stretched at the finite faces that have to be included in this case. The configuration of D3-branes supported at these finite faces can be lifted to the configuration of M5-brane wrapping a coresponding four-cycle in the toric Calabi-Yau manifold and it is natural to conjecture some relation of our glued algebras with those of \cite{Dedushenko:2017tdw} in these particular cases of four-manifolds that can be realized as holomorphic four-cycles in a toric Calabi-Yau threefold.

\paragraph{Action of VOAs on equivariant cohomology of moduli spaces}
The duality with D4-branes intersecting each other inside $CY^3$ provides us with conjectural actions of VOAs on equivariant cohomologies of various moduli spaces as discussed in \cite{Maulik:2012rm}. Note that $Y_{0,0,N}$ associated to a single stack of D4-branes wraping $\mathds{C}^2$ inside $\mathds{C}^3$ correctly reproduces $\mathcal{W}_N$ algebra as an algebra acting on the moduli space of ADHM instantons \cite{Maulik:2012rm,Braverman:2014xca}. The more general algebra $Y_{L,M,N}$ have a dual of three stacks of D4-branes wraping three holomorphic $\mathds{C}^2$ cycles (fixed under the torus action) inside $\mathds{C}^3$ and intersecting each other. This is a special case of the configuration of spiked instantons discussed in \cite{Nekrasov:2016qym,Nekrasov:2016gud}. It is natural to conjecture the existence of the action of Y-algebras on these moduli spaces. This has a natural generalization to D4-branes intersecting in general toric $CY^3$ and we expect the glued algebras to play important role in the corresponding BPS counting problems.

 Moreover, one can realize the configuration of $N$ D4-branes wrapping a resolution of the $\mathds{C}^2/\mathbb{Z}_n$ singularity from the configuration of $n$ D5-branes ending on $(n,1)$ branes from the right and $N$ D3-branes supported at the face on the left. This configuration is indeed dual to the one of \cite{Nishinaka:2011nn} and generalizes it for a stack of D4-branes. Our BRST definition (and the gluing construction in the case of \cite{Alfimov:2013cqa}) seems to produce the same (or closely related) algebras to those discussed in \cite{Nishioka:2011jk,Belavin:2011pp,Belavin:2011tb,Belavin:2011sw,Alfimov:2013cqa}. The simple examples seem to match. It would be nice to explore this relation further. See also \cite{Kimura:2015rgi,Bourgine:2015szm}.

\paragraph{Refined topological vertex}
There are refined versions of the topological vertex \cite{Iqbal:2007ii,Awata:2005fa,Foda:2017tnv}. Because of the infinite set of commuting charges sitting at each vertex of $(p,q)$-web, we can in principle obtain many refined quantities. It would be interested to see if we can obtain the various refined quantities perhaps using the ideas of \cite{Song:2016yfd} or if a $q$-deformed version of the gluing construction works analogously for quantum toroidal algebras \cite{Awata:2011ce,feigin2011quantum,feigin2011quantum2,feigin2012quantum,Mironov:2016yue,Awata:2016riz,Awata:2016mxc,Awata:2016bdm,Bourgine:2017jsi}.

\paragraph{Integrability}
It is known that the $Y_{L,M,N}$ algebra can be (by a change of variables) recast in the Yangian form \cite{Maulik:2012rm,tsymbaliuk2017affine,Prochazka:2015aa}. In this form, there is an infinite set of commuting charges that can be explicitly diagonalized. From our gluing construction, we have such a Yangian structure associated to each vertex. It would be interesting to understand how the gluing picture fits in the Yangian picture, perhaps by extending the Yangian as an associative algebra by additional generators and relations corresponding to the gluing matter. It would be nice to explore the possible relation of our glued algebras with those of \cite{Tsymbaliuk}.

\paragraph{Spin chains and R-matrix}
The $Y_{N,0,0}$ algebras can be obtained from the Maulik-Okounkov $R$-matrix and are associated to a spin chain of length $N$, where at each site we have a free boson \cite{Maulik:2012rm,Zhu:2015nha,Fukuda:2017qki}. There exist also other $R$-matrices in free field representations associated to the other asymptotics. It would be interesting to see if by using a combination of these different $R$-matrix representations one can obtain more general $Y_{L,M,N}$ vertices and also if the gluing discussed in this article can be related to this spin chain picture.

\paragraph{Gluing and $4d$ $\mathcal{N}=2$ SCFTs}
Vertex operator algebras appear also in the context of $4d$ $\mathcal{N}=2$ SCFTs. For example the vertex operator algebra of Argyres-Douglas theories of type $(A_{M-1},A_{N-1})$ contains a double truncation of $\mathcal{W}_{\infty}$ corresponding to $(M,0,0)$ and $(0,N,0)$ truncation curves \cite{Cordova:2015nma}. This connects $\mathcal{W}_N$ minimal models and box counting to counting of BPS states in these theories. For class $S$ theories there is a gluing procedure which associated a vertex operator algebra to a given $4d$ theory \cite{Beem:2013sza,Beem:2014rza}. The elements of this gluing construction are formally very similar to the ones used here, namely the Drinfeld-Sokolov reduction modifying the punctures and the BRST coset construction used for gluing the basic building blocks. It would be very interesting to see how these two approaches to gluing are related.

\paragraph{Grassmannian coset}
There are well-known GKO cosets
\begin{equation}
\frac{SU(N)_k \times SU(N)_l}{SU(N)_{k+l}} \simeq \frac{SU(k+l)_N}{SU(k)_N \times SU(l)_N}
\end{equation}
related by the level-rank duality which specialize to $\mathcal{W}_N$ algebra for $l=1$. These cosets appear as a subalgebras of the algebra associated to the configuration of $l$ D5-branes ending on $(n,1)$ branes from the right and $N$ D3-branes ending from the left but they do not seem to be realizable directly by the gluing construction. The reason is that there is only one stress-energy tensor in the coset, while the algebras that we construct have one spin $2$ field for each vertex. It is possible that we could obtain a larger class of $\mathcal{W}$-algebras by using this Grassmannian coset as a one-parameter deformation of the basic vertex used in the gluing construction.

\section*{Acknowledgements} 
We are thankful to Davide Gaiotto for numerous discussions and suggestions throughout the work. We would like to thank Mina Aganagic, Christopher Beem, Lakshya Bhardwaj, Kevin Costello, Mykola Dedushenko, Tudor Dimofte, Ying-Hsuan Lin, Lesha Litvinov, Sylvain Ribault, Ivo Sachs, Piotr Su\l{}kowski and Oleksandr Tsymbaliuk for useful discussions. We thank Kris Thielemans for his Mathematica package OPEdefs. The research of TP was supported by the DFG Transregional Collaborative Research Centre TRR 33 and the DFG cluster of excellence Origin and Structure of the Universe and in part by Perimeter Institute for Theoretical Physics where this project was initiated. The research of MR was supported by the Perimeter Institute for Theoretical Physics. Research at Perimeter Institute is supported by the Government of Canada through the Department of Innovation, Science and Economic Development and by the Province of Ontario through the Ministry of Research, \& Innovation and Science.

\appendix

\section{Free fields}
\label{conventions}

We use the notation $U(N|M;\Psi)$ for the Kac-Moody algebra associated to the Lie superalgebra $U(N|M)$ that contains $SU(N|M)_{\Psi-N+M}$ and $U(1)_{(N-M)\Psi}$ as subalgebras. The OPE of the currents is given by
\begin{align}
J^A_B(z) J^C_D(0) &\sim \frac{(-1)^{p(B)p(C)}(\Psi -M+N)\delta^A_D \delta^C_B+ \delta^A_B \delta^C_D}{z^2}+ \cr &+\frac{ (-1)^{p(A)p(B) + p(C) p(D)+p(C)p(B)} \delta^A_D J^C_B(0)-(-1)^{p(B)p(C)}\delta^C_B J^A_D}{z}
\end{align}
where $p(a)=0$ for $a=1,\dots, M$ and $p(a)=1$ otherwise.

Note that for $U(N|0;\Psi)$ the algebra reduces to the Kac-Moody algebra $SU(N)_{\Psi-N}\times U(1)_{N\Psi}$ with OPEs given by
\begin{eqnarray}
 J^A_B(z)  J^C_D(w) &\sim&  \frac{(\Psi-N)\delta^A_D \delta^C_B+\delta_B^A\delta_D^C}{(z-w)^2}+ \frac{\delta_D^A J^C_B(w)-\delta^C_B J^A_D(w)}{z-w}
\end{eqnarray}
and similarly $U(0|M;\Psi)$ reduces to the algebra $SU(M)_{-\Psi+N}\times U(1)_{-N\Psi}$. In the expressions above, we choose a certain normalization of the diagonal $U(1)$ current in $U(N|0;\Psi)$ or $U(0|N;\Psi)$. This normalization is such that modules associated to the electric modules have integral charges.

We introduce the notation $\mathcal{S}^{N|M}$ for a system of $N$ pairs of symplectic bosons $(X_a,Y^a)$ where $a=1,\dots N$ and $M$ free fermions $(\chi_i,\psi^i)$ for $i=1,\dots, M$ with OPEs given by
\begin{eqnarray}
X_a(z)Y^b(w)\sim \frac{\delta_a^b}{z-w},\qquad  \chi_i(z) \psi^j(w)\sim \frac{\delta_i^j}{z-w}.
\end{eqnarray} 
Note that the algebra $\mathcal{S}^{N|M}$ contains $U(N|M;N-M-1)$ subalgebra generated by bilinears
\begin{eqnarray}
J=\begin{pmatrix}
X_a Y^b & X_a \psi^i\\
\chi_j Y^b & \chi_j \psi^i
\end{pmatrix}.
\end{eqnarray}
Similarly, exchanging the role of bosons and fermions, we introduce a notation $\overline{\mathcal{S}}^{N|M}$ for a system $M$ symplectic bosons $(X_i,Y^i)$ and $N$ free fermions  $(\chi_a,\psi^a)$ with OPEs given by
\begin{eqnarray}
X_i(z)Y^j(w)\sim \frac{\delta_i^j}{z-w},\qquad  \chi_a(z)\psi ^b(w)\sim \frac{\delta_a^b}{z-w}.
\end{eqnarray}
From their bilinears, we can construct $U(N|M;N-M+1)$ algebra generated by bilinears
\begin{eqnarray}
J=\begin{pmatrix}
\chi_a \psi^b & \chi_a Y^i \\
X_j \psi^b & X_j Y^i
\end{pmatrix}.
\end{eqnarray}
We also introduce a notation $\mathcal{F}$ for the algebra of a free fermion $\mathcal{S}^{0|1}$ and $\mathcal{B}$ for the algebra of a free symplectic boson $\mathcal{S}^{1|0}$.  

Occasionally, we also use notation $\mathcal{S}^{N|M}_i$ for the system $\mathcal{S}^{N|M}$ with a choice of the stress-energy tensor such that the fields have the shifted conformal dimension $(\frac{1}{2}+i,\frac{1}{2}-i)$. Corresponding stress-energy tensor have central charge
\begin{eqnarray}
(N-M)(12(1+i)^2-1).
\end{eqnarray}
We analogously define $\overline{\mathcal{S}}^{N|M}_i$, $\mathcal{B}_i$ and $\mathcal{S}_i$.

\section{Examples of BRST constraints}
\label{BRSTexamples}

\subsection{Y-algebras}

For $L=1, M=1, N=4$, one needs to impose the following constraints in the DS-reduction part of the BRST definition of Y-algebras
\begin{eqnarray}
\begin{blockarray}{ccccccl}
\begin{block}{c(ccc|c|c) l}
\BAmulticolumn{1}{c}{\multirow{5}{*}{}} &*&1&0&0&0&\BAmulticolumn{1}{c}{\multirow{3}{*}{$N-M$}}\\ 
&*&*&1&*&*&\\
&*&*&*&*&*&\\
\cline{2-6}
&*&*&0&&&\BAmulticolumn{1}{l}{\multirow{1}{*}{$M$}}\\
\cline{2-6}
\BAmulticolumn{1}{c}{\multirow{1}{*}{}}&*&*&0&& &\BAmulticolumn{1}{l}{\multirow{1}{*}{$L$}}\\
\end{block}
\end{blockarray}.
\end{eqnarray}
This constraint is implemented by the BRST charge given by
\begin{eqnarray}\nonumber
\oint dz \big [ (J_{12}-1)c_{12}+(J_{23}-1)c_{23}+J_{13}c_{13}+J_{14}c_{14}+J_{15}\gamma_{15}+J_{43}c_{43}+J_{53}\gamma_{53}\\
+b_{31}c_{12}c_{23}+\beta_{45}c_{44}\gamma_{45}-
\beta_{54}c_{44}\gamma_{54}-\beta_{45}c_{55}\gamma_{45}+\beta_{54}c_{55}\gamma_{45}+(c_{44}+c_{55})\gamma_{45}\big ].
\end{eqnarray}
In the second step of the construction of the corresponding Y-algebra one needs to sew the $U(1|1;\Psi-1)$ subalgebra with an extra $U(1|1;-\Psi+1)$ Kac-Moody algebra.

In the case of $N-M$ odd, one fixes the upper-triangular part of the first $(N-M)\times(N-M)$ block, $\frac{N-M-1}{2}$ rows in the upper right off-diagonal blocks of size $(N-M) \times M$ and $(N-M) \times L$ and $\frac{N-M-1}{2}$ columns in the lower left off-diagonal blocks of size $M \times (N-M)$ and $L \times (N-M)$.

For $N-M$ even, we can choose the Lagrangian subspace by constraining $\frac{N-M}{2}-1$ rows in the upper right off-diagonal blocks and $\frac{N-M}{2}$ columns in lower left off-diagonal blocks (or vice versa). In both cases, the expression for the BRST charge contains also cubic terms in ghosts that mix the components in the diagonal and the off-diagonal blocks.

\subsection{X-algebras}

For $N=5,M=2,K=7,L=3$ in the BRST definition of X-algebras, one needs to impose following constraints
\begin{eqnarray}
\begin{blockarray}{cccccccccccccl}
\begin{block}{c(ccc|cc|cccc|ccc) l}
\BAmulticolumn{1}{c}{\multirow{5}{*}{$N$}} &*&1&0&0&0&0&0&0&0&0&0&0&\BAmulticolumn{1}{c}{\multirow{3}{*}{$N-M$}}\\ 
&*&*&1&*&*&*&*&*&*&*&*&*&\\
&*&*&*&*&*&*&*&*&*&*&*&*&\\
\cline{2-13}
&*&*&0&\BAmulticolumn{2}{c|}{\multirow{2}{*}{$$}}&*&*&*&0&\BAmulticolumn{3}{l|}{\multirow{2}{*}{$$}}&\BAmulticolumn{1}{l}{\multirow{2}{*}{$M$}}\\
&*&*&0&&&*&*&*&0&&&&\\
\cline{2-13}
\BAmulticolumn{1}{c}{\multirow{7}{*}{$K$}}&*&*&0&0&0&*&1&0&0&0&0&0&\BAmulticolumn{1}{l}{\multirow{4}{*}{$K-L$}}\\
&*&*&0&0&0&*&*&1&0&0&0&0&\\
&*&*&0&*&*&*&*&*&1&*&*&*&\\
&*&*&0&*&*&*&*&*&*&*&*&*&\\
\cline{2-13}
&*&*&0&\BAmulticolumn{2}{c|}{\multirow{3}{*}{$$}}&*&*&*&0&\BAmulticolumn{3}{c}{\multirow{3}{*}{}}&\BAmulticolumn{1}{l}{\multirow{3}{*}{$L$}}\\
&*&*&0&&&*&*&*&0&&&&\\
&*&*&0&&&*&*&*&0&&&&\\
\end{block}
\end{blockarray}.
\end{eqnarray}
Note that DS-reduction in the $3\times 3$ block is performed first and then DS-reduction in the $4\times 4$ block of inside the $U(2|7)$ algebra.

\section{Characters}
\label{Characters}

\subsection{Building blocks}
This section contains explicit formul\ae\, for various terms appearing in the calculation of the characters using BRST construction of the algebras discussed in the text. The vacuum character of $\mathcal{W}_N$ algebra is given by 
\begin{eqnarray}
\chi_{\mathcal{W}_N}(q)=\prod_{m=0}^\infty \prod_{n=1}^N \frac{1}{1-q^{n+m}}.
\end{eqnarray}
The characters of the complex $\mathcal{S}_m^{M|L}$ of $M$ symplectic bosons and $L$ free fermions with the level shifted by $m$ is
\begin{eqnarray}
\chi^{M|L}_{m}(q,x_i,y_j)=\prod_{n=0}^\infty \prod_{i=1}^M\prod_{j=1}^L \frac{\left (1+y_j q^{n+\frac{1}{2}+m}\right )\left (1+y_j^{-1} q^{n+\frac{1}{2}+m}\right )}{\left (1-x_i q^{n+\frac{1}{2}+m}\right )\left (1-x_i^{-1} q^{n+\frac{1}{2}+m}\right )}.
\end{eqnarray}
The character of $\overline{\mathcal{S}}_m^{M|L}$ with $M$ fermionic and $L$ bosonic components has analogous character with $x_i \leftrightarrow y_i$ together with $M \leftrightarrow L$ interchanged.

The projection onto $U(M|L)$ invariant combinations is performed by integration with the Vandermonde measure
\begin{eqnarray}
dV_{M,L}=\frac{1}{M! L!}\prod_{i=1}^M\frac{dx_i}{x_i}\prod_{j=1}^L\frac{dy_j}{y_j}\frac{\prod_{i_1>i_2}\left (1-\frac{x_{i_1}}{x_{i_2}}\right )\prod_{j_1>j_2}\left (1-\frac{y_{j_1}}{y_{j_2}}\right )}{\prod_{i}\prod_{j}\left ( 1+\frac{x_i}{y_j} \right )\left ( 1+\frac{y_j}{x_i} \right )}.
\end{eqnarray}
In the generic $U(M|L)$ case, the denominator needs to be expanded and regularized. In all the examples of this paper, we restrict to the case of $M=0$ or $L=0$ when the denominator vanishes and we do not have to deal with these technicalities.

In later sections, we also need the character of free fermions and symplectic bosons with shifted dimension
\begin{eqnarray}\nonumber
\chi_{m}^{\mathcal{F}}=\prod_{n=0}^{\infty}\left (1+q^{n+\frac{1}{2}+m}\right ),\\
\chi_{m}^{\mathcal{B}}=\prod_{n=0}^{\infty}\frac{1}{1-q^{n+\frac{1}{2}+m}}.
\end{eqnarray}

\subsection{S-duality transformations of modules}

Triality transformation of modules of Y-algebras is given by following diagram:
\begin{figure}[h]
\centering
\includegraphics[width=0.7\textwidth]{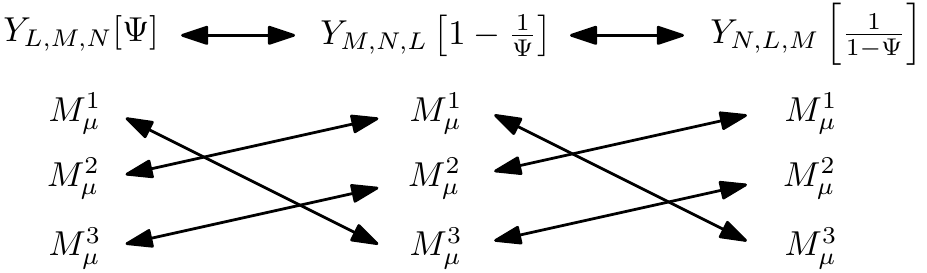}
\end{figure}

S-duality then acts as
\begin{eqnarray}
Y_{L,M,N}[\Psi]\ \leftrightarrow \ Y_{M,L,N}\left [\frac{1}{\Psi}\right ],\qquad M^1_\mu \ \leftrightarrow \ M^2_\mu.
\end{eqnarray}

\subsection{Examples with one asymptotics}
\label{ap:dimensions}

\paragraph{Example 1: $\mathcal{W}_N$}

The first example are simply characters of $\mathcal{W}_N$ algebras. The character of the $M^3_\mu$ representations can be identified (via S-duality) with $M^2_\mu$ characters of $Y_{0,0,N}\left [1-\frac{1}{\Psi}\right ]$ that are computed by specializing (\ref{charm2}). The result is given by
\begin{eqnarray}
\chi \left [Y_{N,0,0}[\Psi]\right ]\left (M^3_\mu\right )=q^{\frac{1}{2}\left (1-\frac{1}{1-\Psi}\right )\left (\sum_{i}\mu_i^2+\sum_i(N-2i+1)\mu_i\right )}\prod_{n=0}^{\infty}\prod_{m=1}^{N}\frac{s_\mu \left (x_i=q^{\frac{1}{2}(N-2i+1)}\right )}{1-q^{n+m}}
\end{eqnarray} 
where the denominator is the character of $U(1) \times \mathcal{W}_N$ and the numerator is sometimes called the quantum dimension of the representation $\mu$. The conformal dimension of the representation is given by formula
\begin{eqnarray}
h \left (M^3_{\mu}\right )=\frac{1}{2}\left (1-\frac{1}{1-\Psi}\right )\sum_{i}\mu_i^2+\frac{1}{2}\frac{1}{1-\Psi}\sum_i(2i-1-N)\mu_i.
\end{eqnarray}
In particular, specializing to $N=1$ one gets
\begin{eqnarray}
\chi \left [Y_{1,0,0}[\Psi]\right ] \left (M^3_\nu \right )=q^{\frac{1}{2}\left (1-\frac{1}{1-\Psi}\right )\nu^2}\prod_{n=0}^{\infty}\frac{1}{1-q^{n+1}}
\end{eqnarray}
and for $N=2$ we have
\begin{eqnarray}
\chi \left [Y_{2,0,0}[\Psi]\right ] \left (M^3_{\mu_1,\mu_2}\right )=q^{\frac{1}{2}\left (1-\frac{1}{1-\Psi}\right )(\mu_1^2+\mu_1+\mu_2^2-\mu_2)}q^{-\frac{\mu_1}{2}+\frac{\mu_2}{2}}\frac{1-q^{\mu_1-\mu_2+1}}{\prod_{n=0}^{\infty}(1-q^{n+1})}.
\end{eqnarray}

\paragraph{Example 2: $Y_{0,1,1}[\Psi]$}

For the $Y_{0,1,1}[\Psi]$ algebra, one gets from the reduction of the $U(1)_{\Psi-1}$ module the expression
\begin{eqnarray}
\chi \left [ Y_{0,1,1}[\Psi] \right ] \left (M^3_\nu\right )=q^{\frac{\nu^2}{2(\Psi-1)}}\prod_{n=0}^{\infty}\frac{1}{(1-q^{n+1})^2}\sum_{n=|m|}^{\infty}(-1)^{n-m}q^{\frac{n(n+1)}{2}}
\end{eqnarray}
The conformal dimension of the representation is given by
\begin{eqnarray}
h\left (M^3_\nu\right )=\frac{\nu^2}{2(\Psi-1)}+\frac{\nu^2}{2}+\frac{|\nu|}{2}.
\end{eqnarray}

\paragraph{Example 3: $Y_{1,0,2}[\Psi]$}

The character of the representation $M^3_\nu$ is given by
\begin{eqnarray}
\chi \left [Y_{1,0,2}[\Psi]\right ] \left (M^3_\nu\right )&=&q^{\frac{1}{2}\left (1-\frac{1}{(1-\Psi)}\right )n^2}\prod_{n=0}^{\infty}\frac{1}{(1-q^{n+1})^3} \times \\ \nonumber
&& \times \left (\sum_{n=m}^{\infty}(-1)^{n+m}q^{\frac{n(n+1)-m(m-1)}{2}}+\sum_{n=m+1}^{\infty}(-1)^{n+m}q^{\frac{n(n+1)-m(m+1))}{2}}\right )
\end{eqnarray}
Note that the conformal dimensions of such representations are
\begin{eqnarray}
h(M^3_\nu)=\frac{\nu^2}{2(\Psi-1)}+\frac{\nu^2}{2}+|\nu|.
\end{eqnarray}

\paragraph{Example 4: $Y_{2,0,1}[\Psi]$}

The character of the representation $M^3_{\mu_1,\mu_2}$ is given by
\begin{eqnarray}
\chi \left [Y_{2,0,1}[\Psi]\right ] \left (M^3_{\mu_1,\mu_2}\right )&=&q^{\frac{\mu_1^2+\mu_1}{2(\Psi-1)}+\frac{\mu_2^2-\mu_2}{2(\Psi-1)}}\prod_{n=0}^{\infty}\frac{(-1)^{\mu_1+\mu_2} }{(1-q^{n+1})^3}\cdot\\ \nonumber
&&\cdot \left (\sum_{n=|\mu_1|}^{\infty}\sum_{m=|\mu_2|}^{\infty}-\sum_{n=|\mu_1+1|}^{\infty}\sum_{m=|\mu_2-1|}^{\infty}\right )(-1)^{n+m}q^{\frac{n(n+1)+m(m+1)}{2}}.
\end{eqnarray}
The conformal dimensions is
\begin{eqnarray}
h(M^3_{\mu_1,\mu_2})=\frac{\mu_1^2+\mu_1}{2(\Psi-1)}+\frac{\mu_2^2-\mu_2}{2(\Psi-1)}+\frac{\mu_1^2}{2}+\frac{\mu_2^2}{2}+\frac{|\mu_1|+|\mu_2|}{2}.
\end{eqnarray}

\paragraph{Example 5: $Y_{0,1,3}[\Psi]$}

By a small modification of the calculation from \cite{Gaiotto:2017euk}, one gets for the $M^3_\nu$ modules of the $Y_{0,1,3}[\Psi]$ algebra the expression
\begin{eqnarray}
&&q^{\frac{\nu^2}{2}\frac{1}{1-\Psi}}\prod_{n=0}^\infty \frac{1}{(1-q^{n+1})(1-q^{n+2})}\oint \frac{dz}{z}\frac{z^\nu}{\prod_{n=0}^\infty \left (1-zq^{n+\frac{3}{2}}\right )\left (1-z^{-1}q^{n+\frac{3}{2}}\right )}\\ \nonumber
&=&q^{\frac{\nu^2}{2}\frac{1}{1-\Psi}}\prod_{n=0}^\infty \frac{1}{(1-q^{n+1})(1-q^{n+2})}\oint \frac{dz}{z}\frac{z^\nu \left (1-zq^{n+\frac{1}{2}}\right )\left (1-z^{-1}q^{n+\frac{1}{2}}\right )}{\prod_{n=0}^\infty \left (1-zq^{n+\frac{1}{2}}\right )\left (1-z^{-1}q^{n+\frac{1}{2}}\right )}\\ \nonumber
&=&(-1)^{\nu}q^{\frac{\nu^2}{2}\frac{1}{1-\Psi}}\prod_{n=0}^\infty \frac{1}{(1-q^{n+1})^3(1-q^{n+2})}\bigg [(1+q)\sum_{n=|\nu|}^\infty (-1)^{n}q^{\frac{n(n+1)-\nu^2}{2}}\\
&&+q^{\frac{1}{2}}\sum_{n=|\nu+1|}^\infty (-1)^{n}q^{\frac{n(n+1)-(\nu+1)^2}{2}}+q^{\frac{1}{2}}\sum_{n=|\nu-1|}^\infty (-1)^{n}q^{\frac{n(n+1)-(\nu-1)^2}{2}}\bigg ]
\end{eqnarray}
where we used the formula from the appendix \ref{form} to expand the product in the denominator and then performed the contour integral. Note that a similar calculation holds for $M^3_\nu$ modules of any $Y_{0,1,N}[\Psi]$. The conformal dimension of these modules is given by
\begin{eqnarray}
h \left ( M^3_\nu \right )=\frac{\nu^2}{2}\frac{1}{1-\Psi}+\frac{3}{2}|\nu|.
\end{eqnarray}

\paragraph{Example 6: $Y_{0,2,3}[\Psi]$}

Our last example with one asymptotic is the character of $M^3_{\mu_1,\mu_2}$ representation of the $Y_{0,2,3}[\Psi]$ algebra
\begin{eqnarray}\nonumber&&q^{\frac{1}{2}\frac{\mu_1^2+\mu_1+\mu_2^2-\mu_2}{1-\Psi}}\frac{1}{2}\prod_{n=0}^\infty \frac{1}{1-q^{n+1}}\oint \frac{dx_1}{x_1} \frac{dx_2}{x_2}\frac{(x_2-x_1)(x_1^{\mu_1}x_2^{\mu_2-1}-x_1^{\mu_2-1}x_2^{\mu_1})}{\prod_{n=0}^\infty \prod_{i=1,2} \left (1-x_iq^{n+1}\right )\left (1-x_i^{-1}q^{n+1}\right )}\\ \nonumber
&=&\frac{q^{\frac{1}{2}\frac{\mu_1^2+\mu_1+\mu_2^2-\mu_2}{1-\Psi}}}{2}\prod_{n=0}^\infty \frac{1}{(1-q^{n+1})^5}\sum_{n_1,n_2}^{\infty}(-1)^{n_1+n_2}(1-q^{n_1+1})(1-q^{n_2+1})q^{\frac{n_1(n_1+1)+n_1(n_1+1)}{2}}\\ \nonumber
&&\big [q^{(n_1+1)\mu_2+(n_2+1)\mu_1}-q^{(n_1+1)(\mu_2-1)+(n_2+1)(\mu_1+1)}\\
&&-q^{(n_1+1)(\mu_1+1)+(n_2+1)(\mu_2-1)}+q^{(n_1+1)\mu_1+(n_2+1)\mu_2}\big ].
\end{eqnarray}
The conformal dimension of this module are given by
\begin{eqnarray}
h \left ( M^3_{\mu_1,\mu_2} \right )=\frac{1}{2}\frac{\mu_1^2+\mu_1+\mu_2^2-\mu_2}{1-\Psi}-\frac{1}{2}\mu_1+\frac{1}{2}\mu_2+\frac{3}{2}(|\mu_1|+|\mu_2|).
\end{eqnarray}

\subsection{Example with two asymptotics}
In this section, we derive the character of modules with two asymptotics for the example of $Y_{0,1,2}[\Psi]$ associated to the two Wilson lines supported at the NS5- and $(1,1)$ interfaces. Such modules are labeled by an integer $\nu$ and a pair of integers $(\mu_1,\mu_2)$ satisfying $\mu_1\geq \mu_2$. These two label representations of $U(1)$ and $U(2)$ respectively. Inserting the two Schur polynomials, one gets
\begin{eqnarray}\nonumber
&=&q^{\frac{\mu_1^2+\mu_2^2+\mu_1-\mu_2}{2\Psi}+\frac{\nu^2}{2}\left (1-\frac{1}{1-\Psi}\right )}\prod_{n=0}^\infty \frac{1}{1-q^{n+1}}\oint \frac{dz}{z}\frac{z^{-\nu}(z^{\mu_2}+z^{\mu_1+1})}{(1-z)(1-zq^{n+1})(1-z^{-1}q^{n+1})}\\ \nonumber
&=&q^{\frac{\mu_1^2+\mu_2^2+\mu_1-\mu_2}{2\Psi}+\frac{\nu^2}{2}\left (1-\frac{1}{1-\Psi}\right )}\oint \frac{dz}{z}\frac{z^{\mu_2-\nu}-z^{\mu_1+1-\nu}}{(1-q^{n+1})^3}\sum_{n=0}^{\infty}\sum_{m=-n}^{n}z^m(-1)^{n-m}q^{\frac{n(n+1)-m(m+1)}{2}}\\ \nonumber
&=&q^{\frac{\mu_1^2+\mu_2^2+\mu_1-\mu_2}{2\Psi}+\frac{\nu^2}{2}\left (1-\frac{1}{1-\Psi}\right )}\prod_{n=0}^\infty \frac{1}{(1-q^{n+1})^3}\cdot\\
&&\cdot\left (\sum_{n=|\mu_1-\nu|}^\infty (-1)^{n+\mu_1-\nu}q^{\frac{n(n+1)-\mu_1(\mu_1+1)}{2}}+\sum_{n=|\mu_2-\nu|}^\infty (-1)^{n+\mu_2-\nu}q^{\frac{n(n+1)-\mu_2(\mu_2-1)}{2}} \right )
\end{eqnarray}

\section{Representations of $\mathcal{W}_{\infty}$}

In this section we will demonstrate few properties of $\mathcal{W}_{1+\infty}$ using the Yangian generators of the algebra \cite{Schiffmann:2012gf,tsymbaliuk2017affine,Fukuda:2015ura,arbesfeld2013presentation}. For many more details and examples see \cite{Prochazka:2015aa}. The main piece of information that we need is that there is a change of variables of $\mathcal{W}_{1+\infty}$ which maps the algebra to an associative algebra with generators $\psi_j, e_j $ and $f_j$ where $j=0,1,2,\ldots$. $\psi_j$ generators all commute and can be explicitly diagonalized in representations while $e_j$ and $f_j$ generators act as raising and lowering operators. The relations between these generators are generated by certain quadratic and cubic relations spelled out explicitly in \cite{tsymbaliuk2017affine,Prochazka:2015aa}. While the local properties of $\mathcal{W}_{1+\infty}$ as a vertex operator algebra are obscure in the Yangian language, the representation theory considerably simplifies and the underlying integrable structure is also visible in this picture.

\subsection{Representations from Yangian point of view}
The advantage of the Yangian basis of $\mathcal{W}_{1+\infty}$ is that many representations have a simple combinatorial interpretation. The representation space is spanned by a certain generalization of Young diagrams (typically plane partitions) with possible additional geometrical conditions (like a truncation to a fixed number of layers in one of the directions, more general pit condition or non-trivial asymptotics as in the case of the topological vertex). In the following, we will denote these basis vectors by $|\Lambda\rangle$. The Cartan generators $\psi_j$ are diagonal in this basis with explicitly known eigenvalues,
\begin{equation}
\left(1 + h_1 h_2 h_3 \sum_{j=0}^{\infty} \psi_j u^{-j-1} \right) |\Lambda\rangle = \frac{u-q+\psi_0 h_1 h_2 h_3}{u-q} \prod_{\Box \in \Lambda} \varphi(u-q-h_\Box) |\Lambda\rangle \equiv \psi_\Lambda(u) |\Lambda\rangle
\end{equation}
Here $h_\Box$ are the weighted Cartesian coordinates of a box in the generalized Young diagram,
\begin{equation}
h_\Box = h_1 x_1(\Box) + h_2 x_2(\Box) + h_3 x_3(\Box).
\end{equation}
The structure function of the algebra $\varphi(u)$ is a rational function
\begin{equation}
\varphi(u) = \frac{(u+h_1)(u+h_2)(u+h_3)}{(u-h_1)(u-h_2)(u-h_3)}.
\end{equation}
The parameters $h_j$ parametrize the algebra and are related to $\lambda_j$ parameters of $\mathcal{W}_{\infty}$ via
\begin{eqnarray}
h_1=-\psi_0 \lambda_2 \lambda_3,\quad h_1=-\psi_0\lambda_2 \lambda_3,\quad h_1=-\psi_0\lambda_2 \lambda_3.
\end{eqnarray}
The satisfy $h_1+h_2+h_3=0$. In the representations that we consider the generator $\psi_0$ takes a non-zero constant value and can be eliminated by a rescaling symmetry of the algebra.

The action of the raising and the lowering operators $e_j$ and $f_j$ can be schematically written as
\begin{eqnarray}\nonumber
e_j |\Lambda\rangle = \sum_{\Box \in \Lambda^+} (q+h_\Box)^j E(\Lambda \rightarrow \Lambda+\Box) |\Lambda+\Box \rangle\\
f_j |\Lambda\rangle = \sum_{\Box \in \Lambda^-} (q+h_\Box)^j F(\Lambda+\Box \rightarrow \Lambda) |\Lambda+\Box \rangle.
\end{eqnarray}
The sum runs over all the boxes that can be added to or removed from the partition $\Lambda$. $E(\Lambda \rightarrow \Lambda+\Box)$ and $F(\Lambda+\Box \rightarrow \Lambda)$ are amplitudes for adding and removing a box at given position and are not arbitrary but are partially fixed by the commutation relations of the algebra \cite{Prochazka:2015aa} (there is a certain degree of arbitrariness coming from rescaling of the vectors in the representation space). In particular, the boxes can be only added or removed at positions where the $\psi_\Lambda(u)$ eigenvalue has a simple pole when considered as a meromorphic function in $u$ plane.

\subsection{Conformal dimension of minimal representations}
\label{appmindim}
We are now in position to determine the dimension of minimal representations from the combinatorics of the plane partitions. The minimal representation has Young diagram asymptotics $(\Box,\cdot,\cdot)$ along the three coordinate axes, i.e. we need to build a tower of an infinite number of boxes in the first direction. The minimal configuration of boxes with fixed asymptotics corresponds to a highest weight state and the descendant states are obtained by adding an additional finite number of boxes in a way compatible with plane partition restrictions. The $\psi_j$ charges of the highest weight state are captured by the product
\begin{equation}
\psi_{\Box,\cdot,\cdot}(u) = \frac{u+\psi_0 h_1 h_2 h_3}{u} \prod_{j=1}^{\infty} \varphi(u-j h_1 - h_2 - h_3).
\end{equation}
We put $q=0$ because we are building the configuration from the uncharged vacuum with $J_0 |0\rangle = \psi_1 |0\rangle = 0$. Now using the constraint $h_1 + h_2 + h_3 = 0$ we see that most of the factors in the product cancel and the limit at finite $u$ is just
\begin{equation}
\psi_{\Box,\cdot,\cdot}(u) = \frac{(u+\psi_0 h_1 h_2 h_3)(u+h_1)}{(u-h_2)(u-h_3)}.
\end{equation}
The conformal dimension of this state corresponds to one half of $\psi_2$ eigenvalue which in this case is
\begin{equation}
\frac{1-h_2 h_3 \psi_0}{2} \equiv \frac{1+\lambda_1}{2}
\end{equation}
exactly as in (\ref{boxdim}). Apart from knowing the conformal dimension of $(\Box,\cdot,\cdot)$ configuration we can similarly use the generating function to extract also all the higher spin charges $\psi_j$ of any state in the representation.

If we were interested in the conformal dimension with respect to $\mathcal{W}_{\infty}$, i.e. not including the contribution from the $U(1)$ charge of the representation, we could use the the same generating function of charges with replacement $u \to u-q$ where $q$ is a parameter which lets us choose an arbitrary $U(1)$ charge of the vacuum state. In particular, for
\begin{equation}
q = \frac{1}{\psi_0 h_1}
\end{equation}
we obtain a generating function of charges for $(\Box,\cdot,\cdot)$ representation with vanishing $U(1)$ charge. Its conformal dimension can be extracted from the third order pole at infinity and we find
\begin{equation}
h_{\infty}(\Box,\cdot,\cdot) = \frac{1}{2} \lambda_1 \left(1-\frac{1}{\lambda_2}\right)\left(1-\frac{1}{\lambda_3}\right).
\end{equation}

\begin{figure}
\centering
\includegraphics[width=0.36\textwidth]{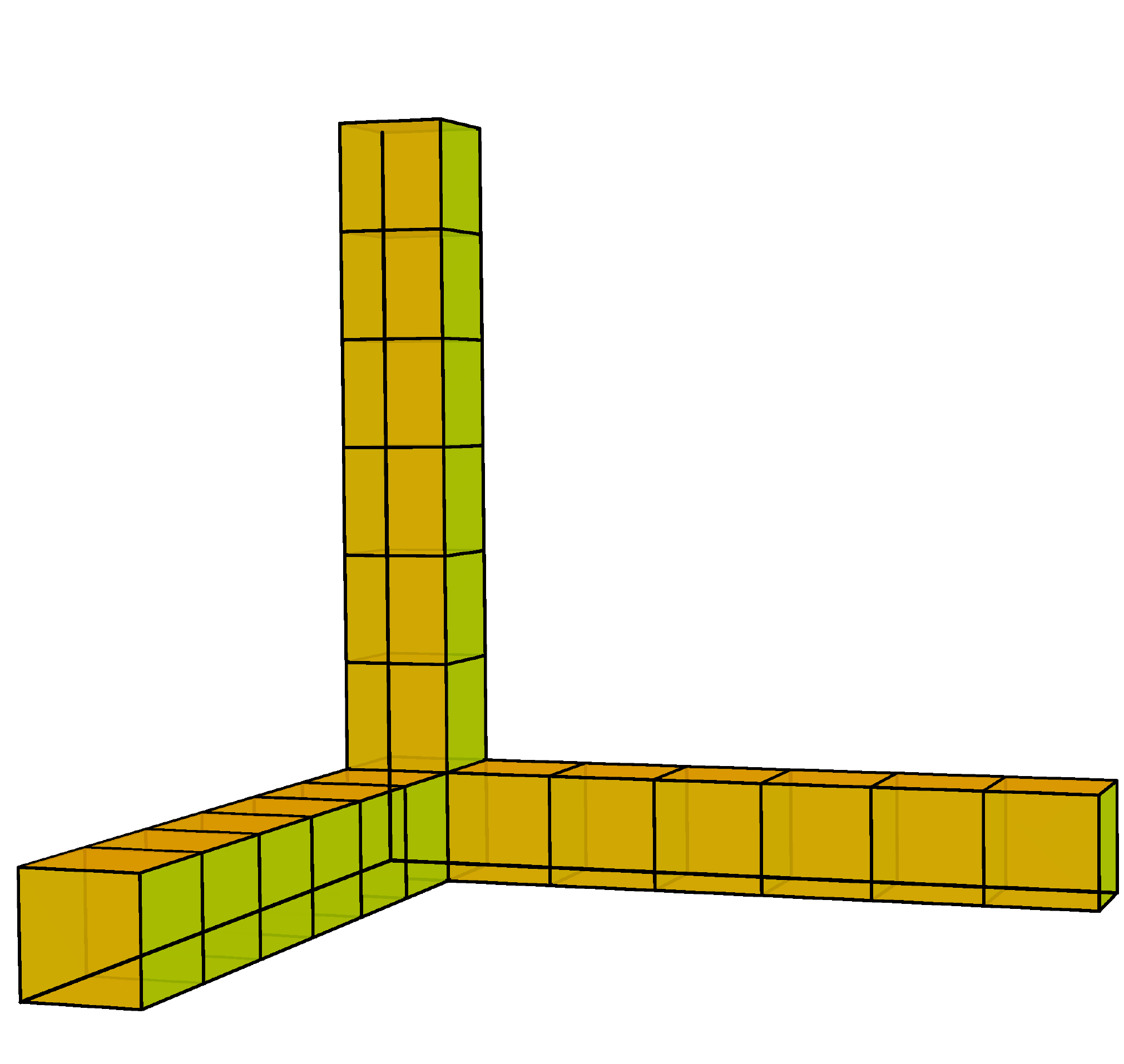}
\caption{An example of the configuration of boxes representing a highest weight state with three non-trivial asymptotics, $(\Box,\Box,\Box)$. There is one box shared by all three piles of boxes which contributes a $-2$ shift of the conformal dimension of this representation. This figure is taken from \cite{Prochazka:2015aa}.}
\label{fig3as}
\end{figure}

\paragraph{More asymptotics}
It is useful to understand how to obtain the conformal dimension of the representations with two non-trivial asymptotics, for example the representation $(\Box,\Box,\cdot)$. This is relatively easy once we know the conformal dimension of representations $(\Box,\cdot,\cdot)$ and $(\cdot,\Box,\cdot)$. The conformal dimension is the eigenvalue of $L_0$ operator which combinatorially counts the number of boxes. In the case of plane partitions with non-trivial asymptotics, there is an infinite number of boxes needed to build the asymptotics, so we need to regularize the infinite sum $1+1+1+\ldots$ 	in certain way. Fortunately, the generating function of charges $\psi_{\Lambda}(u)$ provides a natural regularization: $\psi_{\Lambda}(u)$ as a rational function has a well-defined limit when the number of boxes piled in certain direction goes to infinity, creating a non-trivial asymptotic. From this limiting function we can easily extract the $L_0$ eigenvalue, i.e. the conformal dimension and the result exactly reproduces the $U(N)$ quadratic Casimir known from the representation theory of $\mathcal{W}_N$ algebras.

In the case of two non-trivial asymptotics (see figure \ref{fig3as}), we can relate the conformal dimension of primary $(\Box,\Box,\cdot)$ (which counts the regularized number of boxes with two piles of boxes, one in the first direction and one in the second direction) to the conformal dimension of primaries $(\Box,\cdot,\cdot)$ and $(\cdot,\Box,\cdot)$ (which measure the regularized pile of boxes in first and second direction separately). We see that the difference is only one box at the origin which we are overcounting if we simply add the conformal dimensions of the individual piles, so we have a formula
\begin{equation}
\label{boxboxfusion}
h(\Box,\Box,\cdot) = h(\Box,\cdot,\cdot) + h(\cdot,\Box,\cdot) - 1.
\end{equation}

Analogous discussion applies for any two asymptotic Young diagrams, but we must correctly count the number of boxes in the overlap. For three non-trivial asymptotics, we have simply
\begin{equation}
\label{box3fusion}
h(\Box,\Box,\Box) = h(\Box,\cdot,\cdot) + h(\cdot,\Box,\cdot) + h(\cdot,\cdot,\Box) - 2
\end{equation}
because by combining the three piles in three different directions we counted the box at coordinates $(1,1,1)$ three times.

\subsection{Truncation curves}
\label{truncations}

To derive the truncation curves, we proceed analogously as in the previous section by determining $\psi(u)$ for certain state. The state in question is now represented by a cube of dimensions $(L+1,M+1,N+1)$ and it is the first state that does not lie below the corner shifted by $(L,M,N)$. We want to see for which values of parameters of the algebra is this state singular (a null vector).

First of all, the generating function of charges for a configuration of $L \times M \times N$ boxes is
\begin{eqnarray}
\psi(u) & = & \frac{u+\psi_0 h_1 h_2 h_3}{u} \prod_{l=1}^L \prod_{m=1}^M \prod_{n=1}^N \varphi(u - l h_1 - m h_2 - n h_3) \\
& = & \frac{(u+\psi_0 h_1 h_2 h_3)(u-L h_1-M h_2) (u-L h_1-N h_3) (u-M h_2-N h_3)}{(u-L h_1)(u-M h_2)(u-N h_3)(u-L h_1-M h_2-N h_3)}
\end{eqnarray}
The simple poles in this function are positions where boxes can be added or removed \cite{Fukuda:2015ura,Prochazka:2015aa}. In particular there is a simple pole at
\begin{equation}
u = L h_1 + M h_2 + N h_3
\end{equation}
which means that the box at coordinates $(L,M,N)$ can be generically removed. But for special values of parameters $h_1, h_2$ and $h_3$ this simple pole can be canceled by a simple zero at $u = -\psi_0 h_1 h_2 h_3$ and this is the equation for the truncation curve:
\begin{equation}
L h_1 + M h_2 + N h_3 = -\psi_0 h_1 h_2 h_3.
\end{equation}
Note that for $(L,M,N)$ truncation we should consider the configuration of boxes with $(L+1) \times (M+1) \times (N+1)$ boxes, but because of the condition $h_1 + h_2 + h_3 = 0$ these give us the same truncation curve.

An alternative way to arrive at this result to is to study vanishing of $F$ coefficient. Its vanishing means that the amplitude for removal of the box at coordinates $(L+1,M+1,N+1)$ vanishes which is exactly the condition for this vector to be the highest weight vector of the submodule it generates. We find
\begin{eqnarray}
F(\Lambda+\Box\rightarrow\Lambda)\propto \sqrt{Lh_1+Mh_2+Nh_3 + \psi_0 h_1h_2h_3}
\end{eqnarray}
and the equation (\ref{truncations2}) follows from this immediately.


\section{Calculation of the central charges}

\subsection{Resolved conifold $1|1$ algebra}
\label{cc11}

Let us calculate the central charge of a generic algebra for $K>L$, $N>M$. The central charge of the Kac-Moody algebra that we start with is given by
\begin{eqnarray}
c_0=1+\frac{\Psi-N+K}{\Psi}((N-K)^2-1).
\end{eqnarray}
Contributions from the modification term of the stress-energy tensor gives
\begin{eqnarray}\nonumber
c_{DS}^{(2)}&=&-(\Psi-N+K)(N-M)((N-M)^2-1),\\
c_{DS}^{(1)}&=&(\Psi-M+K-1)(K-L)((K-L)^2-1).
\end{eqnarray}
Ghost contribution to fix fields in the two diagonal blocks is
\begin{eqnarray}\nonumber
c_{1}&=&-(K-L)(K-L-1)((K-L)(K-L-1)-1),\\
c_2&=&-(N-M)(N-M-1)((N-M)(N-M-1)-1).
\end{eqnarray}
In the four off diagonal blocks where fields are graded only with respect to one of the two DS blocks, one gets a contribution
\begin{eqnarray}\nonumber
c_3&=&(K-M)(N-M-1)((N-M-1)^2-2),\\
c_4&=&(M-L)(K-L-1)((K-L-1)^2-2).
\end{eqnarray}
Finally, the coset contributes
\begin{eqnarray}
c_{coset}=-1-\frac{\Psi-M+L}{\Psi}((M-L)^2-1).
\end{eqnarray}
Total central charge is then
\begin{eqnarray}
&&c_0+c_{DS}^{(1)}+c_{DS}^{(2)}+c_1+c_2+c_3+c_4-c_{coset}\\ \nonumber
&=&+\frac{1}{\Psi}\left ((K-N)((K-M)^2-1)-(L-M)((L-M)^2-1) \right )\\ \nonumber
&&+\Psi\left ( (K-L)((K-L)^2-1)-(N-M)((N-M)^2-1) \right )+(K-L+M-N)\cdot\\ \nonumber
&&\cdot (K^2+K L-4 K M+K N-2 L^2+L M+2 L N+M^2+M N-2 N^2+1)
\end{eqnarray}

\subsection{$0|2$ algebra}
\label{cc02}
Starting with the central charge of $U(N|K)_\Psi$ Kac-Moody algebra
\begin{eqnarray}
c_{U(N|K)_\Psi}=1+\frac{\Psi-N+K}{\Psi}((N-K)^2-1),
\end{eqnarray}
we have to modify it by modifying terms contributing to the final central charge by
\begin{eqnarray}\nonumber
c^{(1)}_{DS}&=&-(\Psi-N+K)(N-M)((N-M)^2-1),\\
c^{(2)}_{DS}&=&-(\Psi-M+K-1)(M-L)((M-L)^2-1).
\end{eqnarray}
Ghosts needed to perform a fixing in the two diagonal blocks contribute
\begin{eqnarray}\nonumber
c_1&=&-(N-M)(N-M-1)((N-M)(N-M-1)-1),\\
c_2&=&-(M-L)(M-L-1)((M-L)(M-L-1)-1).
\end{eqnarray}
We have also contributions from the ghosts in the off-diagonal blocks
\begin{eqnarray}\nonumber
c_3&=& (K-M)(N-M-1)((N-M-1)^2-2), \\
c_4&=& (K-L)(M-L-1)((M-L-1)^2-2). 
\end{eqnarray}
Adding the coset term
\begin{eqnarray}
c_{coset}=-1-\frac{\Psi-2-L+K}{\Psi-2}\left ((L-K)^2-1\right )
\end{eqnarray}
and putting everything together one gets the central charge of the final algebra
\begin{eqnarray}
c_{K,L,M,N}[\Psi]&=&c_{U(N|K)_\Psi}+c^{(1)}_{DS}+c_{DS}^{(2)}+c_1+c_2+c_3+c_4+c_{coset}\\ \nonumber
&=&\frac{(L-K)((L-K)^2-1)}{\Psi-2}-\frac{(K-N)((K-N)^2-1)}{\Psi}\\ \nonumber
&&+(((N-M)^2-1)(N-M)+((M-L)^2-1)(M-L))\Psi-M-N\\ \nonumber
&&+(L-M)^2 (-3 K+L+2 M)+(M-N)^2 (-3 K+M+2 N)+2 K\\ \nonumber
&&-\left((L-M)^2-1\right)+(L-M)-M-N(L-M)^2 (-3 K+L+2 M)\\ \nonumber
&&+(M-N)^2 (-3 K+M+2 N)+2 K-\left((L-M)^2-1\right) (L-M).
\end{eqnarray}

\section{Some useful summation formul\ae}
\label{form}

In the examples, we use known summation formul\ae
\begin{eqnarray}
\prod_{n=0}^\infty \left (1+zq^{n+\frac{1}{2}}\right )\left (1+z^{-1}q^{n+\frac{1}{2}}\right )&=&\prod_{n=0}^\infty\frac{1}{1-q^{n+1}}\sum_{n=-\infty}^{\infty}z^nq^{\frac{n^2}{2}}\\ \nonumber
\prod_{n=0}^\infty \left (1+zq^{n+1}\right )\left (1+z^{-1}q^{n+1}\right )&=&\prod_{n=0}^\infty\frac{1}{1-q^{n+1}}\sum_{n=0}^{\infty}\sum_{m=-n}^n (-1)^{m+n}z^n q^{\frac{n(n+1)}{2}}\\ \nonumber
\prod_{n=0}^\infty \frac{1}{(1-zq^{n+\frac{1}{2}}) (1-z^{-1}q^{n+\frac{1}{2}})}&=&\prod_{n=0}^\infty\frac{1}{(1-q^{n+1})^2}\sum_{n=0}^{\infty}\sum_{m=-n}^n (-1)^{n+m} z^mq^{\frac{n(n+1)-m^2}{2}}\\ \nonumber 
\prod_{n=0}^\infty \frac{1}{(1-zq^{n+1}) (1-z^{-1}q^{n+1})}&=&\prod_{n=0}^\infty \frac{1}{(1-q^{n+1})^2}\sum_{n=0}^{\infty}(-1)^n(1-q^{n+1})q^{\frac{n(n+1)}{2}}
\end{eqnarray}
together with special cases when we set $z=1$. We also use integration formul\ae\, derived in \cite{Gaiotto:2017euk} of the form
\begin{eqnarray} 
\oint \prod_{i=1}^N \frac{dx_i}{x_i} \frac{x_i^{s_i}}{\prod_{i,n} (1-q^{n+\frac12} x_i)(1-q^{n+\frac12} x^{-1}_i)}&=&\sum_{n_i=0}^\infty \prod_i (-1)^{n_i} q^{\frac{n_i(n_i+1)}{2}+\left (n_i + \frac12\right )s_i}, \\ \nonumber
\oint \prod_{i=1}^N \frac{dx_i}{x_i} \frac{x_i^{s_i} }{\prod_{i,n} (1-q^{n+1} x_i)(1-q^{n+1} x^{-1}_i)}&=&\frac{\sum_{n_i=0}^\infty \prod_i (-1)^{n_i} (1-q^{n_i + 1}) q^{ \frac{n_i(n_i+1)}{2}+(n_i + 1)s_i} }{ \prod_{n=0}^\infty (1-q^n)^{2N}}.
\end{eqnarray}

\section{Romans $\mathcal{N}=2$ $\mathcal{W}_3$ algebra}
\label{secn2w3}

Another example of the gluing construction is the $\mathcal{N}=2$ supersymmetric version of $\mathcal{W}_3$ algebra constructed by Romans \cite{Romans:1991wi}. This algebra is obtained by extending the $\mathcal{N}=2$ stress-energy tensor supermultiplet $(J,G^{\pm},T)$ by another multiplet $(W_2^0,W_2^{\pm},W_2^1)$ with spins $\left(2,\frac{3}{2},\frac{3}{2},3\right)$. The operator product expansions of spin $2$ supermultiplet with itself contains a priori two structure constants not fixed by $\mathcal{N}=2$ superconformal algebra,
\begin{equation}
W_2 W_2 \sim C_{22}^0 \mathbbm{1} + C_{22}^2 W_2
\end{equation}
but one of them is fixed by the Jacobi identities to
\begin{equation}
C_{22}^0 = -\frac{(c-15)(c-1)c(c+6)(2c-3)}{4(c+3)^2(5c-12)^2} (C_{22}^2)^2
\end{equation}
and the remaining structure constant $C_{22}^2$ can be chosen at will by rescaling the fields of spin $2$ supermultiplet (this is analogous to what happens in construction of Zamolodchikov $\mathcal{W}_3$ algebra). This means that $\mathcal{N}=2$ $\mathcal{W}_3$ algebra has one free continuous parameter with is the central charge.

We can construct two stress-energy tensors commuting with the $U(1)$ current $J$ and with each other
\begin{eqnarray}
T^{(1)} & = & \frac{2(c+6)(2c-3)}{3(c-1)(c+12)} T - \frac{2(c+3)(5c-12)}{3(c-1)(c+12) C_{22}^2} W_2^0 - \frac{(c+6)(2c-3)}{(c-1)c(c+12)} (JJ) \\
T^{(2)} & = & \frac{(c-15)c}{3(c-1)(c+12)} T + \frac{2(c+3)(5c-12)}{3(c-1)(c+12) C_{22}^2} W_2^0 + \frac{(c-15)}{c(c-1)(c+12)} (JJ)
\end{eqnarray}
and such that
\begin{equation}
T = T^{(1)} + T^{(2)} + \frac{3}{2c} (JJ).
\end{equation}
The central charges of these stress-energy tensors are
\begin{eqnarray}
c_{\infty}^{(1)} & = & \frac{2(c+6)(2c-3)}{3(c+12)} \\
c_{\infty}^{(2)} & = & -\frac{(c-15)c}{3(c+12)}.
\end{eqnarray}
There are no primary fields commuting with $T^{(1)}$ and $J$ which indicates that $\mathcal{W}_\infty^{(2)}$ is truncated at the level of  the Virasoro subalgebra and so it has $\lambda$-parameters
\begin{equation}
\lambda^{(2)} = \left(2, \frac{c-6}{9}, -\frac{2(c-6)}{c+12} \right).
\end{equation}
On the other hand, we can construct the unique (up to rescaling) spin $3$ and $4$ primaries with respect to $T^{(1)}$ and commuting with $T^{(2)}$ and $J$,
\begin{eqnarray}
W_3^{(1)} & = & W_2^1 - \frac{3C_{22}^2}{2(c+3)} (G^+ G^-) + \ldots \\
W_4^{(1)} & = & (G^- W_2^+) - (G^+ W_2^-) + \frac{12}{c} (J W_2^1) + \ldots
\end{eqnarray}
such that the $\mathcal{W}_\infty^{(1)}$ structure constants are
\begin{eqnarray}
C_{33}^0 & = & -\frac{(c-24)(c-1)^2(c+6)(2c-3)(C_{22}^2)^2}{4c(c+3)^2(5c-12)} \\
C_{44}^0 & = & -\frac{(c-24)(c-3)(c-1)^2(c+6)(c+18)(2c-3)(C_{22}^2)^2}{c^2(5c-12)(5c^2+39c+153)} \\
C_{33}^4 & = & \frac{54(c-1) C_{22}^2}{(c+3)(5c-12)}
\end{eqnarray}
From this we find the $\lambda$-parameters of $\mathcal{W}_{\infty}^{(1)}$ to be
\begin{equation}
\lambda^{(1)} = \left(-\frac{c}{6}, -\frac{3c}{c-6}, \frac{3c}{c+12} \right).
\end{equation}

\begin{wrapfigure}{l}{0.18\textwidth}
\vspace{-15pt}
\begin{center}
\includegraphics[width=0.17\textwidth]{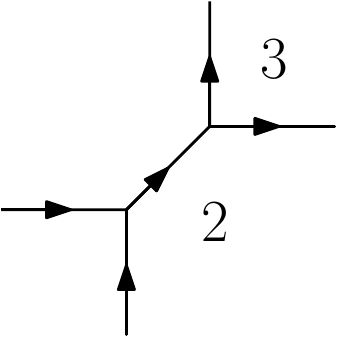}
\end{center}
\vspace{-20pt}
\end{wrapfigure}
To find the corresponding gluing diagram we need first to understand the truncation curves. Using (\ref{truncations2}) we see that $\mathcal{W}_\infty^{(1)}$ has $\lambda$-parameters compatible with $(0,2,3)$ truncation while as already mentioned for $\mathcal{W}_{\infty}^{(2)}$ we have the $(2,0,0)$ truncation. Since we want the same spin of the gluing matter as in the case of $\mathcal{N}=2$ SCA, we associate to $\mathcal{N}=2$ $\mathcal{W}_3$ the same conifold diagram with $\rho = \frac{1}{2}$. It is easy to check that this is compatible with (\ref{lambdafromdiag}) which confirms the identification of Romans $\mathcal{N}=2$ $\mathcal{W}_3$ times a commuting $U(1)$ factor with the algebra associated to this diagram.

\bibliography{Gluing}
\bibliographystyle{JHEP}

\end{document}